\documentclass[a4paper,12pt]{article}
\usepackage{jheppubm}

\usepackage[utf8]{inputenc}
\usepackage[english]{babel}
\usepackage{amsmath}
\usepackage{amsthm}
\usepackage{amsfonts}
\usepackage{amssymb}
\usepackage{bm}
\usepackage{amsbsy}
\usepackage{hyperref}
\usepackage{textcomp}
\usepackage{graphicx}
\usepackage{mathrsfs}
\usepackage{epsfig}
\usepackage{color}
\usepackage{euscript}
 \usepackage{multirow}
 \usepackage{ulem}

\DeclareMathOperator{\const}{const}

\newcommand{\nn}{\nonumber}
\newcommand{\be}{\begin{equation}}
\newcommand{\ee}{\end{equation}}
\newcommand{\bea}{\begin{eqnarray}}
\newcommand{\eea}{\end{eqnarray}}

\newcommand{\fb}{\mathfrak{b}}
\newcommand{\fc}{\mathfrak{c}}

\newcommand{\fg}{\mathfrak{g}}

\newcommand{\fm}{\mathfrak{m}}

\newcommand{\fz}{\mathfrak{z}}

\newcommand{\fF}{\mathfrak{F}}

\newcommand{\cL}{\cal L}
\newcommand{\cA}{\cal A}
\newcommand{\cS}{\cal S}

\definecolor{orange}{rgb}{0.89,0.32,0.05}

\definecolor{dgreen}{rgb}{0,0.6,0}

\definecolor{dblue}{rgb}{0., 0, 1}

\definecolor{dblue}{rgb}{0., 0, 1}
\definecolor{brown}{rgb}{0.64, 0.16, 0.16}

\numberwithin{equation}{section}

\title{
Spatial Wilson Loops and Energy Loss for Heavy Quarks in Magnetized HQCD Model
}

\author{Irina Ya. Aref'eva$^a$, Ali Hajilou$^{a,b}$, Kristina Rannu$^c$ and Pavel Slepov$^a$}

\affiliation{$^a$Steklov Mathematical Institute, Russian Academy of
  Sciences,\\ Gubkina str. 8, 119991, Moscow, Russia  \\
  $^b$School of Particles and Accelerators, Institute for Research in Fundamental \\ Sciences (IPM), P.O. Box 19395-5531, Tehran, Iran \\
  $^c$Peoples Friendship University of Russia,\\ Miklukho-Maklaya str. 6, 117198, Moscow, Russia
}

\emailAdd{arefeva@mi-ras.ru}
\emailAdd{hajilou@mi-ras.ru}
\emailAdd{rannu-ka@rudn.ru}
\emailAdd{slepov@mi-ras.ru}

\abstract{We investigate the effective potential and the string tension for the spatial Wilson loop (SWL) in hot dense QGP with two types of anisotropy, i.e. external magnetic field and spatial anisotropy, employing a holographic approach for the heavy quark model. In this approach, the string is extended in the 5th, holographic direction and has a turning point either on a dynamical wall (DW) configuration or on the horizon configuration in the 5th direction.  We obtain the magnetic catalysis behavior
for a phase transition between DW and horizon configuration of the string. The structure of the phase diagram does not depend on the boundary conditions choice for the dilaton field.  
Inclusion of the external magnetic field and spatial anisotropy enhance the string tension in the horizon configuration, namely drag force.
For the spatially isotropic case $\nu = 1$ at different magnetic field values the string tension is proportional to $T^2$ and is qualitatively consistent with lattice results. However, for the anisotropic case, $\nu = 4.5$, it deviates from the quadratic term. \\ 
}


\begin{document}

\maketitle


\newpage

\section{Introduction}

The holographic approach is a non-perturbative method for investigating the strongly coupled regime of quantum chromodynamics (QCD). There are two different classes of holographic models: ``top-down'' and ``bottom-up''. Holographic QCD is a suitable candidate for studying the quark-gluon plasma (QGP) observed in ultrarelativistic heavy-ion collisions (HIC) \cite{Casalderrey-Solana:2011dxg,Arefeva:2014kyw,DeWolfe:2013cua,Arefeva:2018bnh,Arefeva:2016rob,Arefeva:2020vhf}. In particular, the phase diagram of QCD has  received a lot of attention to study. In the ``bottom-up'' holographic approach the choice of the metric warp factor has a crucial influence on the phase transition structure of QCD \cite{Cai:2012xh,He:2013qq,Yang:2014bqa,Yang:2015aia,Fang:2015ytf,Chelabi:2015gpc,Li:2016smq,Arefeva:2018hyo,Fang:2018axm,Arefeva:2018cli,Chen:2018msc,Arefeva:2020byn}. This choice is different for the light- and heavy-quark models and should correspond to the lattice results known as the Columbia plot \cite{Brown:1990ev,Philipsen:2019rjq}. In \cite{Andreev:2006nw,Gursoy:2008za,Gursoy:2009jd,He:2010ye,Gursoy:2010fj,Colangelo:2011sr,Li:2012ay,Li:2013oda,Li:2014dsa,Chelabi:2015cwn,Li:2017tdz,Arefeva:2018jyu,Chen:2019rez,Li:2020hau} different isotropic and anisotropic models were proposed to study QCD using holography.\\

Taking into account the anisotropy in the holographic model \cite{Giataganas:2012zy,Giataganas:2017koz, BitaghsirFadafan:2017tci,Bohra:2019ebj,Bohra:2020qom,Giataganas:2013zaa,Finazzo:2016mhm,Arefeva:2019yzy,Asadi:2021nbd,Brehm:2017dmt,Arefeva:2023jjh,Arefeva:2022avn} is important because of QGP, which is spatially anisotropic after HIC. The first Maxwell field in the model supports the chemical potential in the boundary theory. In addition, to consider two types of anisotropy due to the external magnetic field and spatial anisotropy, two Maxwell fields need to be introduced into the model \cite{Arefeva:2023jjh}. The model that we use in this research was developed in \cite{Arefeva:2023jjh} with modified warp-factor. It produces the direct magnetic catalysis phenomenon for heavy quarks that is expected from lattice calculations. The Wilson loops calculations for the heavy-quarks model with a more simple warp-factor \cite{Arefeva:2020vae} was done in \cite{Arefeva:2020bjk}. The second and third Maxwell fields support the magnetic field and the spatial anisotropy defined by $\nu$ in the boundary theory, respectively. An anisotropic model with $\nu = 4.5$ \cite{Arefeva:2014vjl} has been shown to be able to reproduce the energy dependence of the total multiplicity of particles created in HIC \cite{ALICE:2015juo}. For the light-quark model one can see \cite{Arefeva:2020byn} and for the heavy-quark model \cite{Arefeva:2018hyo}.\\

The experiments in the relativistic HIC have addressed the possibility of new parameters in the QCD phase diagram. A strong magnetic field $eB\sim 0.3$ GeV$^2$ is created in the early stages of non-central HIC \cite{Skokov:2009qp,Bzdak:2011yy,Deng:2012pc}. Therefore, an external magnetic field can be considered the second anisotropy affecting the characteristics of QCD. Magnetized QCD has received a lot of attention in recent years. In particular, recently the magnetic field effect on the running coupling of QCD was investigated in \cite{Deur:2023dzc,Arefeva:2024xmg,Arefeva:2025okg}. Holographic models \cite{Gursoy:2017wzz,Gursoy:2020kjd,Arefeva:2020vae,Jena:2024cqs,Slepov:2024vjt,Li:2016gfn,Rougemont:2015oea,Shahkarami:2021gzl} studied the effect of magnetic field on the different properties of QCD.\\

Temporal Wilson loops and SWLs in the holographic approach have been used to study different properties of QCD \cite{Ageev:2016gtl,Golubtsova:2019zta,Bohra:2019ebj}. Temporal Wilson loops are used to investigate the confinement/deconfinement phase transition in QCD. The drag force acting on the heavy quark, moving in QGP, comes from its interaction with the plasma, that was investigated in holographic QCD models \cite{Herzog:2006gh,Gubser:2006bz,Casalderrey-Solana:2006fio}. For isotropic models the SWL string tension is proportional to the corresponding drag force \cite{Sin:2006yz,Andreev:2017bvr}. For anisotropic models this relation takes place with a small modification, see  \cite{Arefeva:2020vhf}. Therefore, one can
calculate SWL to investigate the drag forces for a heavy quark moving in QGP. The drag forces defining energy loss are studied intensively in holographic models \cite{Gubser:2006qh,Gursoy:2009kk,Chernicoff:2012iq,Fadafan:2012qu,Rougemont:2015wca,Zhu:2019ujc}.\\

In this research we consider the Einstein-three-Maxwell-dilaton action, set up in \cite{Arefeva:2023jjh}, to investigate the effective potential, string tension and to compare our results on the string tension with the drag force and lattice calculations. This model possesses two types of anisotropy, i.e. spatial anisotropy and external magnetic field. We study the effect of these anisotropies on the phase transition temperature obtained by Wilson loops. To do so, we considered  SWL in three different orientations, i.e. ${\cal{W}}_{xY_{1}}$, ${\cal{W}}_{xY_{2}}$, and ${\cal{W}}_{y_{1}Y_{2}}$, referred to as the first, second, and third orientations, respectively.\\

This paper is organized as follows. In Sect.~\ref{Sect:backg} we introduce a fully anisotropic 5-dimensional holographic set up for heavy quarks model. Sect.~\ref{Sect:SWL} presents the Born-Infeld action to get the DW equations. In Sect.~\ref{Sect:solution} we obtain numerical results for particular SWL orientations. Sect.~\ref{drag} investigates the connection between the string tension and the drag force and compares it with lattice results. In Sect.~\ref{Sect:conclusion} we summarize our results. The paper is complemented by an appendix~\ref{appendixA}, which describes the explicit forms of the equations of motion (EOMs). In appendix~\ref{appendixB} arbitrary orientations of the Wilson loop are studied, and in appendix~\ref{appendixC} the string tension at the DW and horizon configurations are derived.

\newpage

\section{Holographic set up for heavy quarks}
\label{Sect:backg}

In this section we consider the background with two types of anisotropy, i.e. spatial anisotropy and external magnetic field, that are found in \cite{Arefeva:2023jjh}, and the thermodynamical properties of this model.

\subsection{Background}
\label{Subsect:Model}

We consider the Lagrangian with three Maxwell fields in Einstein frame, used in \cite{Arefeva:2023jjh}:
\begin{gather}
  {\cL} = \sqrt{-\fg} \left[ R 
    - \cfrac{f_0(\phi)}{4} \, F_0^2 
    - \cfrac{f_1(\phi)}{4} \, F_1^2
    - \cfrac{f_3(\phi)}{4} \, F_3^2
    - \cfrac{1}{2} \, \partial_{\mu} \phi \, \partial^{\mu} \phi
    - V(\phi) \right], \label{eq:2.01} 
\end{gather}
where $\fg$ is the determinant of the metric tensor, $R$ is Ricci scalar, $\phi$ is the dilaton field,  $V(\phi)$ is the dilaton field potential, and  $f_0(\phi)$, $f_1(\phi)$ and $f_3(\phi)$ are the coupling functions associated with stresses $F_0$, $F_1$ and $F_3$ of Maxwell fields, respectively. The electromagnetic tensor is defined as $F_{\rho\sigma} = \partial_{\rho} A_{\sigma} -\partial_{\sigma} A_{\rho}$ and the indexes $\rho$ and $\sigma$ numerate the spacetime coordinates $(t, x_1, x_2, x_3, z)$, with $z$ being the radial holographic coordinate. Maxwell fields $F_0$, $F_1$ and $F_3$ are considered the first, the second, and the third, respectively.

We take the following ansatz for the metric \cite{Arefeva:2023jjh}, that in the Einstein frame is 
\begin{gather}
      ds^2 = \cfrac{L^2}{z^2} \ \fb(z) \left[
    - \ g(z) dt^2 + dx^2 + \left(
      \cfrac{z}{L} \right)^{2-\frac{2}{\nu}} \hspace{-5pt} dy_1^2
    + e^{c_B z^2} \left( \cfrac{z}{L} \right)^{2-\frac{2}{\nu}}
    \hspace{-5pt} dy_2^2
    + \cfrac{dz^2}{g(z)} \right], \label{eq:2.02} \\
  \fb(z) = e^{{2{\cA}(z)}}, \label{eq:2.03}
\end{gather}
and for the matter fields:
\begin{gather}
  \phi = \phi(z), \label{eq:2.04} \\
  \begin{split}
    \mbox{ $F_0$ -- electric anzats } \quad
    & A_0 = A_t(z), \quad A_{i} = 0, \ i = 1,2,3,4, \\
    \mbox{$F_k$ -- magnetic ansatz } \quad
    &F_1 = q_1 \, dy^1 \wedge dy^2, \quad 
    F_3 = q_3 \, dx \wedge dy^1, 
  \end{split}\label{eq:2.05}
\end{gather}
where $L$ is the AdS-radius, $\fb(z)$ is the warp factor in Einstein frame, that is specified by a scale factor ${\cA}(z)$, $g(z)$ is the blackening function, and $\nu$ is
the parameter of primary anisotropy created by the non-symmetry of HIC, such that $\nu = 1$ corresponds to spatially isotropic background and choice $\nu = 4.5$ for anisotropic background is in agreement with the experimental data on the energy dependence of the total multiplicity of particles produced in HIC \cite{ALICE:2015juo}.
The coefficient of secondary anisotropy $c_B$ is related to the external magnetic field $F_3$. Different choices of ${\cA}(z)$ determine the description of the model with heavy/light quarks. For heavy quarks we considered ${\cA}(z) = - \, \fc z^2/4$  \cite{Arefeva:2018hyo,Arefeva:2018cli,Arefeva:2020vae} and for light quarks ${\cA}(z) = - \, a \, \ln (b z^2 + 1)$ \cite{Arefeva:2020byn}. 
In this research our choice of ${\cA}(z)$  for the heavy quarks model is \cite{Arefeva:2023jjh}
\be \label{warpnew}
{\cA}(z) =  - \, c z^2/4
    \, - (p-c_B \, q_3) z^4\,,
\ee
where $c = 4 R_{gg}/3$, $R_{gg} = 1.16$ GeV$^2$, $c=1.547$ GeV$^2$, and $p = 0.273$ GeV$^4$ are parameters, that can be fixed with lattice and experimental data for zero magnetic field, i.e. for $c_B = 0$ case \cite{Yang:2015aia}.

Varying the Lagrangian 
(\ref{eq:2.01}) by using the ansatz (\ref{eq:2.02})-(\ref{eq:2.05}) leads to the EOMs for Einstein, dilaton and Maxwell fields, that can be found in appendix~\ref{appendixA} (see equations \eqref{eq:2.17}-\eqref{eq:2.22}).
The EOMs are solved with the boundary conditions
\begin{gather}
  A_t(0) = \mu, \quad A_t(z_h) = 0, \label{eq:4.24} \\
  \hspace{7pt} g(0) = 1, \hspace{20pt} g(z_h) = 0. \label{eq:4.25} 
\end{gather} 
For the dilaton field, different boundary conditions can be used. We denoted the boundary condition for $\phi(z,z_0)$, so that
\be\label{phi-z0}
  \phi(z,z_0)\Big|_{z=z_0} = 0.
\ee
We consider two types of boundary conditions  by taking different $z_0$ \cite{Arefeva:2020byn,Slepov:2021gvl,Arefeva:2024vom,Arefeva:2024poq,Arefeva:2025xtz}. The zero-boundary condition is chosen as 
\bea \label{zerobc}
  z_0 = \epsilon, 
\eea 
where $0 < \epsilon \ll 1$, and another boundary condition is 
\bea
  z_0 = \fz(z_h), \label{bch}
\eea
where $\fz (z_h)$ is a smooth function of $z_h$, different for light and heavy quarks.
For heavy quarks, we take 
\be\label{phi-fz-LQ}
  z_0 = \fz(z_h) = e^{(-\frac{z_h}{4})}+0.1 \,,
\ee
known as a physical-boundary condition, for more details, see \cite{Arefeva:2020byn,Slepov:2021gvl,Arefeva:2024vom,Arefeva:2024poq,Arefeva:2025xtz}.
We note that in this paper we considered the zero-boundary conditions, i.e. \eqref{zerobc} in our calculations, and the physical-boundary condition \eqref{phi-fz-LQ} has just been used in subsection \ref{mbc}.

\subsection{Thermodynamics}
\label{SubSect:thermo}

To investigate thermodynamics we use the following formulas for temperature and entropy  
\begin{gather}
  T = \cfrac{\sqrt{{g_{tt}}' \, {g^{zz}}'}}{4 \pi} \, \Bigl|_{z=z_h}
  = \cfrac{\sqrt{{g_{00}}' \, {g^{44}}'}}{4 \pi} \,
  \Bigl|_{z=z_h} = \cfrac{|g'(z)|}{4 \pi} \, \Biggl|_{z=z_h}, \label{eq:4.33} \\
  s = \cfrac{\sqrt{g_{xx} \, g_{y_1y_1} \, g_{y_2y_2}}}{4 G_5} \,
  \Bigl|_{z=z_h} = \cfrac{\sqrt{g_{11} \, g_{22} \, g_{33}}}{4 G_5} \,
  \Bigl|_{z=z_h}\,, \label{eq:4.34}
\end{gather}
where $G_5$ is the gravitational constant that in our calculations we set $8\pi G_5=1$ and $g(z)$ is the blackening function given by
\begin{gather}
  g(z) = e^{c_B z^2} \left[ 1 - \cfrac{\Tilde{I}_1(z)}{\Tilde{I}_1(z_h)}
    + \cfrac{\mu^2 \bigl(2 R_{gg} + c_B (q_3 - 1) \bigr)
      \Tilde{I}_2(z)}{L^2 \left(1 - e^{(2 R_{gg}+c_B(q_3-1))\frac{z_h^2}{2}}
      \right)^2} \left( 1 - \cfrac{\Tilde{I}_1(z)}{\Tilde{I}_1(z_h)} \,
      \cfrac{\Tilde{I}_2(z_h)}{\Tilde{I}_2(z)} \right)
  \right], \label{eq:4.42} \\
  \Tilde{I}_1(z) = \int_0^z
  e^{\left(2R_{gg}-3c_B\right)\frac{\xi^2}{2}+3 (p-c_B \, q_3) \xi^4}
  \xi^{1+\frac{2}{\nu}} \, d \xi, \qquad \ \label{eq:4.43-1} \\
  \Tilde{I}_2(z) = \int_0^z
  e^{\bigl(2R_{gg}+c_B\left(\frac{q_3}{2}-2\right)\bigr)\xi^2+3 (p-c_B
    \, q_3) \xi^4} \xi^{1+\frac{2}{\nu}} \, d \xi. \label{eq:4.43} 
\end{gather}
Using the metric (\ref{eq:2.04}) and the scale factor (\ref{warpnew}) with  equations (\ref{eq:4.33}) and (\ref{eq:4.34}) one can
obtain the temperature and entropy:
\begin{gather}
  \begin{split}
    T &=  \left|
     - \, \cfrac{e^{(2R_{gg}-c_B)\frac{z_h^2}{2}+3 (p-c_B \, q_3)
         z_h^4} \, 
      z_h^{1+\frac{2}{\nu}}}{4 \pi \, \Tilde{I}_1(z_h)} \right. \times \\
    &\left. \times \left[ 
      1 - \cfrac{\mu^2 \bigl(2 R_{gg} + c_B (q_3 - 1) \bigr) 
        \left(e^{(2 R_{gg} + c_B (q_3 - 1))\frac{z_h^2}{2}}\Tilde{I}_1(z_h) -
          \Tilde{I}_2(z_h) \right)}{L^2 \left(1 
          - e^{(2R_{gg}+c_B(q_3-1))\frac{z_h^2}{2}}
        \right)^2} \right] \right|, 
    \\
    s &=  \cfrac{1}{4 G_5} \left( \cfrac{L}{z_h} \right)^{1+\frac{2}{\nu}}
    e^{-(2R_{gg}-c_B)\frac{z_h^2}{2}-3 (p-c_B \, q_3) z_h^4} \,.
  \end{split} \label{eq:4.48}
\end{gather}

\begin{figure}[h!]
  \centering
  \includegraphics[scale=0.42]{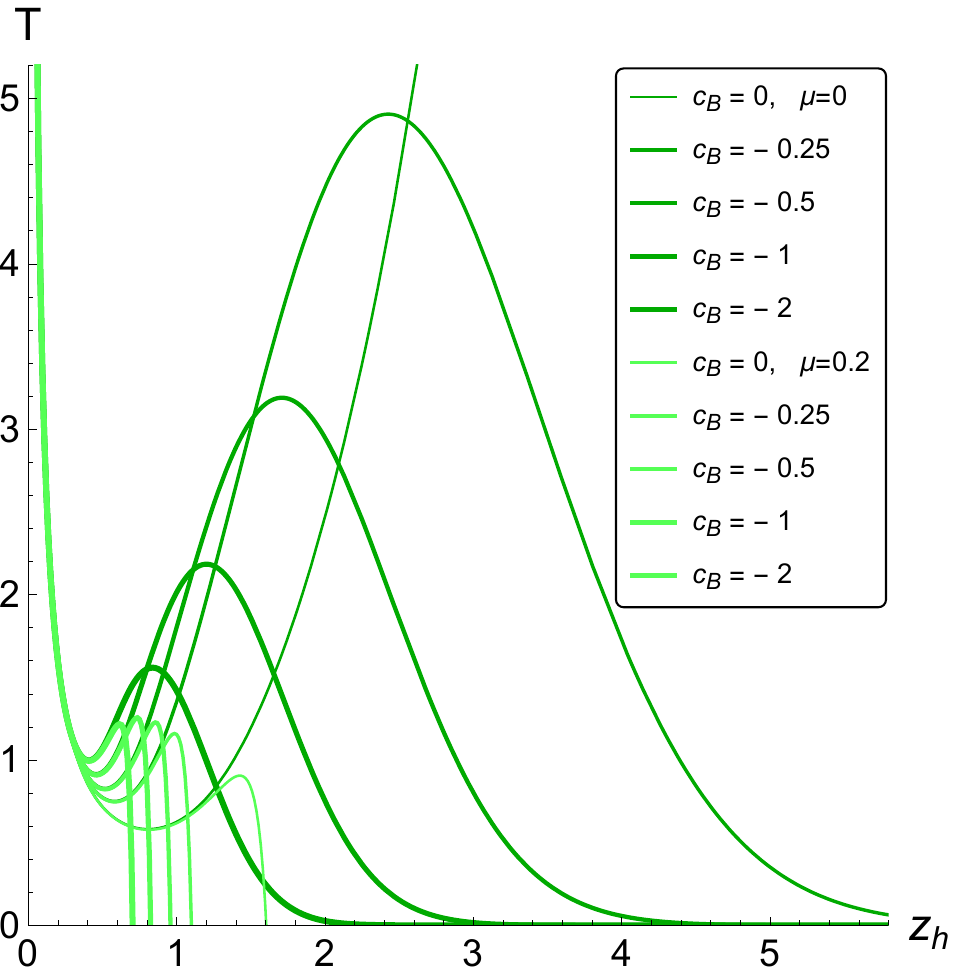} \qquad
  \includegraphics[scale=0.42]{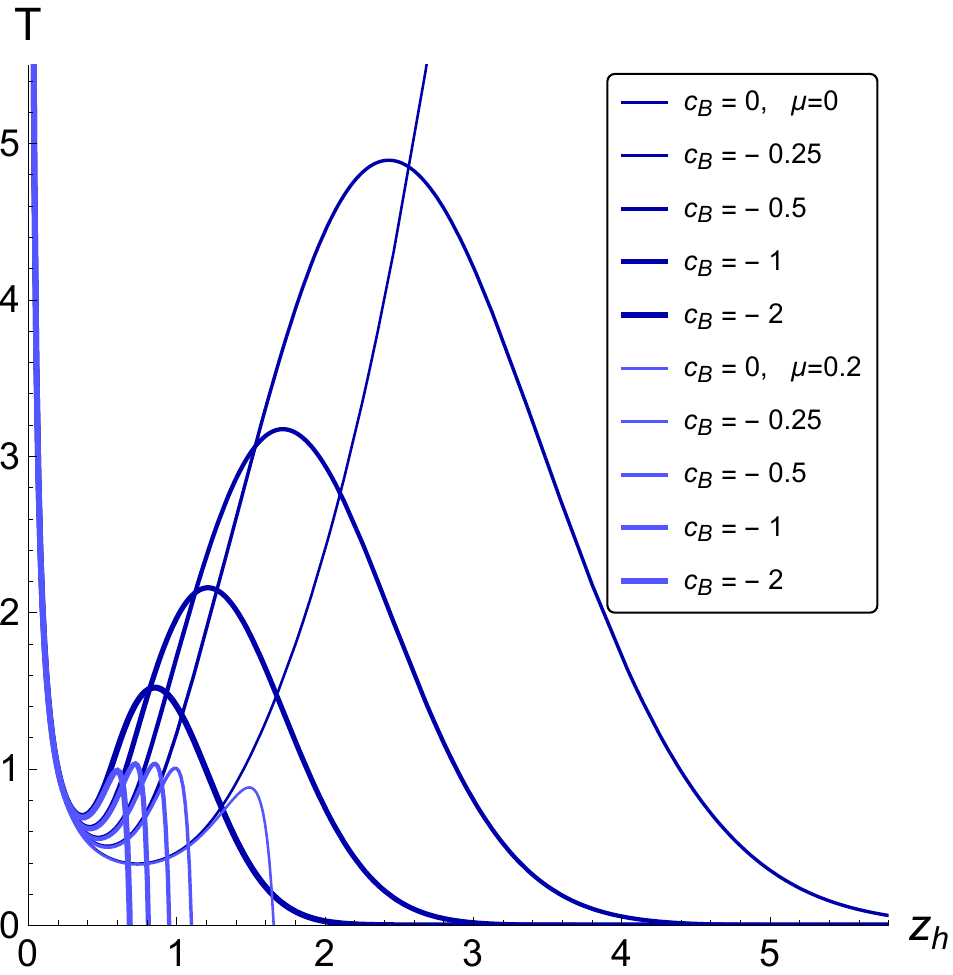} \\
  A \hspace{70mm} B \\ \ \\
  \includegraphics[scale=0.42]{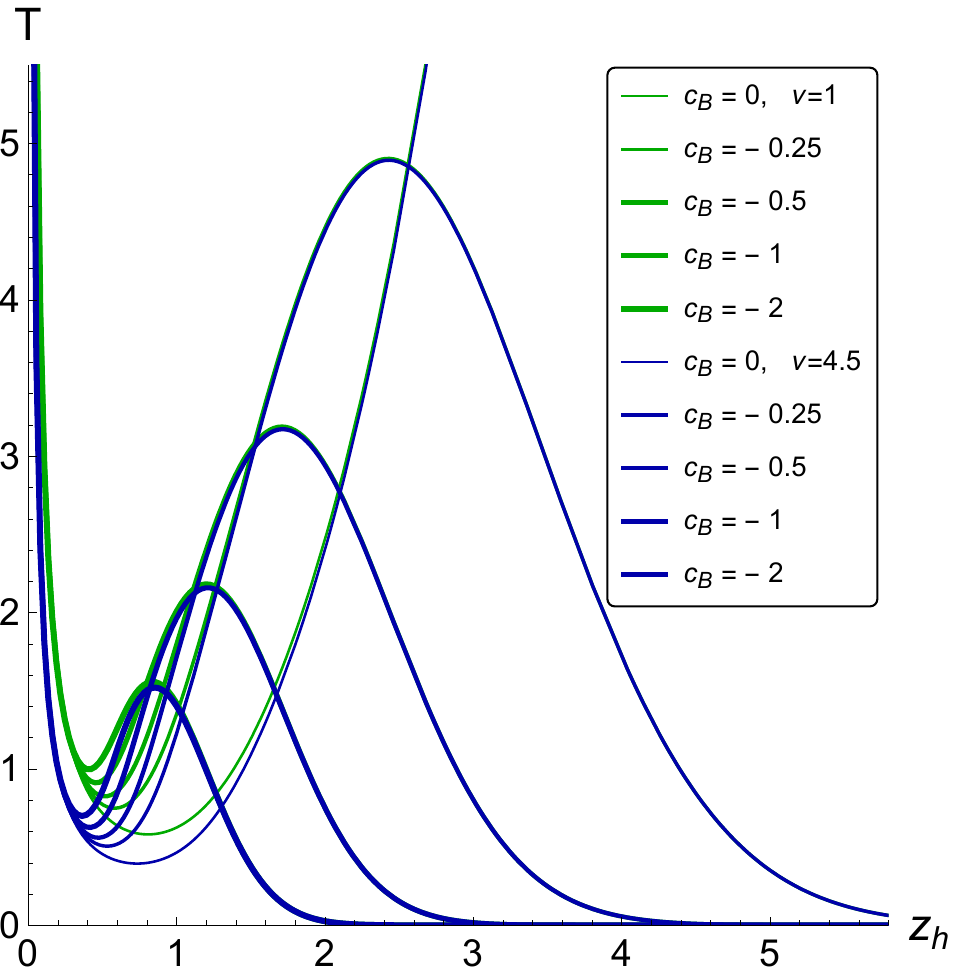} \qquad
  \includegraphics[scale=0.42]{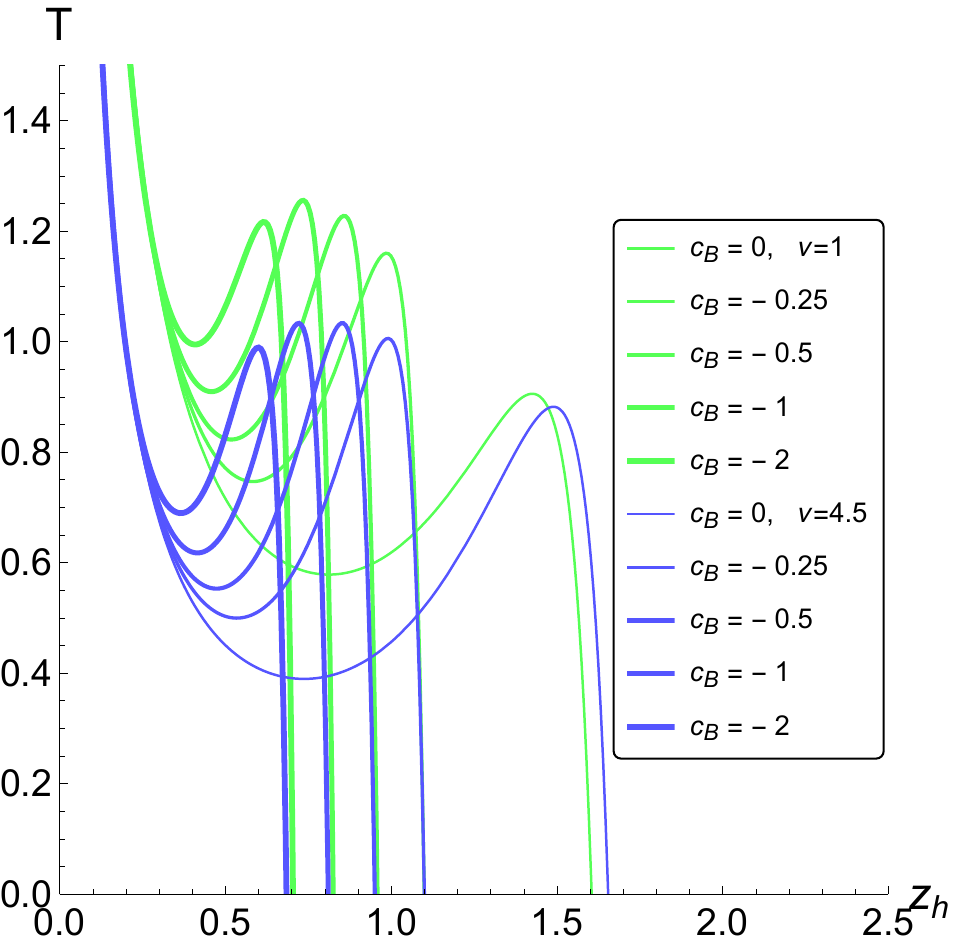} \\
  C \hspace{70mm} D
  \caption{Temperature as a function of horizon $T(z_h)$ for (A) different $\mu$ at fixed $\nu=1$, (B) $\nu = 4.5$, and (C) different $\nu$ and $c_B$ at fixed $\mu=0$, (D) $\mu=0.2$ GeV; $q_3 = 5$, $R_{gg} = 1.16$ GeV${}^2$, $p = 0.273$ GeV${}^4$; $[T]=[\mu]= [z_h]^{-1} = [c_B]^{\frac{1}{2}}=$~GeV.\\
  }
  \label{Fig:Temp1}
\end{figure}

In Fig.~\ref{Fig:Temp1} the temperature as a function of horizon $T(z_h)$ for different $\mu$ at fixed $\nu=1$ (A), $\nu = 4.5$ (B), and different $\nu$ and $c_B$ at fixed $\mu=0$ (C), $\mu=0.2$ GeV (D) is plotted. Figs.~\ref{Fig:Temp1}A and \ref{Fig:Temp1}B
show, that the temperature of the Hawking-Page like phase transition $T_{HP}$ is lower for larger $\mu$ in both isotropic and anisotropic cases.
Figs.~\ref{Fig:Temp1}C and \ref{Fig:Temp1}D show, that primary anisotropy $\nu$ decreases $T_{HP}$ for both zero and non-zero values of chemical potential.
 
\newpage
\,
\newpage

\section{Wilson loops and effective potential}
\label{Sect:SWL}

\subsection{Born-Infeld type action
}
\label{Sect:BI}

The actions that are considered for SWL, temporal Wilson loop, light-like Wilson loop and holographic entanglenment entropy \cite{Arefeva:2018bnh, Arefeva:2020uec, Arefeva:2018cli, Arefeva:2025uym, Arefeva:2020bjk, Slepov:2025uxg} are the specific cases of the Born-Infeld action
\be
\label{BI}
  {\cal S} = \int _{-\ell/2}^{\ell/2} M\bigl(z(\xi)\bigr)
  \sqrt{{\cal {\cal F}}\bigl(z(\xi)\bigr) + \bigl(z^{\prime}(\xi)\bigr)^ 2} \, d\xi,
\ee
where $\ell$ is the interquark distance, $\xi$ denotes the ``time", and ${\cal F}(z)$ is an arbitrary function. The  action \eqref{BI} describes the dynamical system with a dynamic variable $z =z(\xi)$. The associated  effective potential is
\be
\label{EfPot}
  {\cal V}\bigl(z(\xi)\bigr) \equiv M\bigl(z(\xi)\bigr) \sqrt{{\cal F}\bigl(z(\xi)\bigr)}.
\ee

After some algebra (for more detail see appendix~\ref{appendixC}) one can obtain the action $\cal S$ and the length $\ell$ for DW and horizon configurations. The string tension can be obtained for two different configurations:
\begin{itemize}
  \item the DW configuration
  \bea
    \sigma_{DW} = M(z_{DW}) \, \sqrt{{\cal F}(z_{DW})}\,, \label{ten}
  \eea
  \item the horizon configuration
  \be
    \sigma_{z_h} = \fm(z_h)\,\sqrt{\fF(z_h)}\,,
  \ee
\end{itemize}
where $\sigma_{DW}$ and $\sigma_{z_h}$ are the spatial string tension in the DW and horizon configurations. We will investigate the temperature dependence of the string tension for two configurations and three WL orientations in Sect.~\ref{Sect:solution}.

\subsection{Dynamical wall equations}\label{SubSect:DW}

Obtaining general results for spatial Wilson loops and effective potential, we consider a fully anisotropic background
\bea
  ds^2 = G_{\mu\nu}dx^{\mu}dx^{\nu} = \cfrac{L^2 \fb_s(z)}{z^2} \left[
    - \, g(z) dt^2 + \fg_1 dx^2 + \fg_2  dy_1^2 + \fg_3  dy_2^2 +
    \cfrac{dz^2}{g(z)} \right] \label{Gbackgr}
\eea 
and use it to calculate SWLs for different orientations. Here $\fb_s(z) = \fb(z)e^{\sqrt{\frac{2}{3}}\phi(z)}$ is the warp factor in the string frame, and $\fg_1$, $\fg_2$, $\fg_3$ are anisotropic functions depending on $z$.

To get the DW equations, let us take the special cases of \eqref{S_swl} and \eqref{V_swl}, considering the particular orientations of the SWLs.
\begin{itemize}
  \item The first orientation ($xY_{1}$-plane): \quad
  $\phi=0$, $\theta=0$, $\psi=0$; \\
  $a_{11}=a_{22}=a_{33}=1$, $a_{12}=a_{21}=a_{31}=a_{31}=a_{32}=a_{23}=0$:
  \be
    {\cal S}_{xY_{1}}=\frac{1}{2\pi \alpha'} \int_{\cal P}\left(\frac{L^2 \fb_s}{z^2}\right)
    \sqrt{\left(\fg_{1}\fg_{2}+\frac{z'^{2}}{g}\,\fg_{2}\right)} \
    d\xi^{1}d\xi^{2},
  \ee
  \be
    {\cal V}_{1}\equiv\,
    {\cal V}_{xY_{1}}(z(\xi)) = \frac{1}{2\pi \alpha'} \left(\frac{L^2 \fb_s}{z^2}\right) \sqrt{\fg_{1}\fg_{2}}.
  \ee
  \item The first orientation ($Xy_{1}$-plane) \footnote{The Nambu-Goto action for $Xy_{1}$-plane coincides with Nambu-Goto action for  $xY_{1}$-plane.}: \quad
  $\phi=\pi/2$, $\theta=0$, $\psi=0$; \\
  $a_{21}=a_{33}=-\,a_{12}=1$, $a_{11}=a_{13}=a_{22}=a_{23}=a_{31}=a_{32}=0$:
  \be
    {\cal S}_{Xy_{1}}=\frac{1}{2\pi \alpha'} \int_{\cal P} \left(\frac{L^2 \fb_s}{z^2}\right)\sqrt{\left(\fg_{1}\fg_{2}+\frac{z'^{2}}{g}\,\fg_{1}\right)}d\xi^{1}d\xi^{2},
  \ee
  \be
    {\cal V}_{1}\equiv\,
    {\cal V}_{Xy_{1}}(z(\xi))=\frac{1}{2\pi \alpha'} \left(\frac{L^2 \fb_s}{z^2}\right) \sqrt{\fg_{1}\fg_{2}}.
  \ee
  \item The second orientation ($xY_{2}$-plane): \quad
  $\phi=0$, $\theta=\pi/2$, $\psi=0$; \\
  $a_{11}=-\,a_{23}=a_{32}=1$, $a_{12}=a_{13}=a_{21}=a_{22}=a_{31}=a_{33}=0$:
  \be
    {\cal S}_{xY_{2}}=\frac{1}{2\pi \alpha'} \int_{\cal P} \left(\frac{L^2 \fb_s}{z^2}\right)
    \sqrt{\left(\fg_{1}\fg_{3}+\frac{z'^{2}}{g}\,\fg_3\right)}d\xi^{1}d\xi^{2},
  \ee
  \be
    {\cal V}_{2}\equiv\,
    {\cal V}_{xY_{2}}(z(\xi))=\frac{1}{2\pi \alpha'} \left(\frac{L^2 \fb_s}{z^2}\right) \sqrt{\fg_{1}\fg_{3}}.
  \ee
  \item The third orientation ($y_{1}Y_{2}$-plane): \quad
  $\phi=\pi/2$, $\theta=\pi/2$, $\psi=-\,\pi/2$; \\
  $a_{22}=a_{31}=-a_{13}=1$, $a_{11}=a_{12}=a_{21}=a_{23}=a_{32}=a_{33}=0$:
  \be
    {\cal S}_{y_{1}Y_{2}}=\frac{1}{2\pi \alpha'} \int _{\cal P} 
    \left(\frac{L^2 \fb_s}{z^2}\right) \sqrt{\left(\fg_{2}\fg_{3}+\frac{z'^{2}}{g}\,\fg_2\right)}d\xi^{1}d\xi^{2},
  \ee
  \be
    {\cal V}_{3}\equiv\,
    {\cal V}_{y_{1}Y_{2}}(z(\xi))=\frac{1}{2\pi \alpha'} \left(\frac{L^2 \fb_s}{z^2}\right) \sqrt{\fg_{2}\fg_{3}}.
  \ee
\end{itemize}
Note, that the capital letters $Y_{1}$, $X$, $Y_{2}$ in the WL orientations indicate the direction the loop contour has an infinite length along. 

The general form of the DW equation, if DW exists, is given by ${\cal V}'(z)=0$ \cite{Arefeva:2016rob,Andreev:2006ct}. DW coordinates $z_{DW}$ for SWL in particular orientations are given by
\bea
  {\cal DW}_{xY_1} = {\cal DW}_{Xy_1} &\equiv&
  \frac{2 \fb_s'(z)}{\fb_s(z)} + \frac{\fg_1'(z)}{\fg_1(z)} 
  + \frac{\fg_2'(z)}{\fg_2(z)} - \frac{4}{z}\Bigg|_{z = z_{DW}}
  \hspace{-15pt} = 0, \\
  {\cal DW}_{xY_2} &\equiv&
  \frac{2 \fb_s'(z)}{\fb_s(z)} + \frac{\fg_1'(z)}{\fg_1(z)} +
  \frac{\fg_3'(z)}{\fg_3(z)} - \frac{4}{z}\Bigg|_{z = z_{DW}}
  \hspace{-15pt} = 0,\\
  {\cal DW}_{y_1Y_2} &\equiv&
  \frac{2 \fb_s'(z)}{\fb_s(z)} + \frac{\fg_2'(z)}{\fg_2(z)} +
  \frac{\fg_3'(z)}{\fg_3(z)} - \frac{4}{z}\Bigg|_{z = z_{DW}}
  \hspace{-15pt} = 0.
\eea
From the metric \eqref{eq:2.02} 
\be
 \fg_1(z) = 1\,,~~ \fg_2(z) = (z/L)^{2-2/\nu}, \, ~~\fg_3(z) = (z/L)^{2-2/\nu}e^{c_Bz^2}\,.
\ee
In this case spatial string tensions for three particular orientations, i.e. $xY_1$, $xY_2$, and $y_1Y_2$, are given by
\bea\label{sigmaxY1}
  \sigma_1 \equiv \sigma_{xY_1}
  &=& \frac{1}{2\pi \alpha'}\left(\frac{L^2 \fb_s(z)}{z^2}\right)
  \sqrt{\fg_{1}\fg_{2}}=\frac{1}{2\pi \alpha'}\left(\frac{L^{1+1/\nu}
    \fb_s(z)}{z^{1+1/\nu}}\right),\\
  \sigma_{2} \equiv \sigma_{xY_2} 
  &=& \frac{1}{2\pi \alpha'}\left(\frac{L^2 \fb_s(z)}{z^2}\right)
  \sqrt{\fg_{1}\fg_{3}}=\frac{1}{2\pi \alpha'}\left(\frac{L^{1+1/\nu}
    \fb_s(z)}{z^{1+1/\nu}}\right)e^{c_Bz^2/2},
    \label{sigmaxY2}\\
  \sigma_{3} \equiv \sigma_{y_1Y_2}
  &=& \frac{1}{2\pi \alpha'}\left(\frac{L^2 \fb_s(z)}{z^2}\right)
  \sqrt{\fg_{2}\fg_{3}}=\frac{1}{2\pi \alpha'}\left(\frac{L^{2/\nu}
    \fb_s(z)}{z^{2/\nu}}\right)e^{c_Bz^2/2},
  \label{sigmayY2}
\eea
where $z = z_h$ or $z = z_{DW}$.


\newpage
\section{Numerical results}\label{Sect:solution}

We study the position of the DW for effective potentials to investigate the spatial string tension and  phase transition of string at different configurations. The effective potentials depend on the orientation. For particular orientations $xY_1$, $xY_2$ and $y_1Y_2$ the effective potentials denoted by ${\cal V}_1$,  ${\cal V}_2$ and ${\cal V}_3$ are given by equations \eqref{sigmaxY1}, \eqref{sigmaxY2} and \eqref{sigmayY2}, respectively. Although the effective potential depends on the parameters $c$, $z_0$, $z_h$, and  $c_B$, it does not depend on the chemical potential $\mu$.

\subsection{Spatial Wilson loop ${\cal{W}}_{xY_{1}}$
} \label{swl-2bc}
\subsubsection{Zero-boundary condition}\label{swl-0bc}

\begin{figure}[h!]
  \centering
  \includegraphics[scale=0.39]{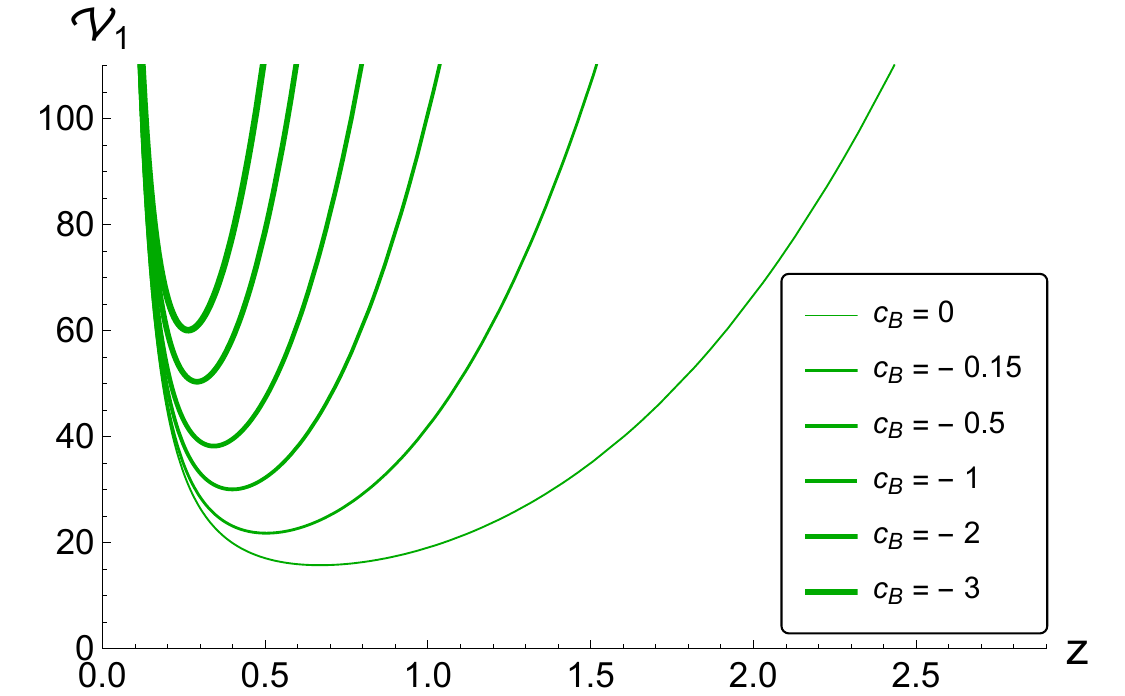} 
  \includegraphics[scale=0.39]{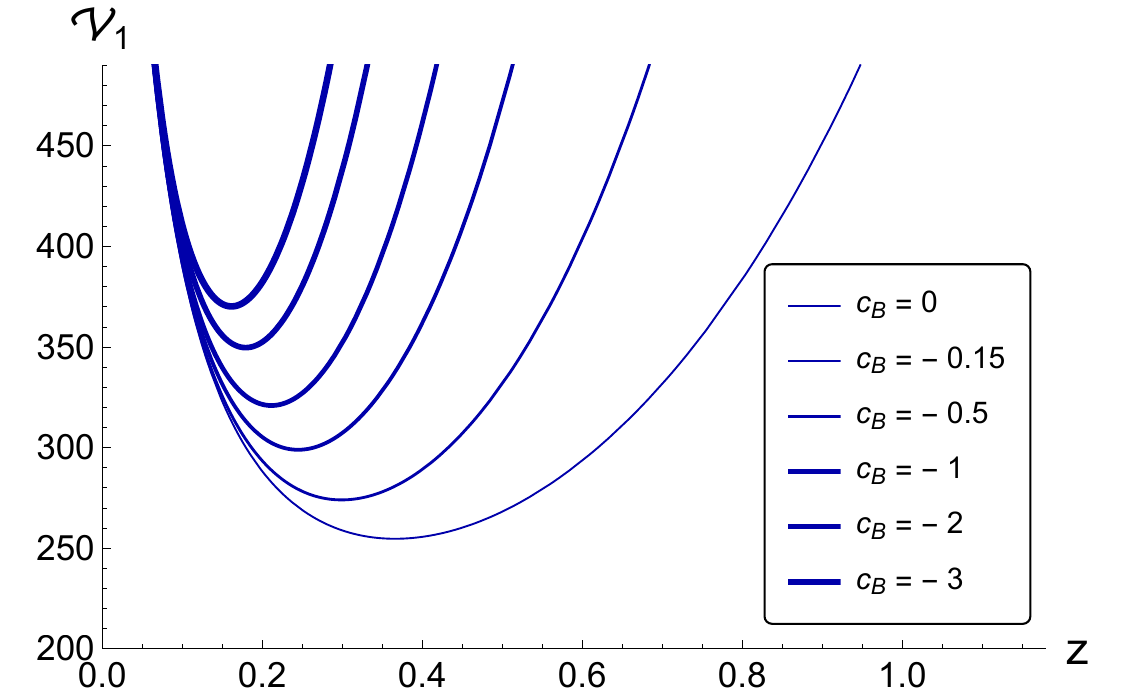} \\
  A \hspace{70mm} B
  \caption{Effective potential ${\cal V}_1(z)$ in the first orientation ${\cal{W}}_{x Y_{1}}$ at (A) $\nu = 1$, and (B) $\nu = 4.5$ for different $c_B$ considering zero-boundary condition \eqref{zerobc}; $[z]^{-1} = [c_B]^{\frac{1}{2}} =$ GeV.
  }
  \label{Fig:V1}
\end{figure}

In this section the behavior of the effective potential ${\cal V}_1$ as a function of the holographic coordinate $z$ corresponding to the first orientation of SWL, i.e. ${\cal{W}}_{xY_{1}}$, for different values of magnetic coefficients $c_B$ is presented in 
Fig.~\ref{Fig:V1} using the zero-boundary condition \eqref{zerobc}. Fig.~\ref{Fig:V1} shows that ${\cal V}_1$ has local minimums, thus confirming ${\cal V}'(z) = 0$, i.e. the  dynamical walls exist in the isotropic case $\nu=1$ and in the anisotropic case $\nu=4.5$ with different values of the magnetic coefficient $c_B$. 

The locations of the dynamical walls for  $\nu=1$ and $\nu=4.5$ with different values of $c_B$ and are given in the Tables~\ref{tab:V1nu1} and~\ref{tab:V1nu45}, respectively.

\begin{table}[h]
  \centering
  \begin{tabular}{|l|c|l|l|l|l|l|l|}
    \hline
    \multicolumn{1}{|c|}{${\cal V}_1$} & \multicolumn{6}{c|}{$\nu=1$} \\
    \hline
    $-\,c_B$ & 0 & 0.15 & \multicolumn{1}{c|}{0.5} & \multicolumn{1}{c|}{1} & \multicolumn{1}{c|}{2}& \multicolumn{1}{c|}{3} \\
    \hline
    $z_{DW}$ & \multicolumn{1}{l|}{0.665} & 0.500 & 0.400 & 0.341 & 0.285 & 0.261 \\ 
    \hline
  \end{tabular}
  \caption{Locations of the dynamical walls $z_{DW}$ for ${\cal V}_1$ in the first orientation ${\cal{W}}_{x Y_{1}}$ at $\nu = 1$, considering zero-boundary condition \eqref{zerobc}; $[z]^{-1} = [c_B]^{\frac{1}{2}} =$ GeV.
  }
  \label{tab:V1nu1}
\end{table}

\begin{table}[h]
  \centering
  \begin{tabular}{|l|c|l|l|l|l|l|l|}
    \hline
    \multicolumn{1}{|c|}{${\cal V}_1$} & \multicolumn{6}{c|}{$\nu=4.5$} \\ 
    \hline
    $-\,c_B$ & 0 & 0.15 & \multicolumn{1}{c|}{0.5} & \multicolumn{1}{c|}{1} & \multicolumn{1}{c|}{2}& \multicolumn{1}{c|}{3} \\
    \hline
    $z_{DW}$ & \multicolumn{1}{l|}{0.366} & 0.300 & 0.245 & 0.212 & 0.180 & 0.162 \\ \hline
  \end{tabular}
  \caption{Locations of the dynamical walls $z_{DW}$ for ${\cal V}_1$ in the first orientation ${\cal{W}}_{x Y_{1}}$ at $\nu = 4.5$, considering zero-boundary condition \eqref{zerobc}; $[z]^{-1} = [c_B]^{\frac{1}{2}} =$ GeV.
  }
  \label{tab:V1nu45}
\end{table}

\begin{figure}[h!]
  \centering
  \includegraphics[scale=0.42]{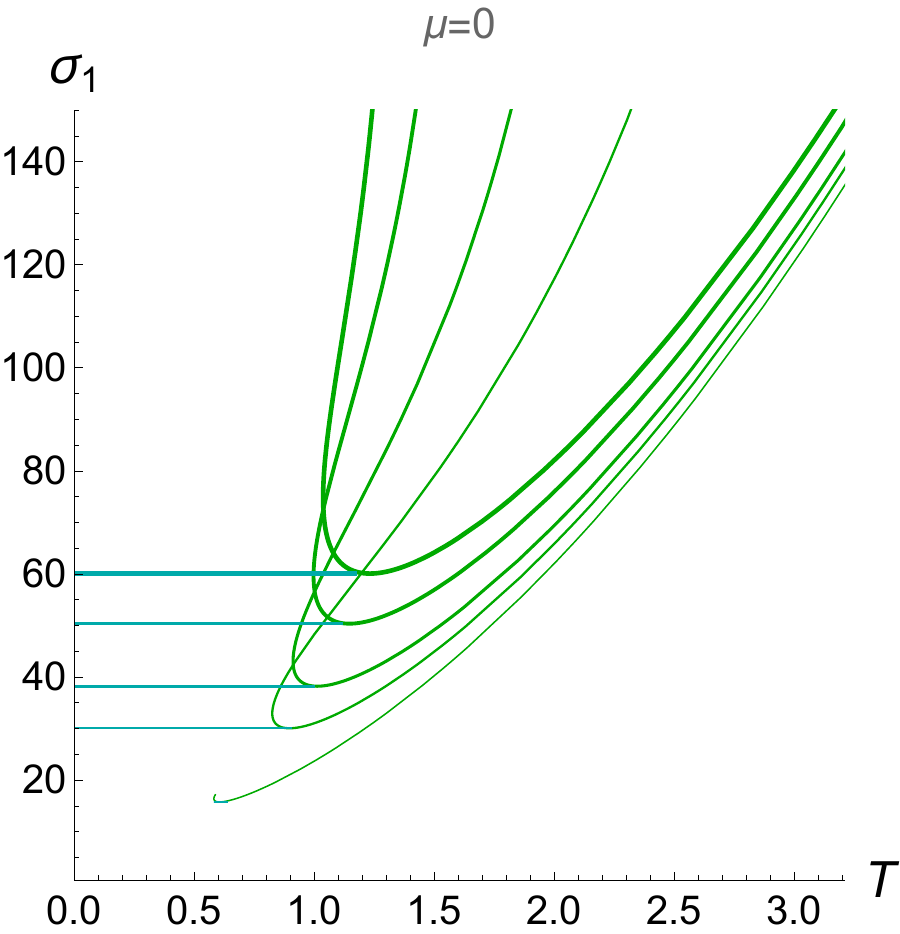}\qquad
  \includegraphics[scale=0.38]{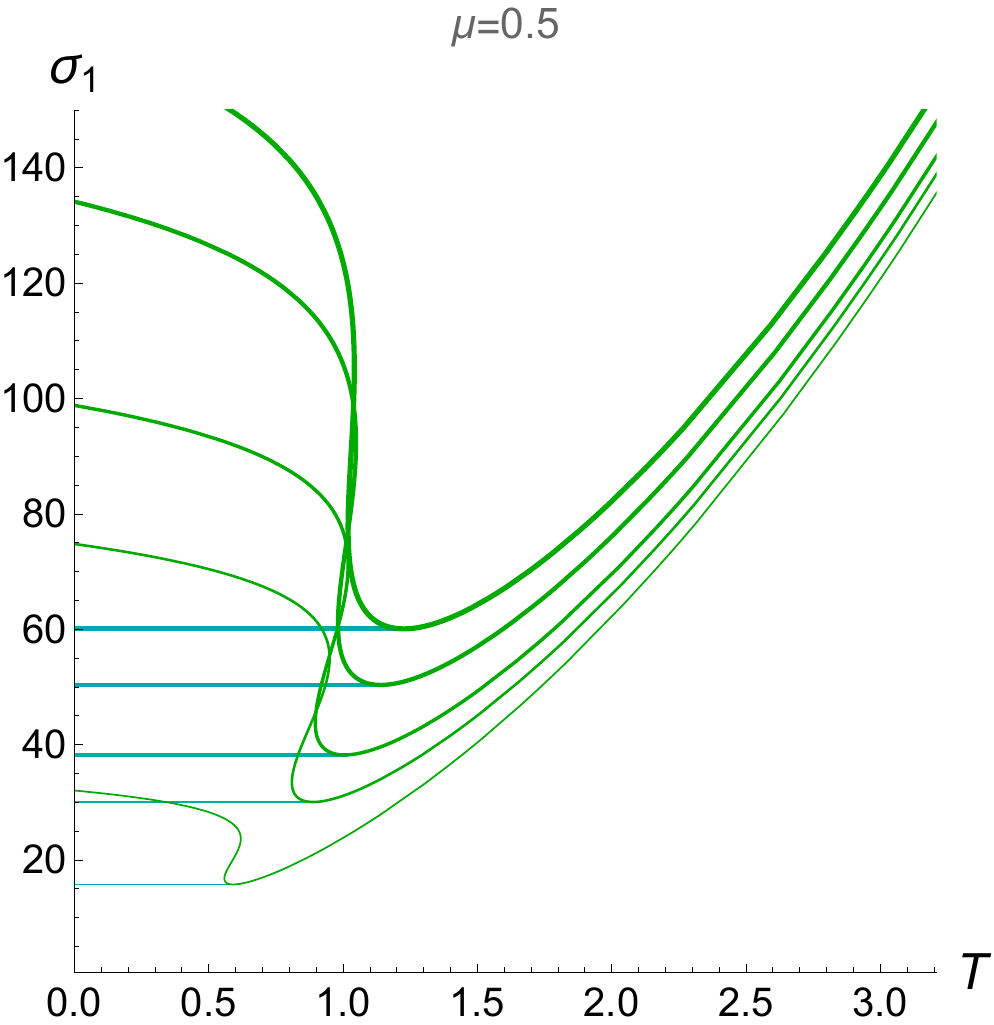}\\
  \quad A \hspace{70 mm} B \\
  \includegraphics[scale=0.41]{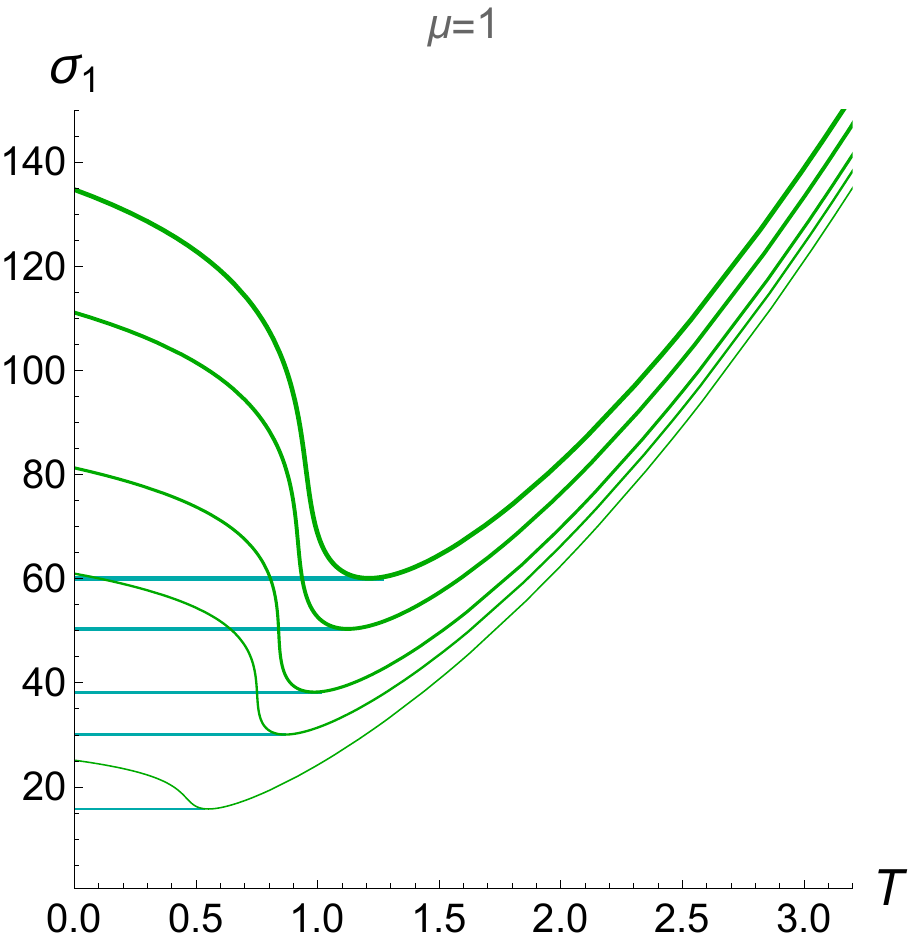}
  \qquad \qquad
  \includegraphics[scale=0.52]{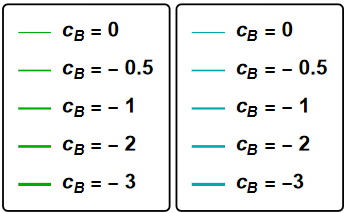}\\
  C \hspace{50 mm} 
  \caption{Spatial string tension $\sigma_1$ in the first orientation ${\cal{W}}_{x Y_{1}}$ as a function of  temperature $T$ for isotropic case $\nu=1$ at (A) $\mu = 0$, (B) $\mu = 0.5$ GeV, and (C) $\mu = 1$ GeV with different $c_B$ considering zero-boundary condition \eqref{zerobc}. The cyan line and green curve show DW and horizon configuration, respectively; $[\sigma]^{\frac{1}{2}}=[T]=[\mu] = [c_B]^{\frac{1}{2}} =$ GeV.  
  }
  \label{Fig:sigma-1-nu1}
\end{figure}

\begin{figure}[h!]
  \centering
  \includegraphics[scale=0.43]{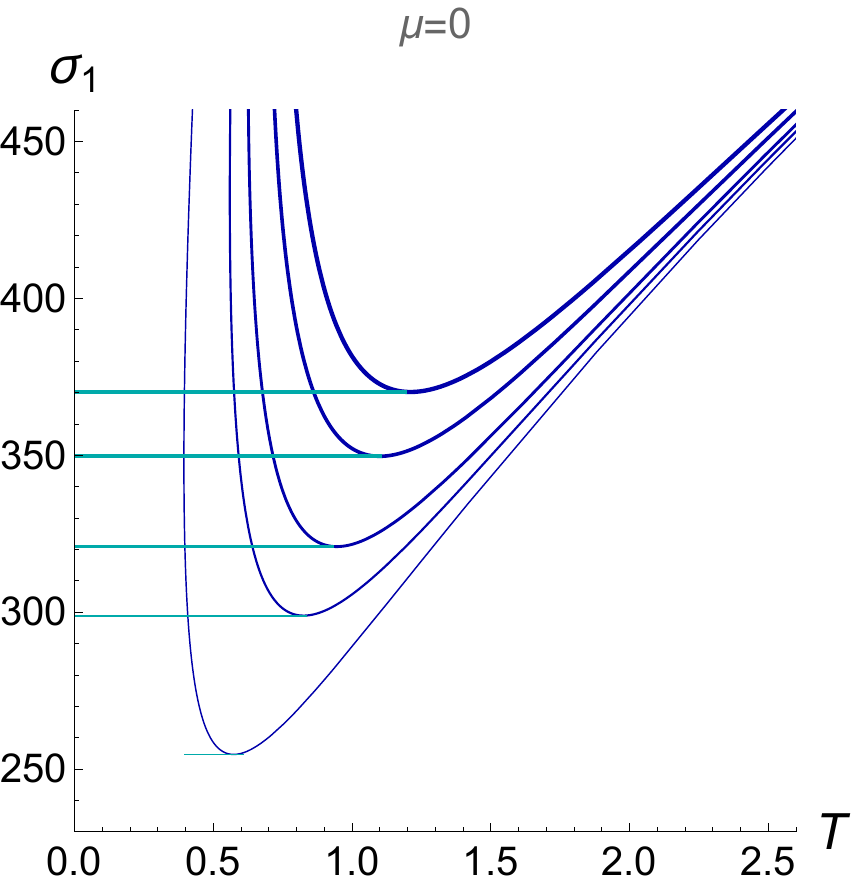}\qquad
  \includegraphics[scale=0.39]{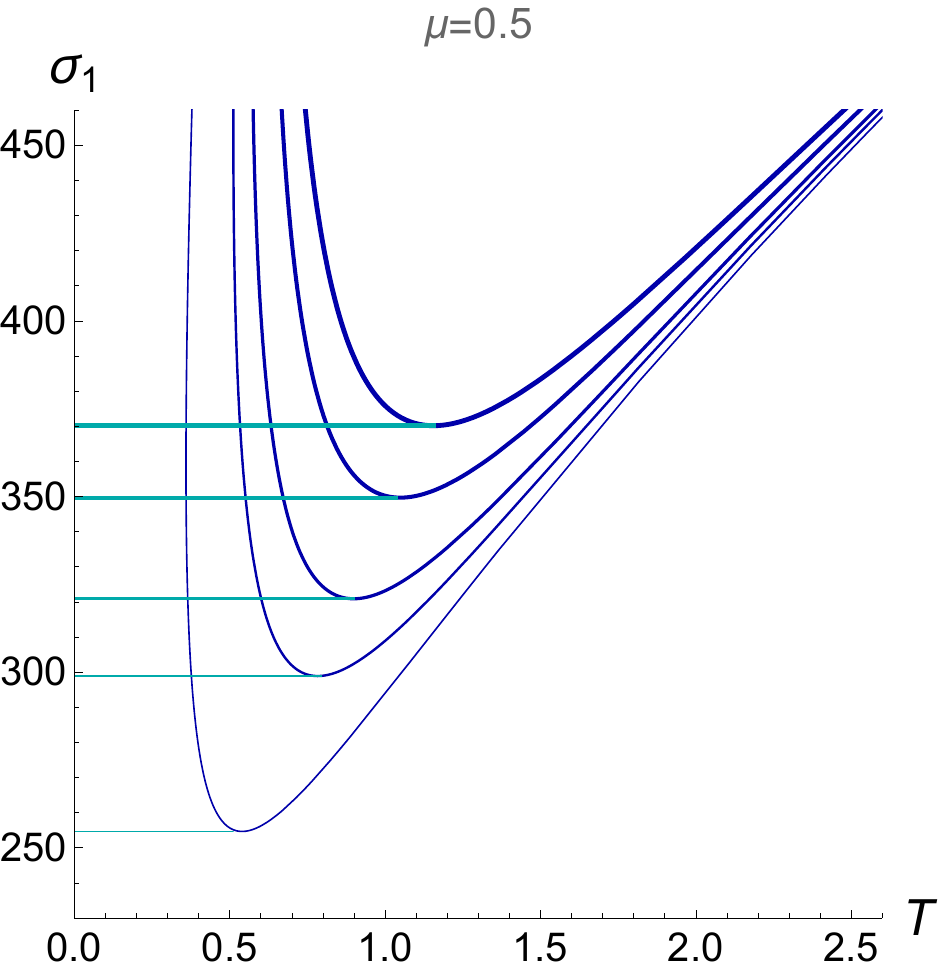}\\
  \quad A \hspace{70 mm} B \\
  \includegraphics[scale=0.42]{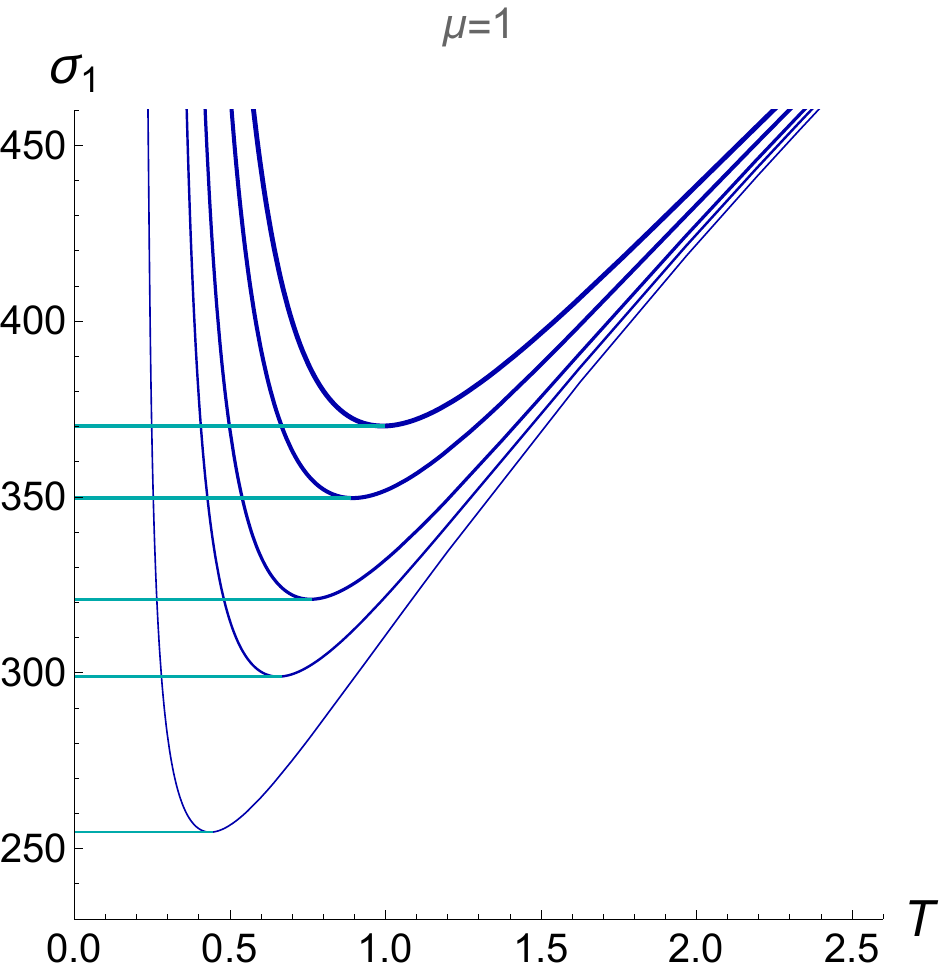}
  \quad \qquad
  \includegraphics[scale=0.5]{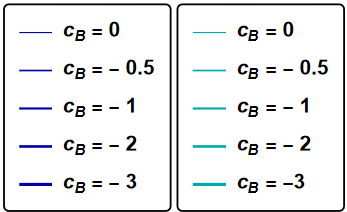}\\
  C \hspace{50 mm} 
  \caption{Spatial string tension $\sigma_1$ in the first orientation ${\cal{W}}_{x Y_{1}}$ as a function of temperature $T$ for anisotropic case $\nu=4.5$ at (A) $\mu = 0$, (B) $\mu = 0.5$ GeV, and (C) $\mu = 1$ GeV with different $c_B$ considering zero-boundary condition \eqref{zerobc}. The cyan line and blue curve show DW and horizon configuration, respectively; $[\sigma]^{\frac{1}{2}}=[T]=[\mu] = [c_B]^{\frac{1}{2}} =$ GeV.
  }
  \label{Fig:sigma-1-nu45}
\end{figure}

The SWL string tension can be obtained by calculating the minimal value of the effective potential at the DW  $z=z_{DW}$ or at the horizon $z=z_h$. Figs.~\ref{Fig:sigma-1-nu1} and \ref{Fig:sigma-1-nu45} show the spatial string tension $\sigma_1$ as a function of temperature $T$ for the first orientation of SWL, i.e. ${\cal{W}}_{x Y_{1}}$, at $\mu = 0$, $\mu = 0.5$ GeV and $\mu = 1$ GeV for different $c_B$ in the isotropic case $\nu = 1$ and the anisotropic case $\nu = 4.5$, respectively. Both figures were obtained considering the zero-boundary condition \eqref{zerobc}, and the cyan lines present the dependence of $\sigma(z_{DW})$ on temperature in the DW configuration, while the green and blue lines present the dependence of $\sigma(z_h)$ on temperature in the horizon configuration. 

The phase transition occurs at critical temperature $T_{cr}$ between two connected string configurations, i.e. DW and horizon configurations (see Fig.~\ref{Fig:SWLs}), with different
values of spatial string tension: $\sigma(z_{DW})$ and $\sigma(z_h)$, respectively (see also Sect.\ref{Sect:BI}). 
Note that this is a continuous phase transition. For this phase transition the string tension receives a large change in its value at the critical temperature $T_{cr}$. To put it in another way, the second derivative of the spatial string tension with respect to temperature   $\partial^2\sigma/\partial T^2$ undergoes a jump. To compare this phase transition obtained from SWL with confinement/deconfinement and 1st order phase transitions, see Sect.~\ref{Sect:conclusion} and Fig.~\ref{fig:PDswl}.

Fig.~\ref{Fig:sigma-1-nu45} shows, that the spatial string tension $\sigma$ value increases with the spatial anisotropy $\nu$ and the external magnetic field $c_B$. Here and below ``larger magnetic field'' means ``larger absolute value of $c_B$''. In addition, Figs.~\ref{Fig:sigma-1-nu1} and \ref{Fig:sigma-1-nu45} show, that at lower temperature $T<T_{cr}$, i.e. in DW configuration, the spatial string tension does not depend on the temperature and gets a constant value such as $\sigma_1=60$ GeV$^2$ for $\mu=0$, $c_B=-\,3$ GeV$^2$. Note also, that the DW coordinates are independent from the chemical potential. However, for $\mu=0$, $c_B=0$ Figs.~\ref{Fig:sigma-1-nu1}A and \ref{Fig:sigma-1-nu45}A show, that for very low $T$ the $\sigma_1$ does not exist due to lack of the DW coordinate.

\begin{figure}[h!]
  \centering
  \includegraphics[scale=0.36]{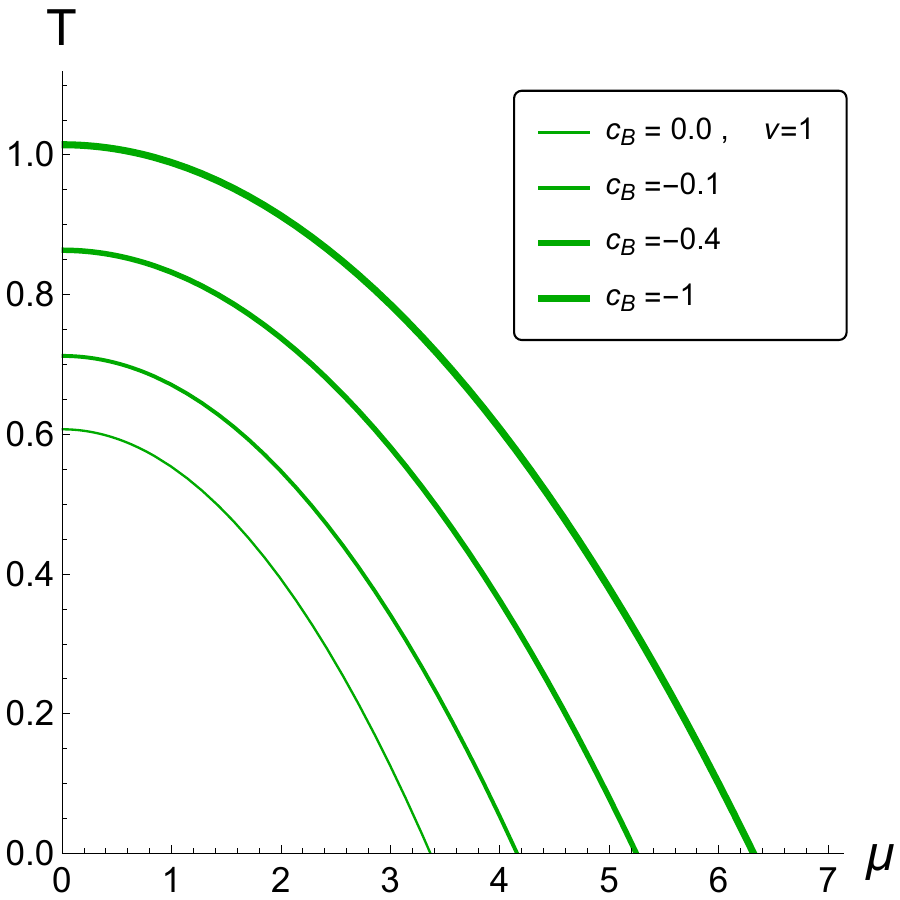}\qquad \qquad
  \includegraphics[scale=0.36]{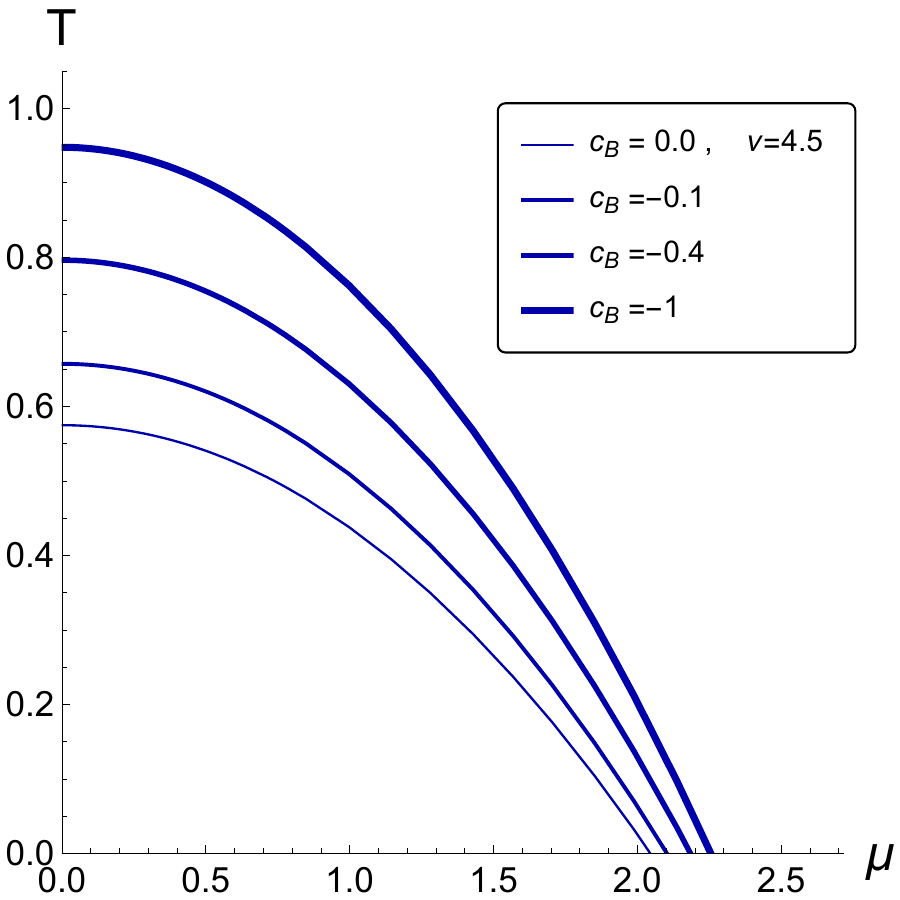}\\
  A \hspace{60 mm} B \\
  \includegraphics[scale=0.38]{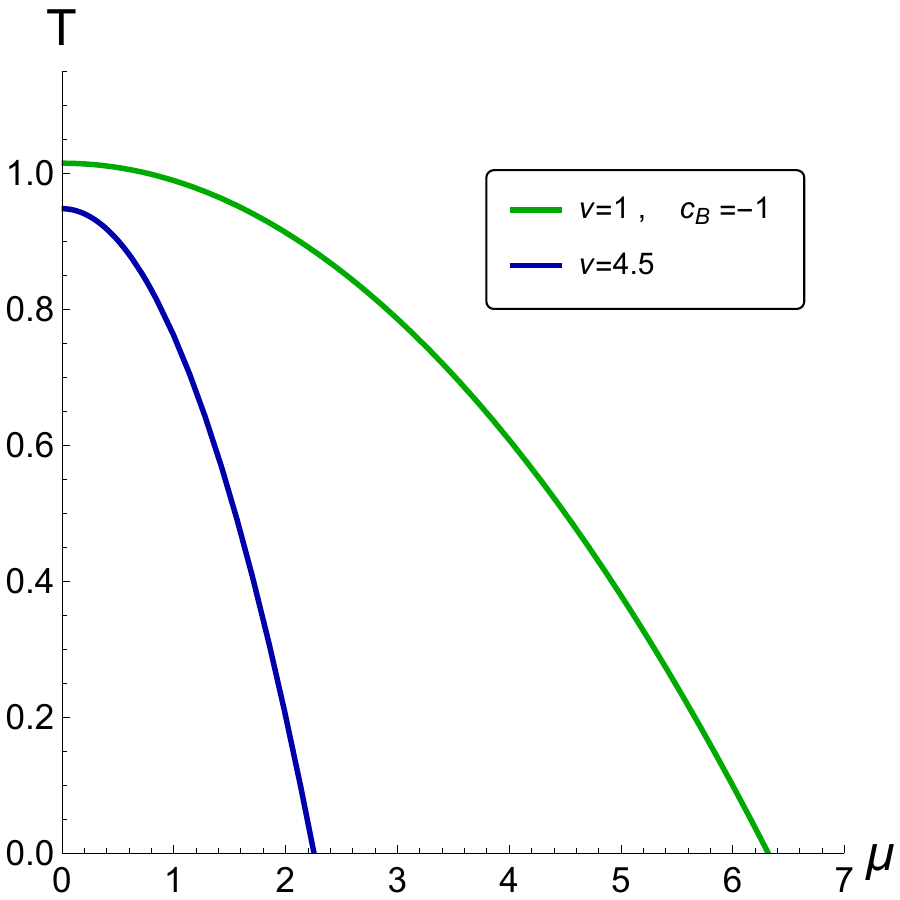} \\
  C
  \caption{Phase transition diagrams in $(\mu,T)$-plane in the first orientation ${\cal{W}}_{x Y_{1}}$ at (A) $\nu = 1$, (B) $\nu = 4.5$, and (C) comparison between $\nu =1$ and $\nu = 4.5$ at fixed $c_B=-\,1$ GeV$^2$ corresponding to the transition from the DW configuration to the horizon configuration of $\sigma_1$ considering zero-boundary condition \eqref{zerobc}; $[T]=[\mu] = [c_B]^{\frac{1}{2}} =$ GeV. 
  }
  \label{Fig:muT-sigma-1-nu45}
\end{figure}

Fig.~\ref{Fig:muT-sigma-1-nu45} shows, that at fixed chemical potential the increasing magnetic coefficient $c_B$ increases the transition temperature between the DW and the horizon configuration, thus confirming the magnetic catalysis phenomenon. Furthermore, inclusion of spatial anisotropy $\nu=4.5$ decreases the critical transition temperature $T_{cr}$ between the DW and the horizon configuration for all values of the magnetic field coefficient $c_B$ in our model.

\newpage
$$\,$$
\newpage
$$\,$$
\newpage

\subsubsection{Physical-boundary condition} \label{mbc}

In this section we apply the physical-boundary condition \eqref{phi-fz-LQ} to the dilaton field to calculate the DW coordinate and the spatial string tension. We would like to compare the results of this subsection and those of subsection \eqref{swl-0bc}.\\

\begin{figure}[h!]
  \centering
  \includegraphics[scale=0.37]{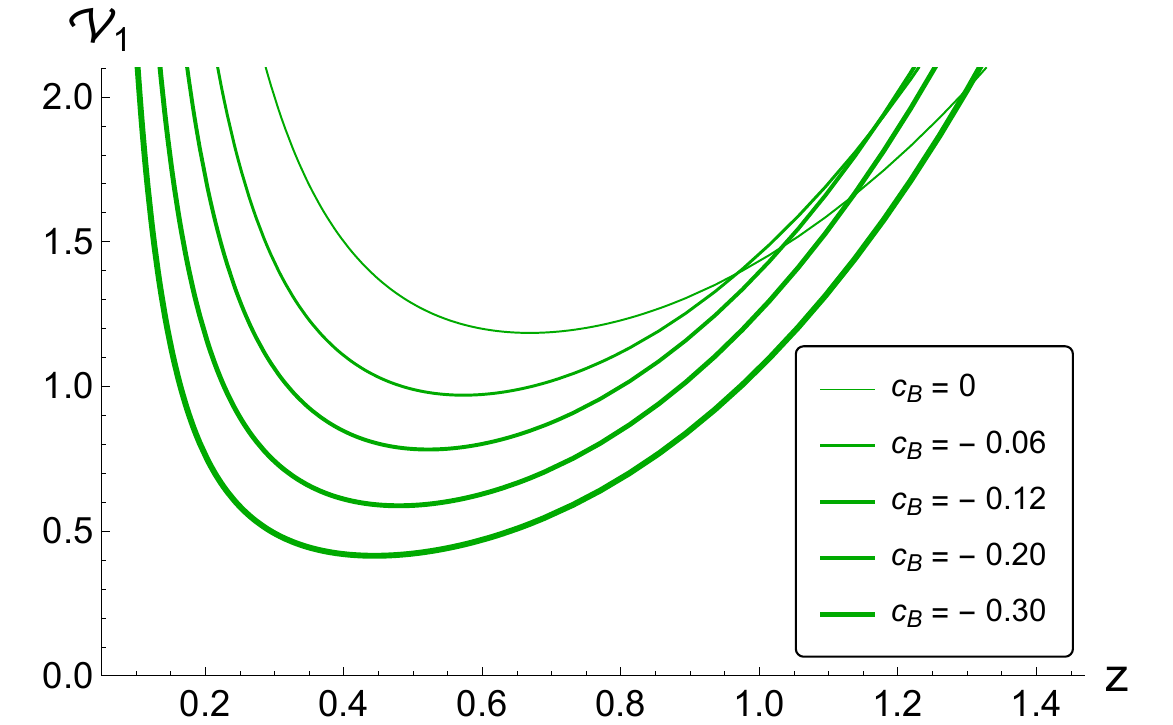} 
  \includegraphics[scale=0.37]{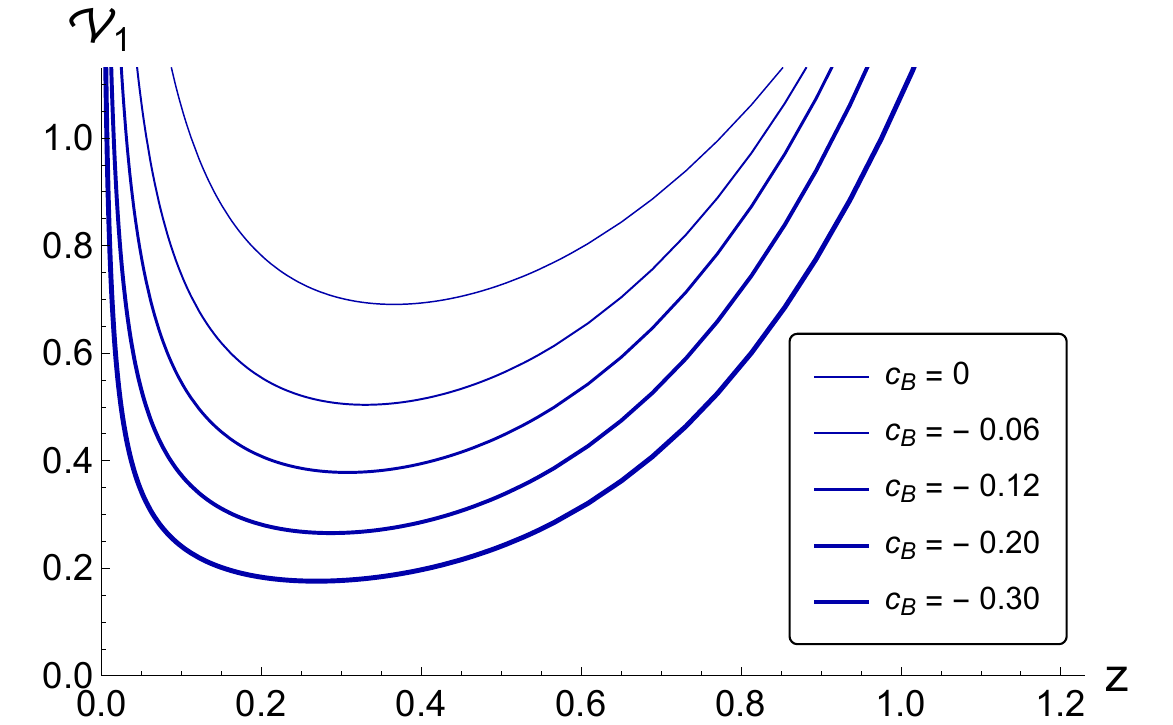} \\
  A \hspace{70mm} B
  \caption{Effective potential ${\cal V}_1(z)$ in the first orientation ${\cal{W}}_{x Y_{1}}$ at (A) $\nu = 1$, and (B) $\nu = 4.5$ for different $c_B$ considering physical-boundary condition \eqref{phi-fz-LQ}; $[z]^{-1} = [c_B]^{\frac{1}{2}} =$ GeV.
  }
  \label{Fig:V1nbc}
\end{figure}

The behavior of the effective potential ${\cal V}_1$ as a function of holographic coordinate $z$ for $\nu = 1$ and $\nu = 4.5$ corresponding to the first SWL orientation ${\cal{W}}_{xY_{1}}$ for different values of the magnetic field coefficients $c_B$ considering the physical-boundary condition \eqref{phi-fz-LQ} is presented in Fig.~\ref{Fig:V1nbc}. This figure shows, that there is a minimum effective potential not only for the zero-boundary condition \eqref{zerobc}, but also for physical-boundary conditions \eqref{phi-fz-LQ}.

The locations of the dynamical walls for different values of the magnetic coefficient $c_B$ in the isotropic case $\nu=1$ and the anisotropic case $\nu=4.5$ for the physical-boundary condition are given in the Tables~\ref{tab:V1nu1nbc} and~\ref{tab:V1nu45nbc}, respectively.

\begin{table}[b]
  \centering
  \begin{tabular}{|l|c|c|l|l|l|l|l|}
    \hline
    \multicolumn{1}{|c|}{${\cal V}_1$} & \multicolumn{5}{c|}{$\nu=1$} \\ 
    \hline
    $-\,c_B$ & 0 & 0.06 & \multicolumn{1}{c|}{0.12} & \multicolumn{1}{c|}{0.2} & \multicolumn{1}{c|}{0.3} \\
    \hline
    $z_{DW}$ & \multicolumn{1}{l|}{0.665} & 0.572 & 0.521 & 0.480 & 0.442 \\
    \hline
  \end{tabular}
  \caption{Locations of dynamical walls $z_{DW}$,  for ${\cal V}_1$ in the first orientation ${\cal{W}}_{x Y_{1}}$ at $\nu = 1$ with different $c_B$ for physical-boundary condition \eqref{phi-fz-LQ}; $[z]^{-1} = [c_B]^{\frac{1}{2}} =$ GeV.\\
  }
  \label{tab:V1nu1nbc}
  \begin{tabular}{|l|c|c|l|l|l|l|l|}
    \hline
    \multicolumn{1}{|c|}{${\cal V}_1$} & \multicolumn{5}{c|}{$\nu=4.5$} \\ 
    \hline
    $-\,c_B$ & 0 & 0.06 & \multicolumn{1}{c|}{0.12} & \multicolumn{1}{c|}{0.2} & \multicolumn{1}{c|}{0.3} \\ 
    \hline
    $z_{DW}$ & \multicolumn{1}{l|}{0.366} & 0.330 & 0.309 & 0.288 & 0.270 \\
    \hline
  \end{tabular}
  \caption{Locations of dynamical walls $z_{DW}$, for ${\cal V}_1$ in the first orientation ${\cal{W}}_{x Y_{1}}$ at $\nu = 4.5$ with different $c_B$ for physical-boundary condition \eqref{phi-fz-LQ}; $[z]^{-1} = [c_B]^{\frac{1}{2}} =$ GeV.
  }
  \label{tab:V1nu45nbc}
\end{table}

\begin{figure}[t]
  \centering
  \includegraphics[scale=0.45]{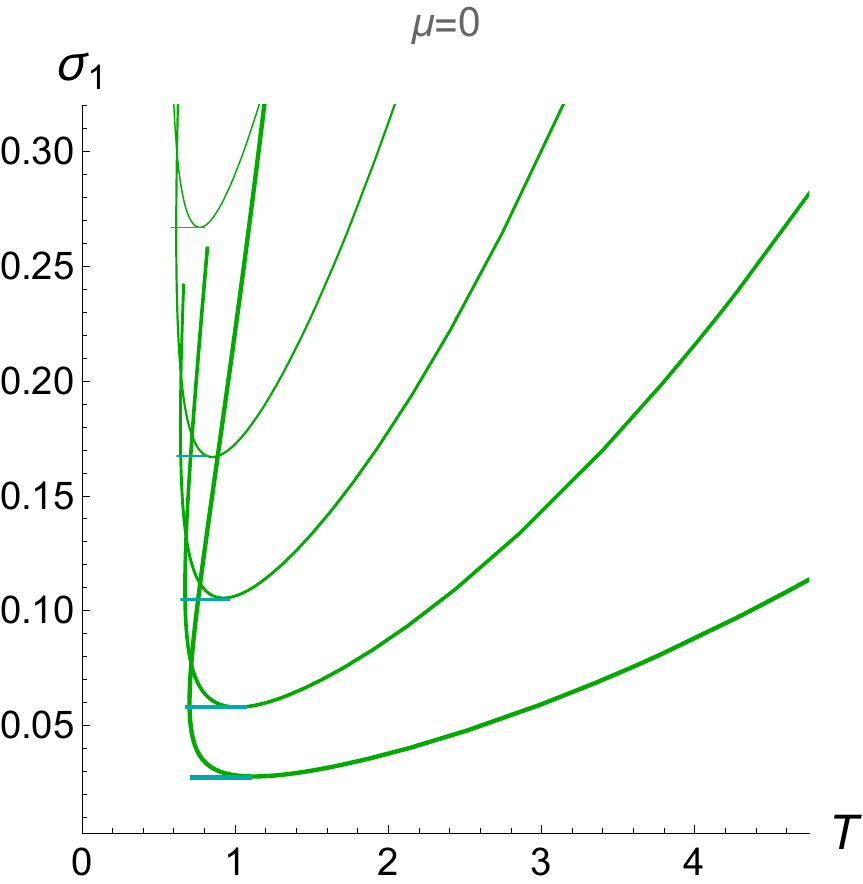}\quad
  \includegraphics[scale=0.44]{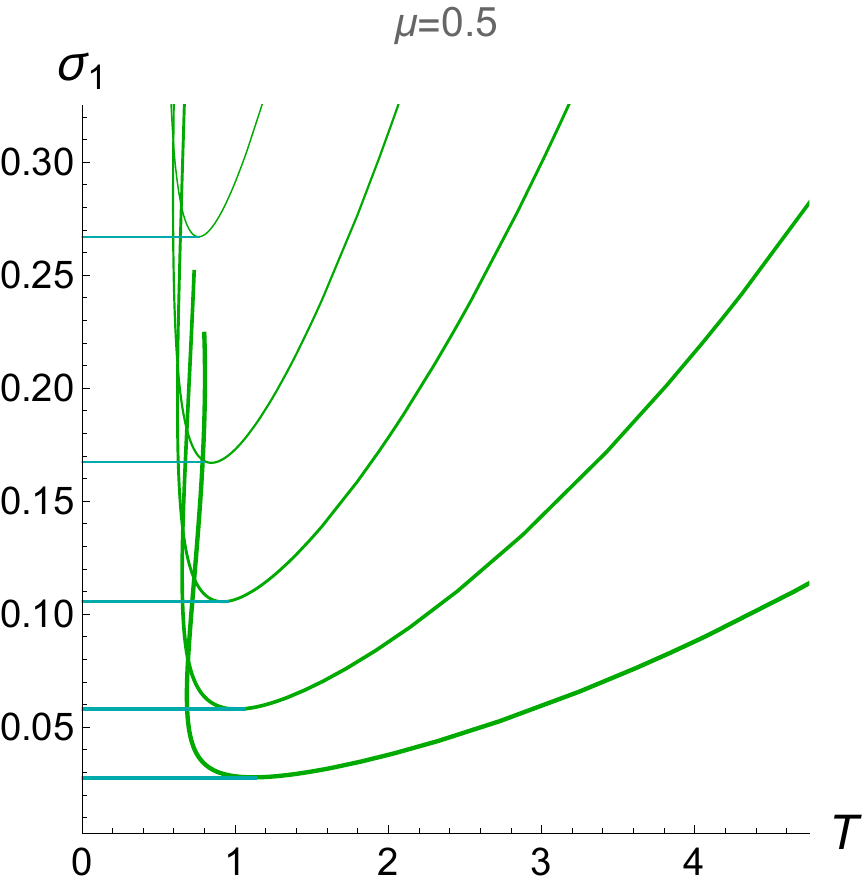}\\
  \quad A \hspace{70 mm} B \\
  \includegraphics[scale=0.45]{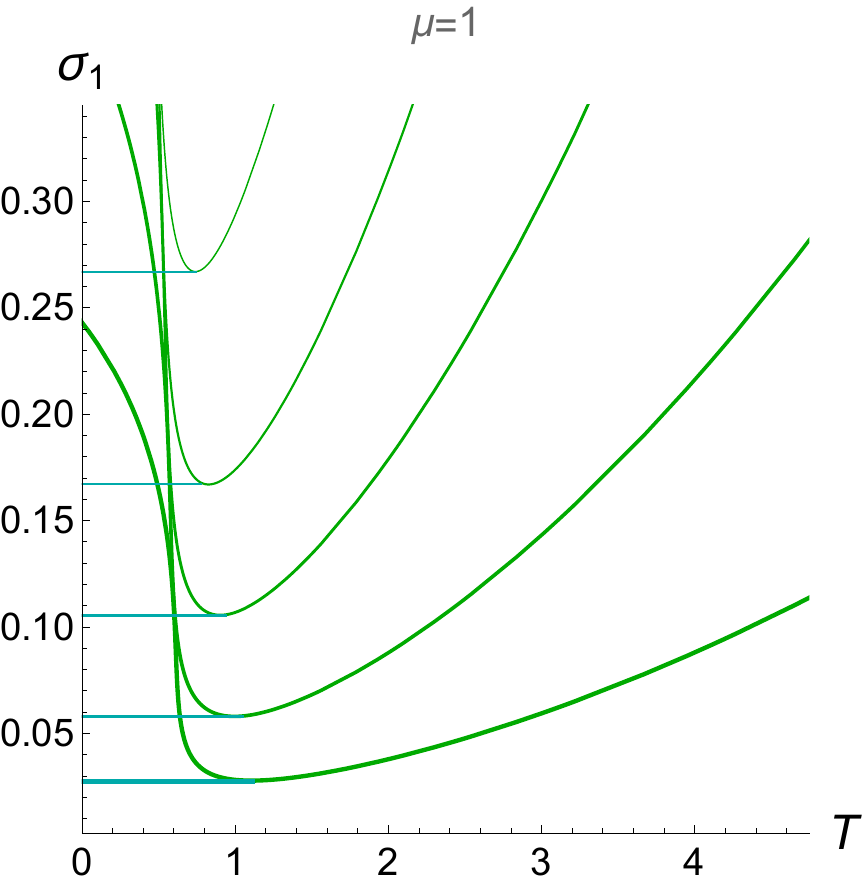} \qquad
  \includegraphics[scale=0.55]{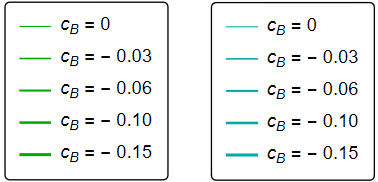}\\
  C \hspace{50 mm} 
  \caption{Spatial string tension $\sigma_1$ in the first orientation ${\cal{W}}_{x Y_{1}}$ as a function of  temperature $T$ for isotropic case $\nu=1$ at (A) $\mu = 0$, (B) $\mu = 0.5$ GeV, and (C) $\mu = 1$ GeV with different $c_B$ considering physical-boundary condition \eqref{zerobc}. The cyan line and green curve show DW and horizon configuration, respectively; $[\sigma]^{\frac{1}{2}}=[T]=[\mu] = [c_B]^{\frac{1}{2}} =$ GeV.
  }
  \label{Fig:sigma-1-nu1pbc}
\end{figure}

\begin{figure}[h!]
  \centering
  \includegraphics[scale=0.43]{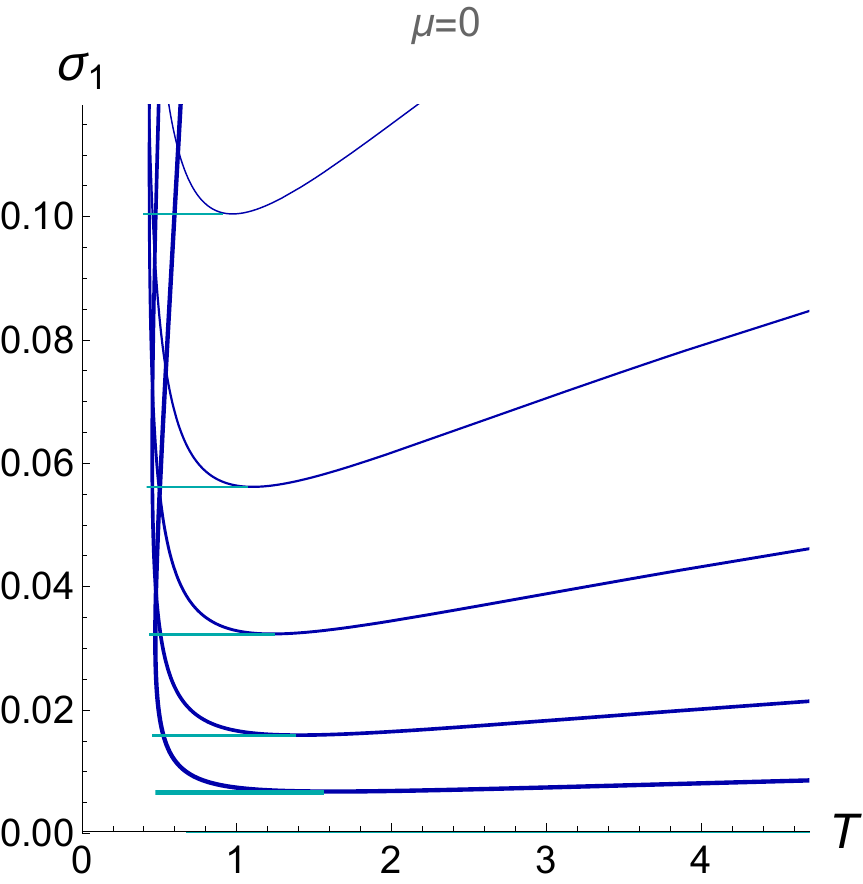}\qquad
  \includegraphics[scale=0.42]{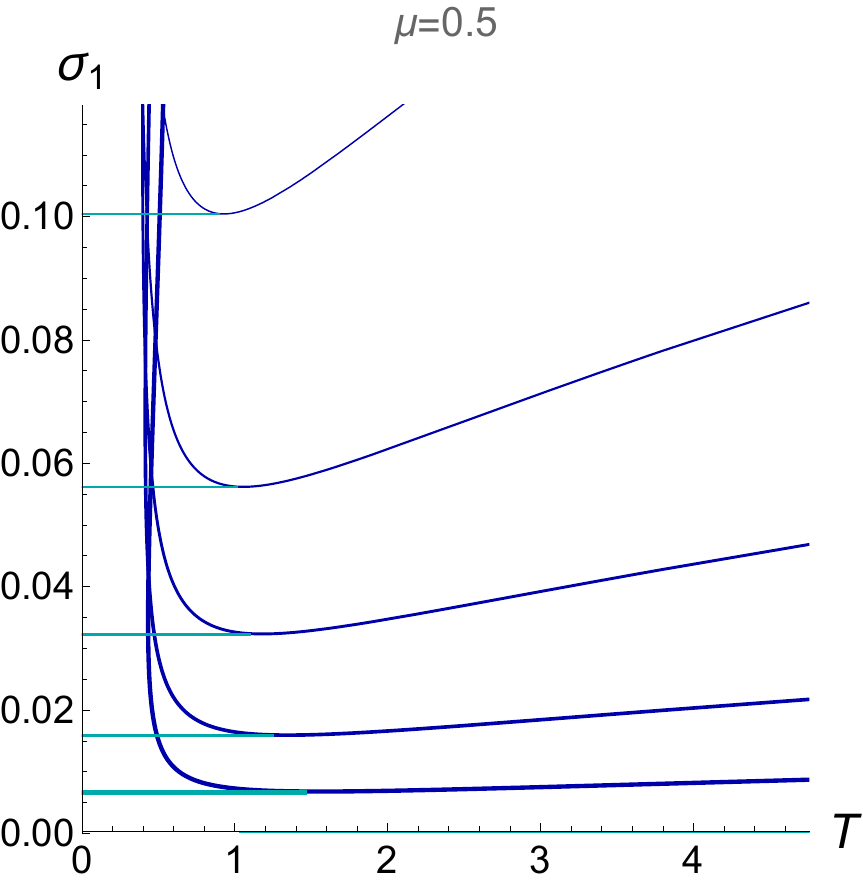}\\
   \quad A \hspace{70 mm} B \\
  \includegraphics[scale=0.45]{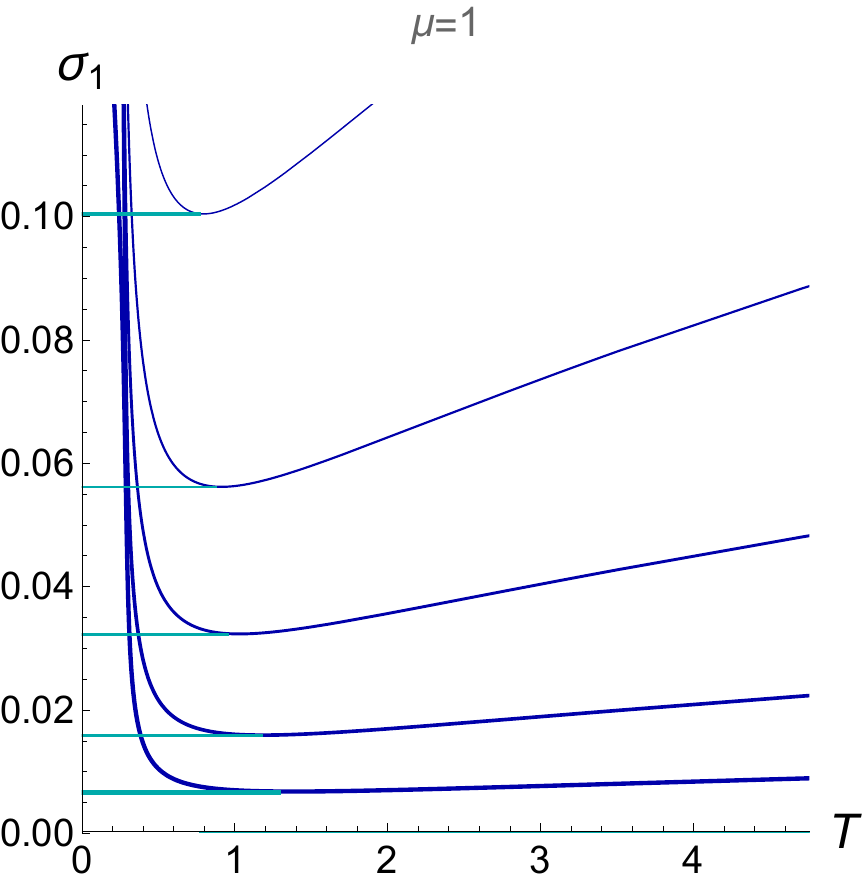} \qquad  \quad
  \includegraphics[scale=0.52]{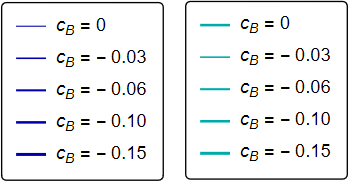}\\
  C \hspace{50 mm} 
 \caption{Spatial string tension $\sigma_1$ in the first orientation ${\cal{W}}_{x Y_{1}}$ as a function of  temperature $T$ for anisotropic case $\nu=4.5$ at (A) $\mu = 0$, (B) $\mu = 0.5$ GeV, and (C) $\mu = 1$ GeV with different $c_B$ considering physical-boundary condition \eqref{zerobc}. The cyan line and blue curve show DW and horizon configuration, respectively; $[\sigma]^{\frac{1}{2}}=[T]=[\mu] = [c_B]^{\frac{1}{2}} =$ GeV.
  }
  \label{Fig:sigma-1-nu45-nbc}
\end{figure}

Figs.~\ref{Fig:sigma-1-nu1pbc} and \ref{Fig:sigma-1-nu45-nbc} show spatial string tension $\sigma_1$ as a function of  temperature $T$ at $\mu = 0$, $\mu = 0.5$ GeV, and $\mu = 1$ GeV with different $c_B$ for the isotropic case $\nu=1$ and anisotropic case $\nu=4.5$, respectively. In this case we considered the physical-boundary condition \eqref{zerobc}. The cyan line shows the dependence of $\sigma$ in the DW configuration $\sigma(z_{DW})$ on temperature, and the green curve shows the dependence of $\sigma(z_h)$ in the horizon configuration on temperature.

In contradistinction to the zero-boundary condition \eqref{zerobc}, for physical-boundary condition Figs.~\ref{Fig:sigma-1-nu1pbc} and \ref{Fig:sigma-1-nu45-nbc} show, that inclusion of spatial anisotropy $\nu=4.5$ and increasing the magnetic field $c_B$ decreases the spatial string tension. However, for both zero- and physical-boundary conditions, the magnetic catalysis behavior was obtained. The DW coordinates are independent from the chemical potential. However, in the case of $\mu=0$ for very low $T$ in all values of $c_B$ Figs.~\ref{Fig:sigma-1-nu1pbc}A and \ref{Fig:sigma-1-nu45-nbc}A show that $\sigma_1$ does not exist due to lack of the DW coordinate. In the horizon configuration $T>T_{cr}$ for both the isotropic case $\nu=1$ and the anisotropic case $\nu=4.5$ the spatial string tension increases monotonically with temperature for zero- and physical-boundary conditions.

\begin{figure}[t!]
  \centering
  \includegraphics[scale=0.40]{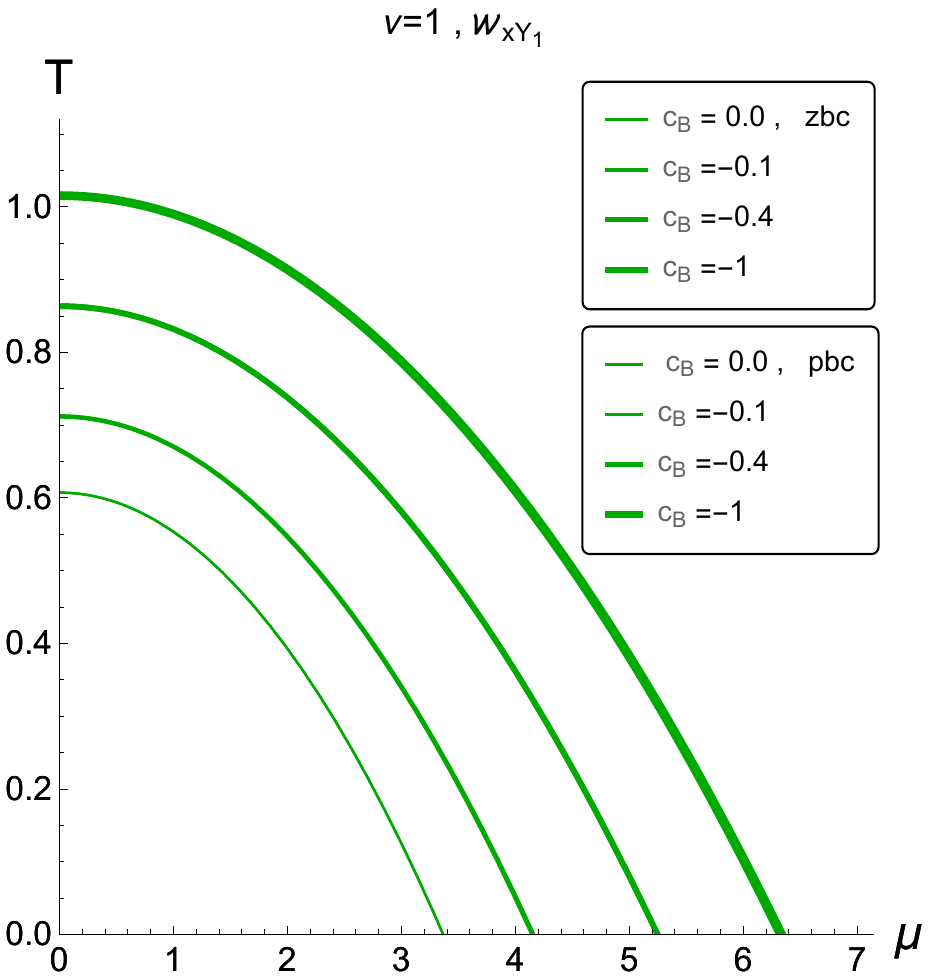}\qquad
  \includegraphics[scale=0.40]{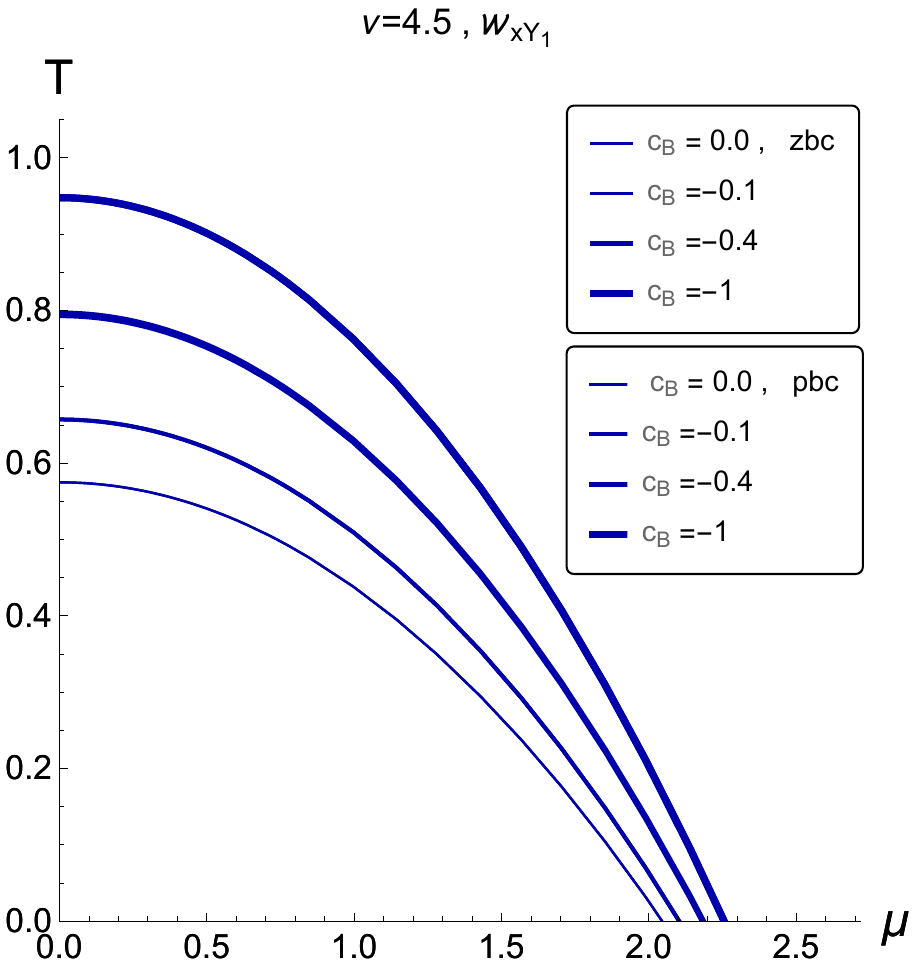}\\
  A \hspace{60 mm} B 
  \caption{Phase transition diagrams in $(\mu,T)$-plane in the first orientation ${\cal{W}}_{x Y_{1}}$ corresponding to the transition from the DW configuration to the horizon configuration of $\sigma_1$ at (A) $\nu = 1$, and (B) $\nu = 4.5$ with  different $c_B$ comparing the zero-boundary condition (zbc) \eqref{zerobc} and physical-boundary condition (pbc) \eqref{phi-fz-LQ}; $[T]=[\mu] = [c_B]^{\frac{1}{2}} =$ GeV.
  }
  \label{Fig:muT-sigma-1-nu45-nbc}
\end{figure}

Comparison of Figs.~\ref{Fig:muT-sigma-1-nu45-nbc} and \ref{Fig:muT-sigma-1-nu45} clearly shows, that the phase diagram does not depend on the boundary conditions. This result is confirmed by other models; see~\cite{Arefeva:2024vom}.

\newpage
$$\, $$
\newpage

\subsection{Spatial Wilson loop ${\cal{W}}_{xY_{2}}$}

\begin{figure}[b!]
  \centering
  \includegraphics[scale=0.41]{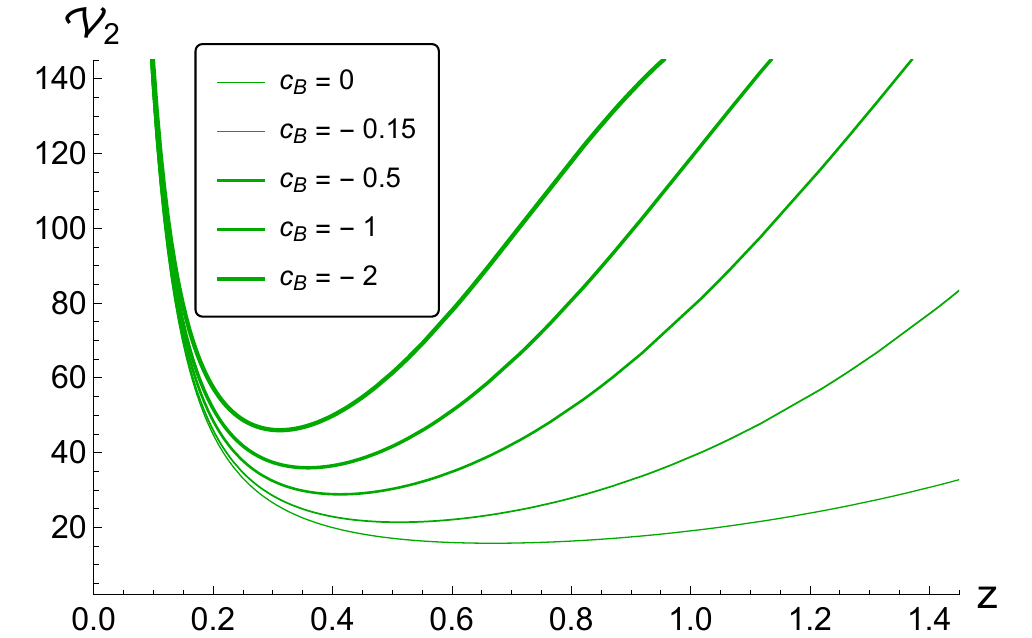} \quad
  \includegraphics[scale=0.41]{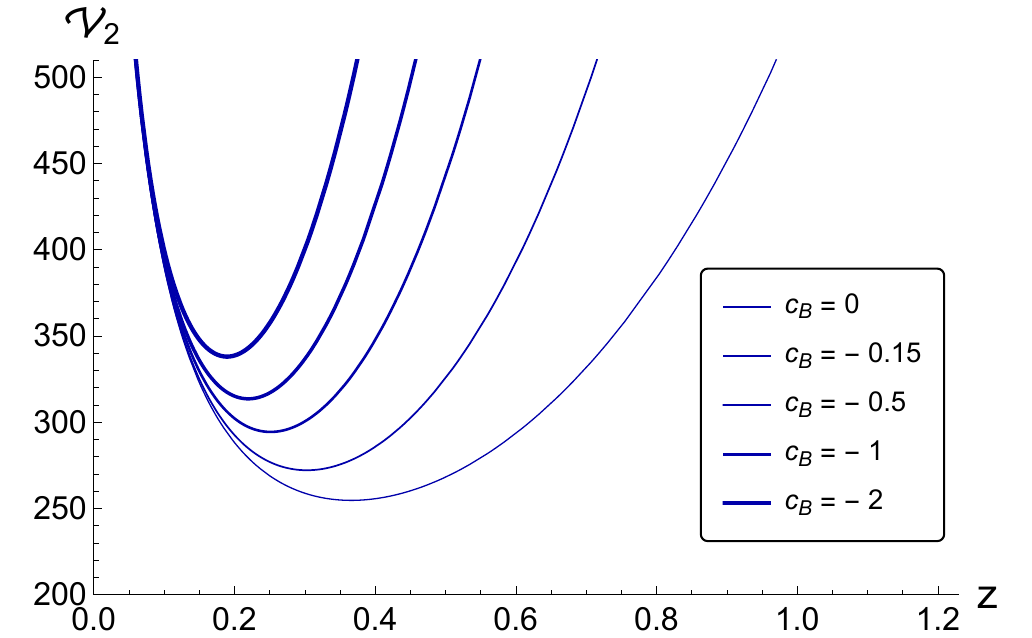} \\
  A \hspace{70mm} B
  \caption{Effective potential ${\cal V}_2(z)$ in the second orientation ${\cal{W}}_{x Y_{2}}$ at (A) $\nu = 1$, and (B) $\nu = 4.5$ for different $c_B$ considering zero-boundary condition \eqref{zerobc}; $[z]^{-1} = [c_B]^{\frac{1}{2}} =$ GeV.
  }
  \label{Fig:V2}
\end{figure}

The behavior of the effective potential ${\cal V}_2$ as a function of the holographic coordinate $z$ corresponding to the second orientation of the SWL, i.e. ${\cal{W}}_{xY_{2}}$ for different values of magnetic coefficients $c_B$ is presented in Fig.~\ref{Fig:V2} using the zero-boundary condition \eqref{zerobc}.  Fig.~\ref{Fig:V2} shows, that ${\cal V}_2$ has local minimums thus confirming ${\cal V}'(z) = 0$, i.e. the  dynamical walls exist at different values of the magnetic coefficient $c_B$ for the isotropic and anisotropic cases $\nu=1$ and $\nu=4.5$, respectively. 

Tables~\ref{tab:V2nu1} and~\ref{tab:V2nu45} show the DW locations at different values of the magnetic field $c_B$ for $\nu=1$ and $\nu=4.5$, respectively.

\begin{table}[h]
  \centering
  \begin{tabular}{|l|c|c|l|l|l|l|l|}
    \hline
    \multicolumn{1}{|c|}{${\cal V}_2$} & \multicolumn{5}{c|}{$\nu=1$} \\ 
    \hline
    $-\,c_B$ & 0 & 0.15 & \multicolumn{1}{c|}{0.5} & \multicolumn{1}{c|}{1} & \multicolumn{1}{c|}{2}
    \\ 
    \hline
    $z_{DW}$ & \multicolumn{1}{l|}{0.665} & 0.515 & 0.415 & 0.360 & 0.315  \\ \hline
  \end{tabular}
  \caption{Locations of dynamical walls $z_{DW}$,  for ${\cal V}_2$ in the second orientation ${\cal{W}}_{x Y_{2}}$ at $\nu = 1$ considering zero-boundary condition \eqref{zerobc}; $[z]^{-1} = [c_B]^{\frac{1}{2}} =$ GeV.
  }
  \label{tab:V2nu1}
\end{table}

\begin{table}[h]
  \centering
  \begin{tabular}{|l|c|c|l|l|l|l|l|}
    \hline
    \multicolumn{1}{|c|}{${\cal V}_2$} & \multicolumn{5}{c|}{$\nu=4.5$} \\ 
    \hline
    $-\,c_B$ & 0 & 0.15 & \multicolumn{1}{c|}{0.5} & \multicolumn{1}{c|}{1} & \multicolumn{1}{c|}{2} \\ 
    \hline
    $z_{DW}$ & \multicolumn{1}{l|}{0.366} & 0.305 & 0.250 & 0.220 & 0.190  \\ 
    \hline
  \end{tabular}
  \caption{Locations of dynamical walls $z_{DW}$, for ${\cal V}_2$ in the second orientation ${\cal{W}}_{x Y_{2}}$ at $\nu = 4.5$ considering zero-boundary condition \eqref{zerobc}; $[z]^{-1} = [c_B]^{\frac{1}{2}} =$ GeV.
  }
  \label{tab:V2nu45}
\end{table}

\begin{figure}[h!]
  \centering
  \includegraphics[scale=0.43]{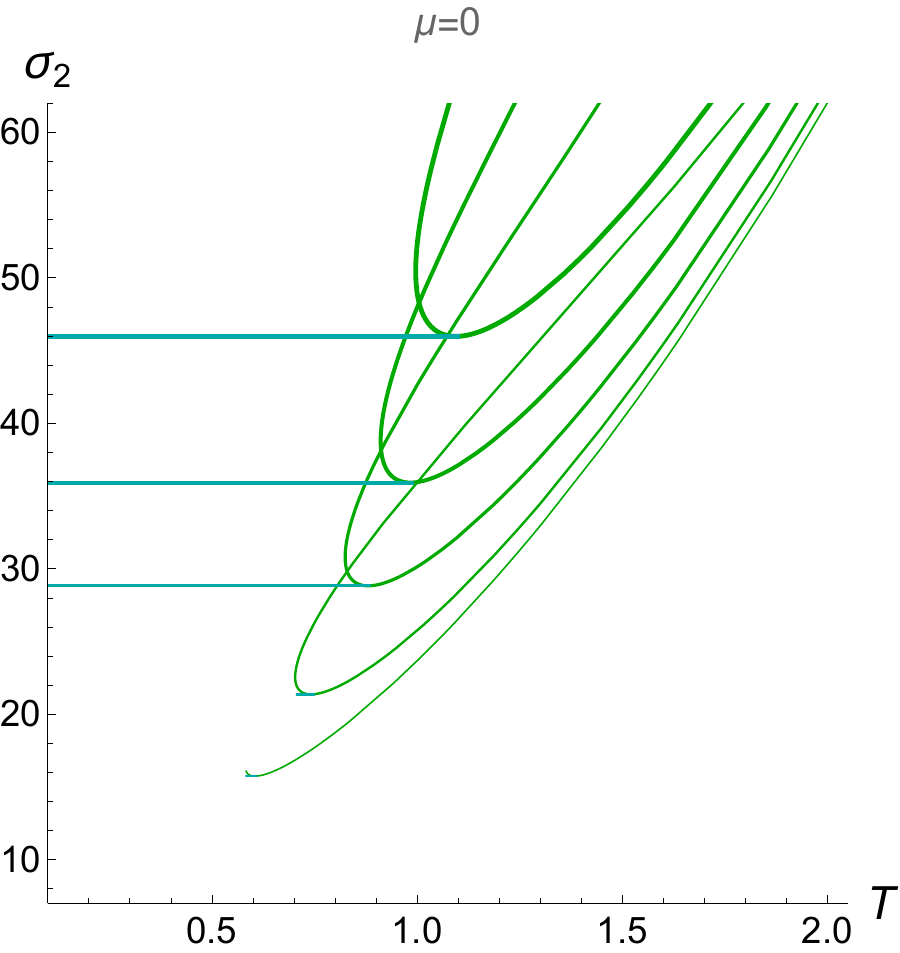}\quad
  \includegraphics[scale=0.50]{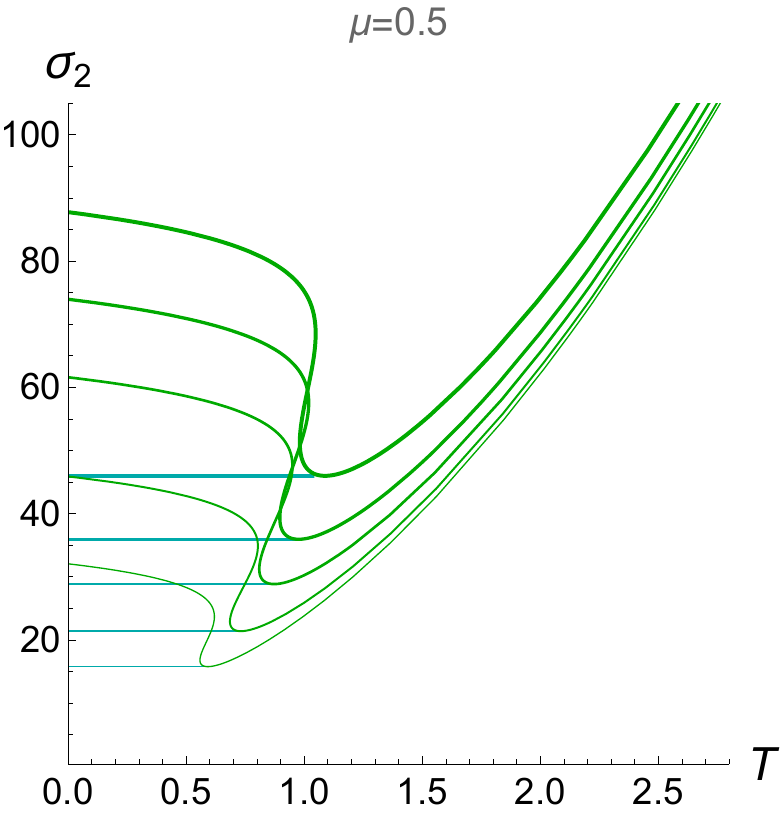}\\
  \quad A \hspace{70 mm} B \\
  \includegraphics[scale=0.43]{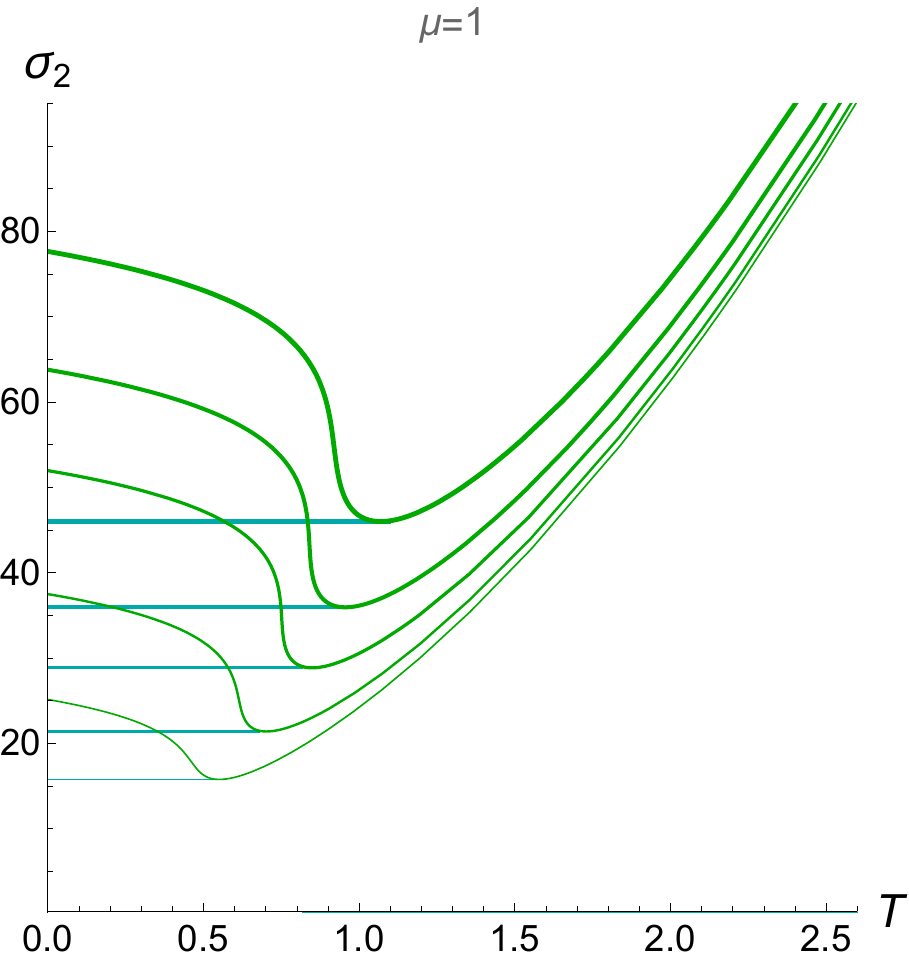} \qquad \quad \qquad
  \includegraphics[scale=0.52]{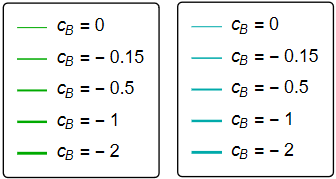}\\
  C \hspace{50 mm} 
  \caption{Spatial string tension $\sigma_2$ in the second orientation ${\cal{W}}_{x Y_{2}}$ as a function of  temperature $T$ for isotropic case $\nu=1$ at (A) $\mu = 0$, (B) $\mu = 0.5$ GeV, and (C) $\mu = 1$ GeV with different $c_B$ considering zero-boundary condition \eqref{zerobc}. The cyan line and green curve show DW and horizon configuration, respectively;
 $[\sigma]^{\frac{1}{2}}=[T]=[\mu] = [c_B]^{\frac{1}{2}} =$ GeV.
  }
  \label{Fig:sigma-2-nu1}
\end{figure}

\begin{figure}[h!]
  \centering
  \includegraphics[scale=0.43]{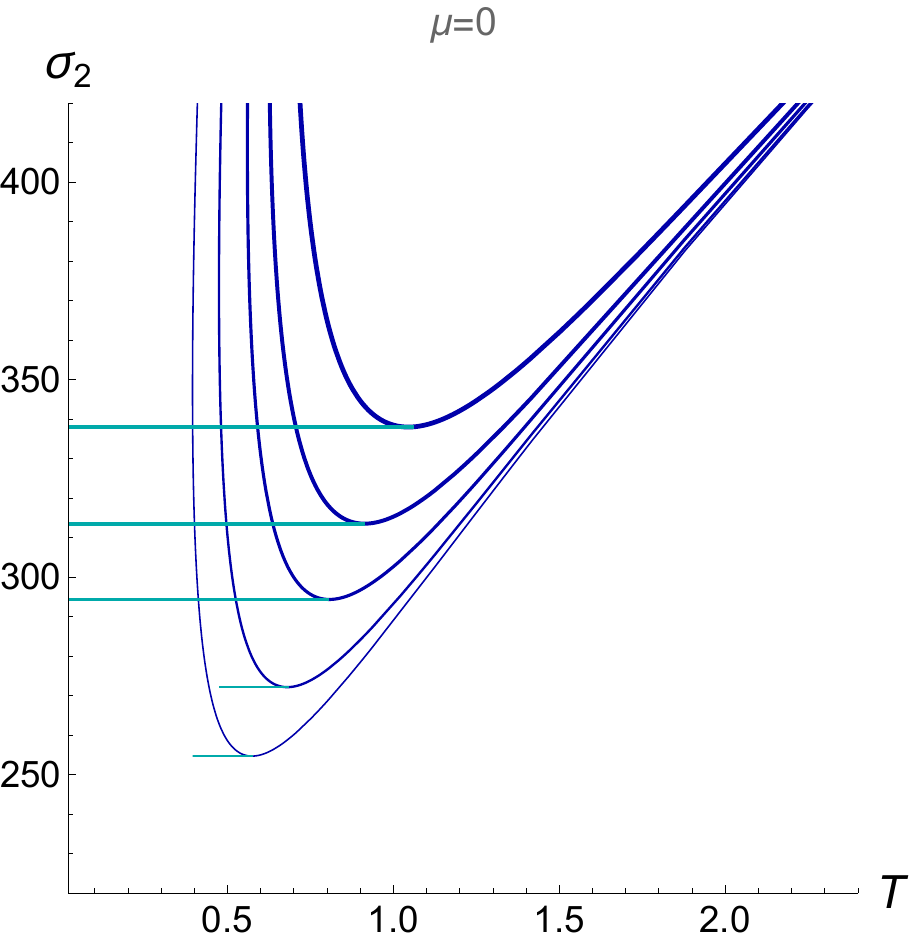}\qquad
  \includegraphics[scale=0.48]{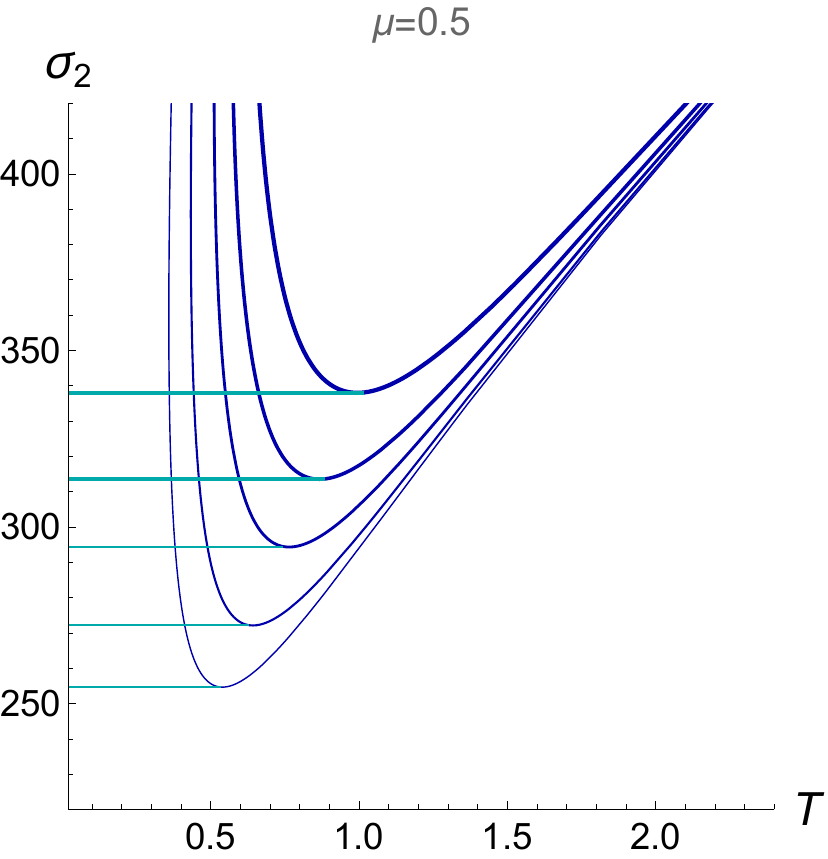}\\
   \qquad A \hspace{70 mm} B \\
  \includegraphics[scale=0.52]{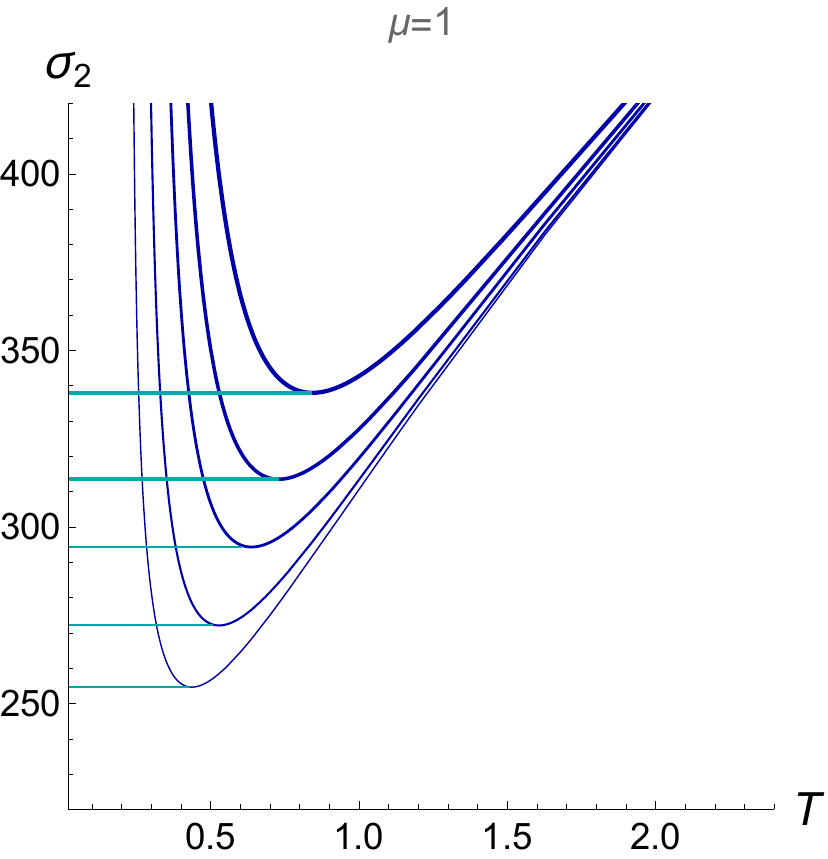}
  \qquad   \qquad
  \includegraphics[scale=0.51]{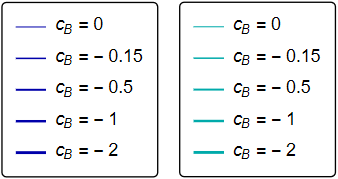}\\
  C \hspace{50 mm} 
  \caption{Spatial string tension $\sigma_2$ in the second orientation ${\cal{W}}_{x Y_{2}}$ as a function of  temperature $T$ for anisotropic case $\nu=4.5$ at (A) $\mu = 0$, (B) $\mu = 0.5$ GeV, and (C) $\mu = 1$ GeV with different $c_B$ considering zero-boundary condition \eqref{zerobc}. The cyan line and blue curve show DW and horizon configuration, respectively; $[\sigma]^{\frac{1}{2}}=[T]=[\mu] = [c_B]^{\frac{1}{2}} =$ GeV.
  }
  \label{Fig:sigma-2-nu45}
\end{figure}

Figs.~\ref{Fig:sigma-2-nu1} and \ref{Fig:sigma-2-nu45} depict the spatial string tension $\sigma_2$ as a function of  temperature $T$ in  $\mu = 0$, $\mu = 0.5$ GeV, and $\mu = 1$ GeV with different $c_B$ for the isotropic case $\nu=1$ and anisotropic case $\nu=4.5$, respectively. The cyan lines present the dependence of $\sigma(z_{DW})$ on temperature in the DW configuration, while the green and blue lines present the dependence of $\sigma(z_h)$ on temperature in the horizon configuration.

\begin{figure}[h!]
  \centering
  \includegraphics[scale=0.40]{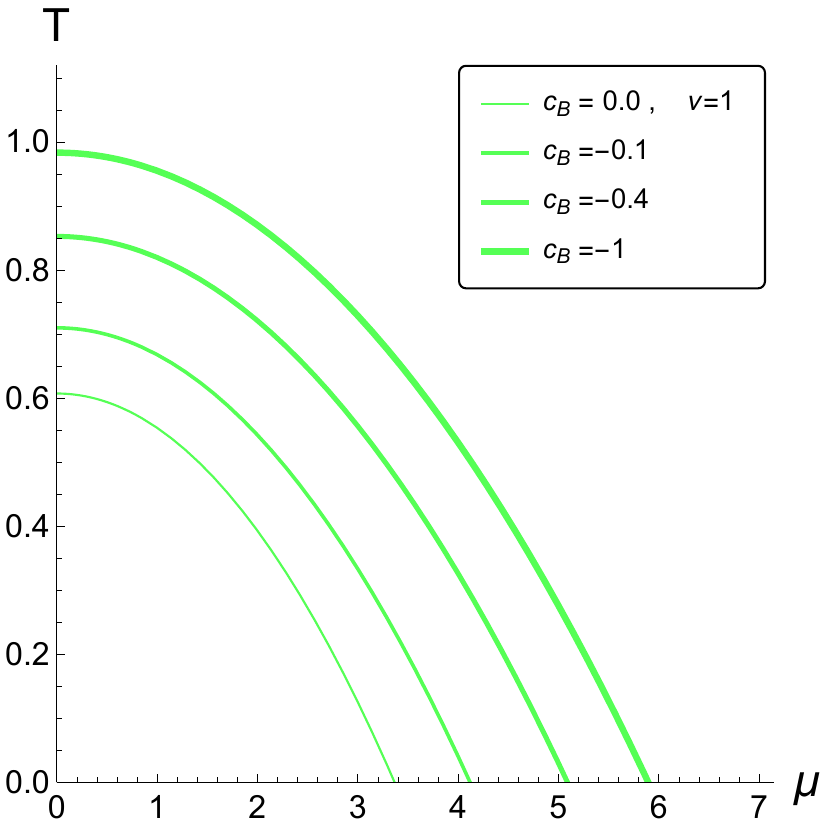}\qquad \qquad
  \includegraphics[scale=0.40]{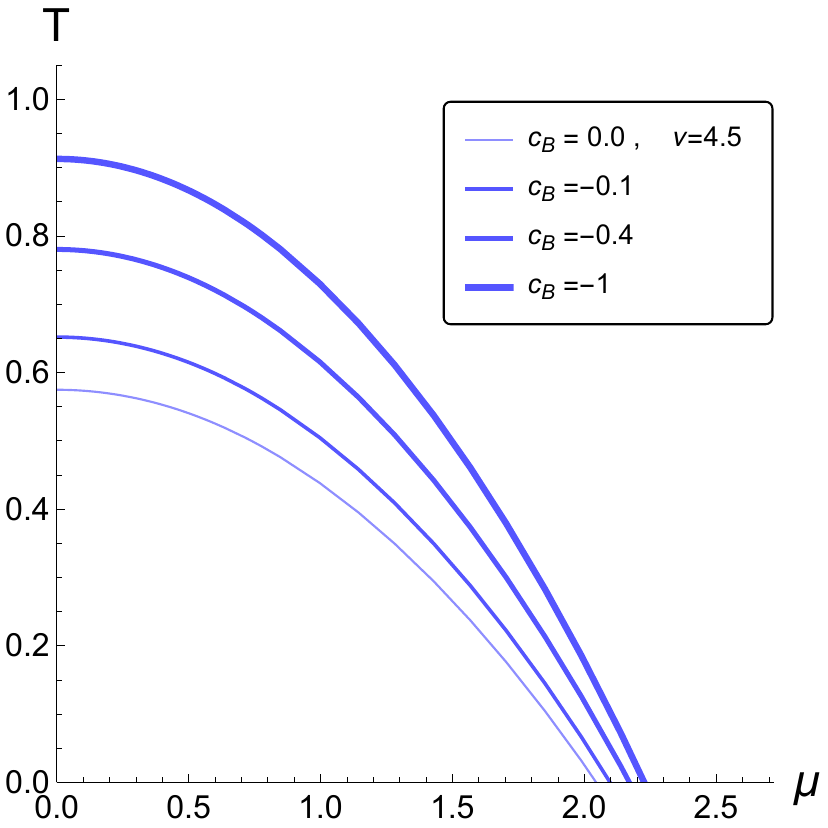}\\
   A \hspace{60 mm} B \\
  \includegraphics[scale=0.40]{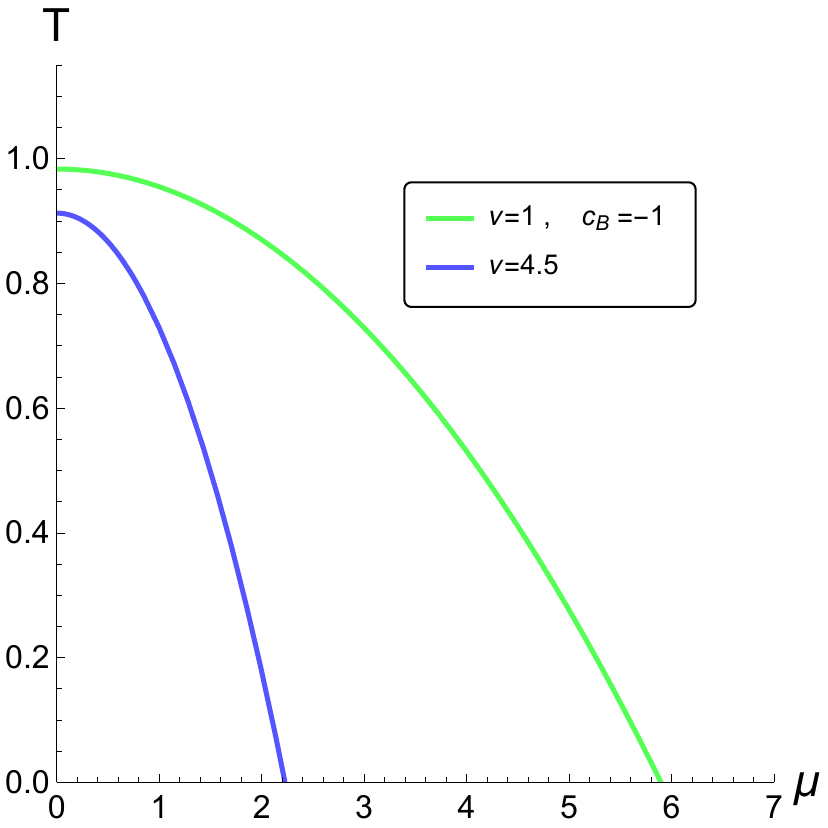} \\
  C
  \caption{Phase diagrams in $(\mu,T)$-plane at (A) $\nu = 1$, (B) $\nu = 4.5$, and (C) comparison between $\nu = 1$ and $\nu = 4.5$ at fixed $c_B=-\,1$ GeV$^2$ in the second orientation ${\cal{W}}_{y_{1}Y_{2}}$, corresponding to the transition from the DW configuration to the horizon configuration of $\sigma_2$ considering zero-boundary condition \eqref{zerobc}; $[T]=[\mu] = [c_B]^{\frac{1}{2}} =$ GeV.
  }
  \label{Fig:sigma-2-nu145}
\end{figure}

Figs.~\ref{Fig:sigma-2-nu145}A and \ref{Fig:sigma-2-nu145}B show the phase diagrams in the $(\mu,T)$-plane for $\sigma_2$ in the second orientation ${\cal{W}}_{x Y_{2}}$ corresponding to the transition from the DW configuration to the horizon configuration at zero-boundary condition \eqref{zerobc} for $\nu = 1$ and $\nu = 4.5$, respectively. Fig.~\ref{Fig:sigma-2-nu145}C shows, that inclusion of spatial anisotropy decreases $T_{cr}$. 

The results of the spatial string tension as a function of $T$  in the second SWL orientation ${\cal{W}}_{xY_{2}}$ shown in Figs.~\ref{Fig:sigma-2-nu1} and \ref{Fig:sigma-2-nu45}
are the following.
\begin{itemize}
  \item Exhibits the magnetic catalysis behavior, which $T_{cr}$ increases by enhancing the magnetic field $c_B$.
  \item The spatial string tension $\sigma$ increases as the spatial anisotropy $\nu$, and the external magnetic field $c_B$ increases.
  \item At lower temperature $T<T_{cr}$ is in the DW configuration, the string tension does not depend on the temperature and gets a constant value such as $\sigma_2=46$~GeV$^2$ at $\mu=0$ and $c_B=-\,2$ GeV${}^2$.
  \item At higher temperature $T>T_{cr}$ in the horizon configuration, the string tension increases monotonically.
  \item The DW coordinates are independent of the chemical potential.
\end{itemize}

Note that comparing the phase diagrams between the first SWL orientation ${\cal{W}}_{x Y_{1}}$ and the second orientation ${\cal{W}}_{x Y_{2}}$ is investigated in Sect.~\ref{com123}.

\newpage
$$\, $$
\newpage
$$\, $$
\newpage

\subsection{Spatial Wilson loop 
${\cal{W}}_{y_{1}Y_{2}}$}

The third SWL orientation, i.e. ${\cal{W}}_{y_{1}Y_{2}}$, has very peculiar characteristics compared to the first orientation ${\cal{W}}_{x Y_{1}}$ and the second orientation ${\cal{W}}_{x Y_{2}}$.\\

\begin{figure}[h!]
  \centering
  \includegraphics[scale=0.41]{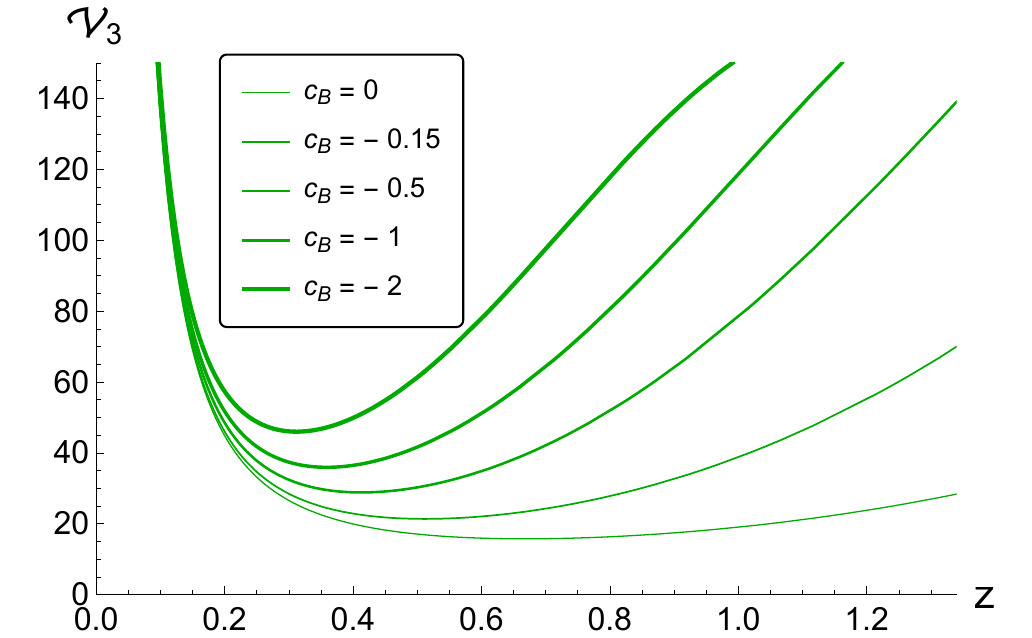} \quad
  \includegraphics[scale=0.41]{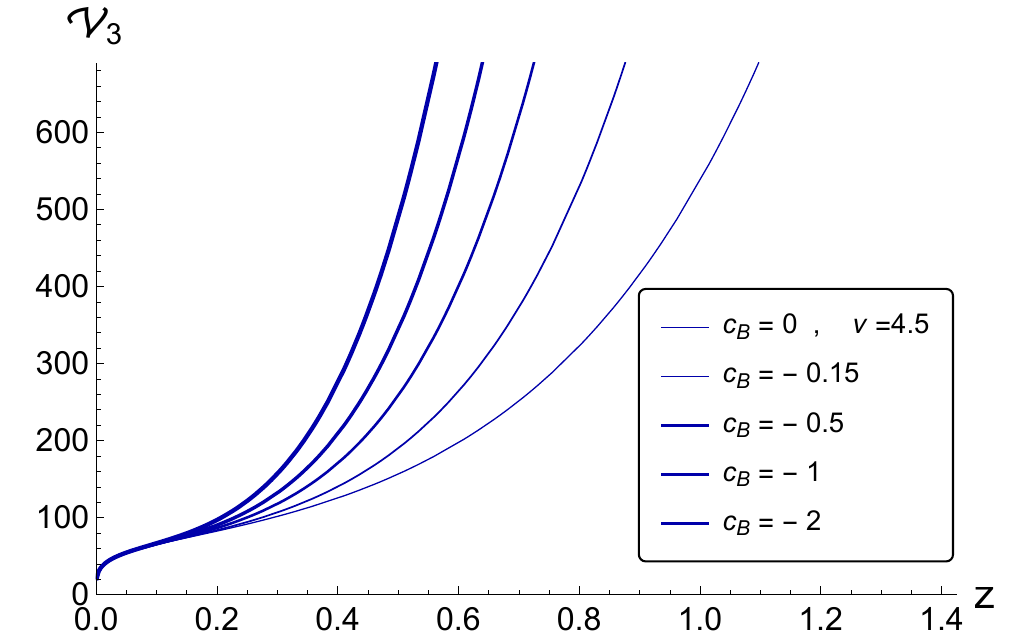} \\
  A \hspace{70mm} B
  \caption{Effective potential ${\cal V}_3(z)$ in the third orientation ${\cal{W}}_{y_{1}Y_{2}}$ for different anisotropies (A) $\nu = 1$, and (B) $\nu = 4.5$ at different $c_B$ considering zero-boundary condition \eqref{zerobc}; $[z]^{-1} = [c_B]^{\frac{1}{2}} =$ GeV.
  }
  \label{Fig:V3}
\end{figure}

The effective potential ${\cal V}_3(z)$ in the third orientation ${\cal{W}}_{y_{1}Y_{2}}$ for $\nu = 1$ and $\nu = 4.5$ in different $c_B$ considering the zero-boundary condition \eqref{zerobc} is shown in Fig.~\ref{Fig:V3}. In $\nu=1$, Fig.~\ref{Fig:V3}A, the effective potential curves and the DW coordinates are exactly the same as the results of the second SWL orientation ${\cal{W}}_{xY_{2}}$, that is clear from equations \eqref{sigmaxY2} and \eqref{sigmayY2}. However, for $\nu=1$ there are DW coordinates, but for $\nu=4.5$, Fig.~\ref{Fig:V3}B, the curves of the effective potential have no minimum and therefore no DW coordinates.

\begin{figure}[b!]
  \centering
  \includegraphics[scale=0.41]{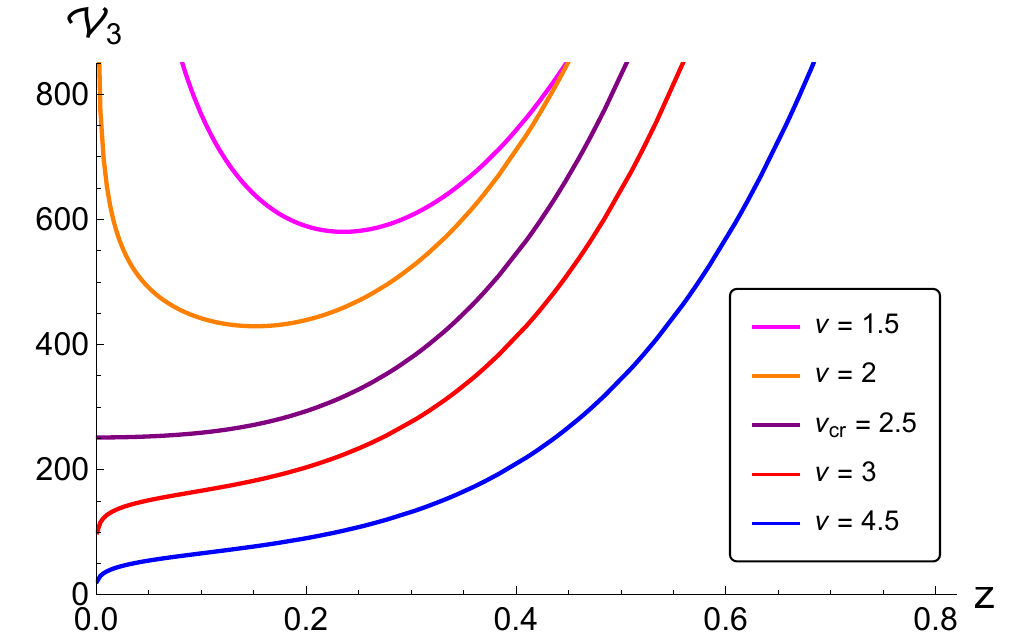}\quad
  \includegraphics[scale=0.43]{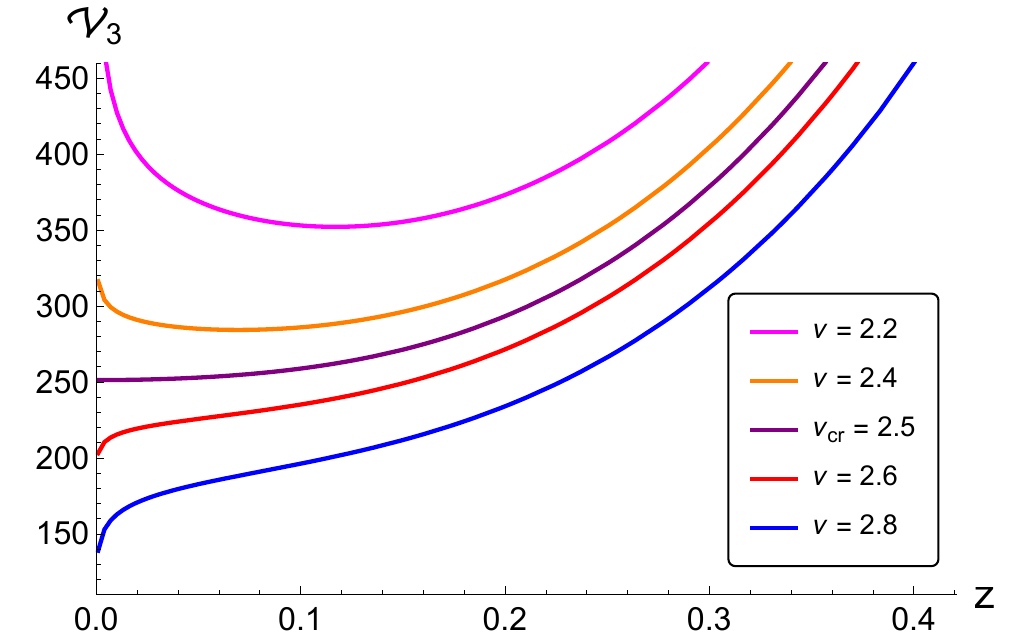}\\
  A \hspace{80 mm} B 
  \caption{Effective potential ${\cal V}_3(z)$ in the third orientation ${\cal{W}}_{y_{1}Y_{2}}$ at fixed value of magnetic field $c_B=-\,1$ for (A) different anisotropies $\nu$, and (B) zoom of panel (A) around the critical value of spatial anisotropy $\nu_{cr}=2.5$ considering zero-boundary condition \eqref{zerobc}; $[z]^{-1} = [c_B]^{\frac{1}{2}} =$ GeV.
  }
  \label{Fig:sigma3-2-nu2253}
\end{figure}

It is important to note that the third SWL orientation ${\cal{W}}_{y_{1}Y_{2}}$ in the anisotropic system has a different characteristic compared to other orientations. We found that there is a critical value of anisotropy $\nu_{cr} = 2.5$, where the DW coordinates disappear. To see this feature, in Fig.~\ref{Fig:sigma3-2-nu2253}A the effective potential ${\cal V}_3(z)$ in the third orientation ${\cal{W}}_{y_{1}Y_{2}}$ for fixed value of the magnetic field $c_B=-\,1$~GeV$^2$ and different $\nu$ considering the zero-boundary condition \eqref{zerobc} is shown. In Fig.~\ref{Fig:sigma3-2-nu2253}B, the critical value of anisotropy $\nu_{cr}=2.5$ is depicted in zoomed mode. For $\nu\geq2.5$ there are no DW coordinates in the system and therefore no spatial string tension.

\begin{figure}[h!]
  \centering
  \includegraphics[scale=0.42]{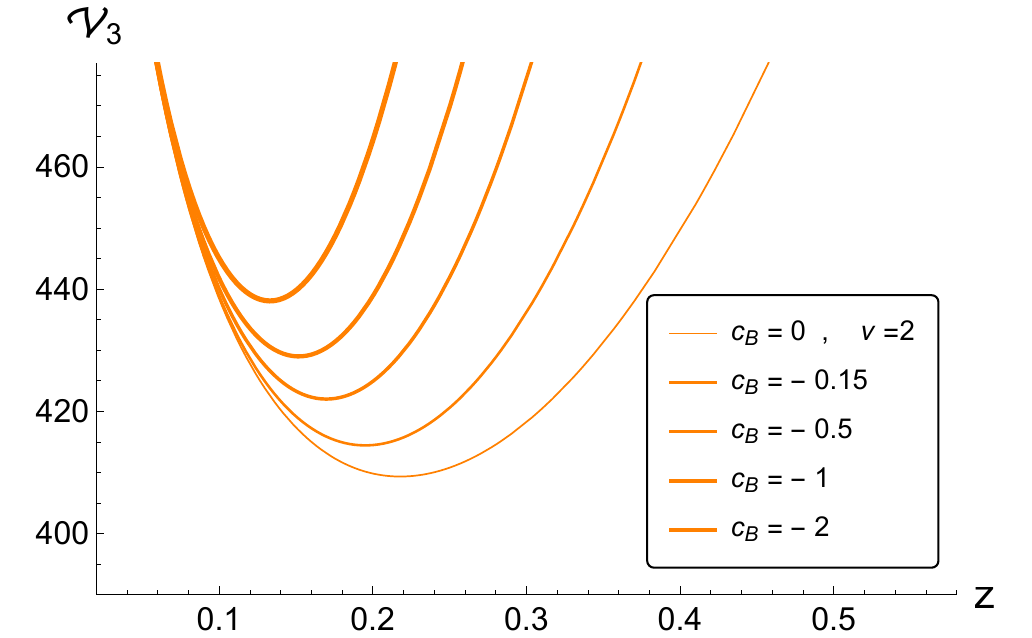} 
  \caption{Effective potential ${\cal V}_3(z)$ in the third orientation ${\cal{W}}_{y_{1}Y_{2}}$ for $\nu = 2$ at different values of magnetic field $c_B$ considering zero-boundary condition \eqref{zerobc}; $[z]^{-1} = [c_B]^{\frac{1}{2}} =$ GeV.
  }
  \label{Fig:sigma3-2-nu452}
\end{figure}

Due to the existence of $\nu_{cr}$ in the third SWL orientation ${\cal{W}}_{y_{1}Y_{2}}$ we should consider anisotropy $\nu<2.5$ to investigate the DW coordinate and spatial string tension. We choose $\nu=2$ and depict the potential ${\cal V}_3(z)$ at different values of magnetic field $c_B$ considering the zero-boundary condition \eqref{zerobc} in Fig.~\ref{Fig:sigma3-2-nu452}. The minimum of the curves in Fig.~\ref{Fig:sigma3-2-nu452} determines the DW coordinates presented in Table~\ref{tab:V3nu2} for $\nu=2$.

\begin{table}[h!]
  \centering
  \begin{tabular}{|l|c|c|l|l|l|l|l|}
    \hline
    \multicolumn{1}{|c|}{${\cal V}_3$} & \multicolumn{5}{c|}{$\nu=2$} \\ 
    \hline
    $-\,c_B$ & 0 & 0.15 & \multicolumn{1}{c|}{0.5} & \multicolumn{1}{c|}{1} & \multicolumn{1}{c|}{2}
    \\
    \hline
    $z_{DW}$ & \multicolumn{1}{l|}{0.218} & 0.195 & 0.170 & 0.152 & 0.133 \\
    \hline
  \end{tabular}
  \caption{Locations of dynamical walls $z_{DW}$, for ${\cal V}_3$ in the third orientation ${\cal{W}}_{y_{1}Y_{2}}$ at $\nu = 2$ considering zero-boundary condition \eqref{zerobc}; $[z]^{-1} = [c_B]^{\frac{1}{2}} =$ GeV.
  }
  \label{tab:V3nu2}
\end{table}

\begin{figure}[h!]
  \centering
  \includegraphics[scale=0.44]{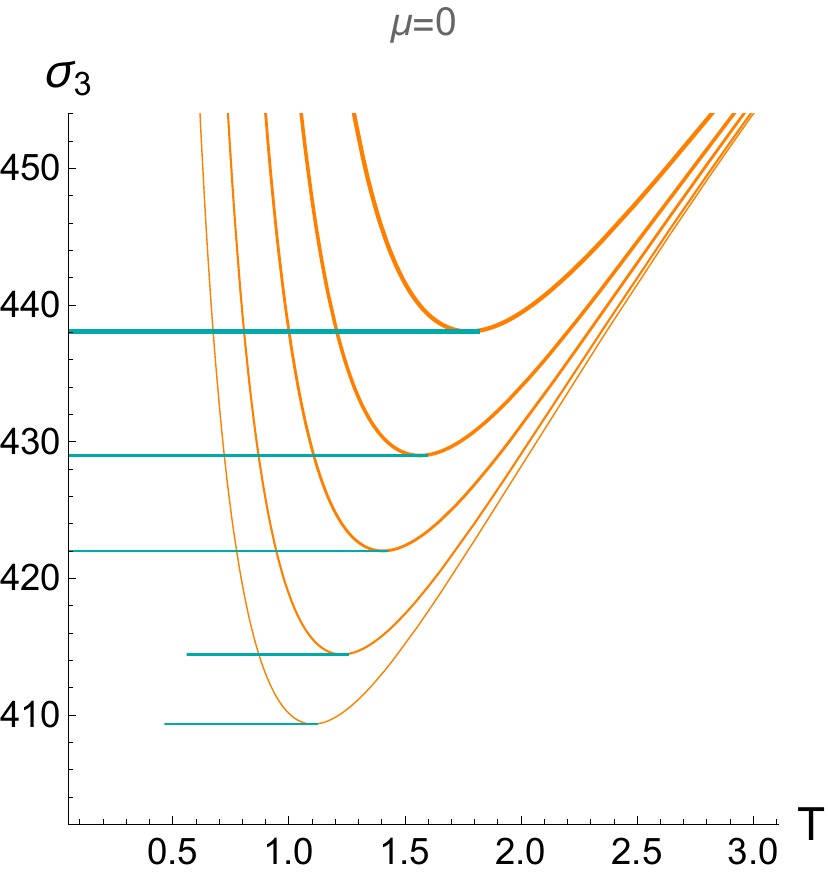}\qquad
  \includegraphics[scale=0.44]{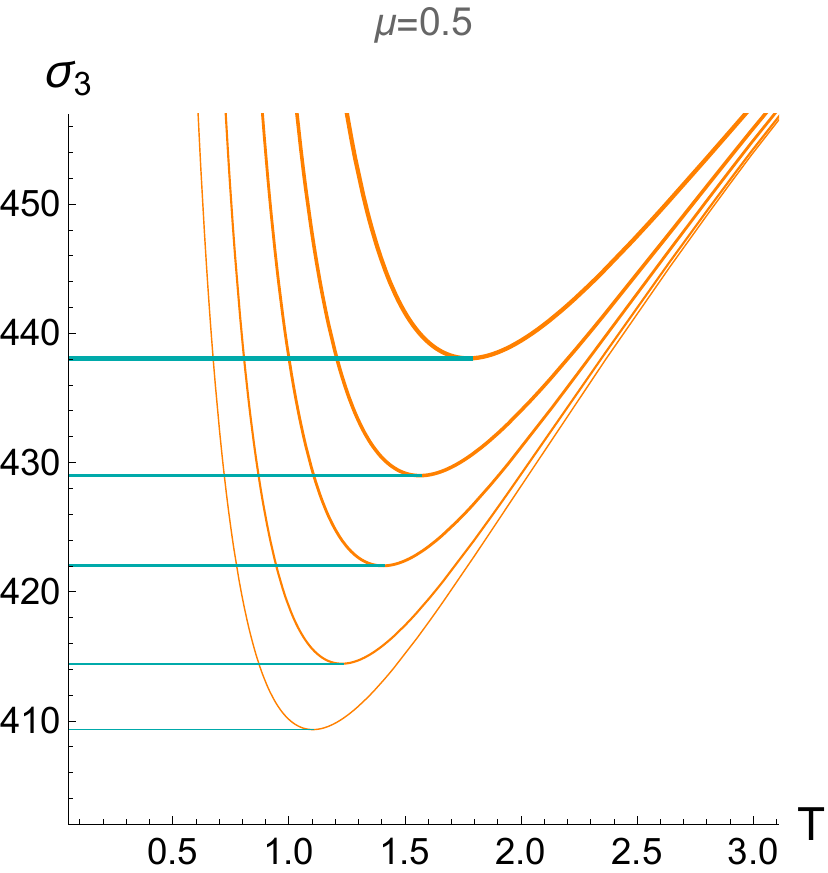}\\
  \qquad A \hspace{70 mm} B \\
  \includegraphics[scale=0.45]{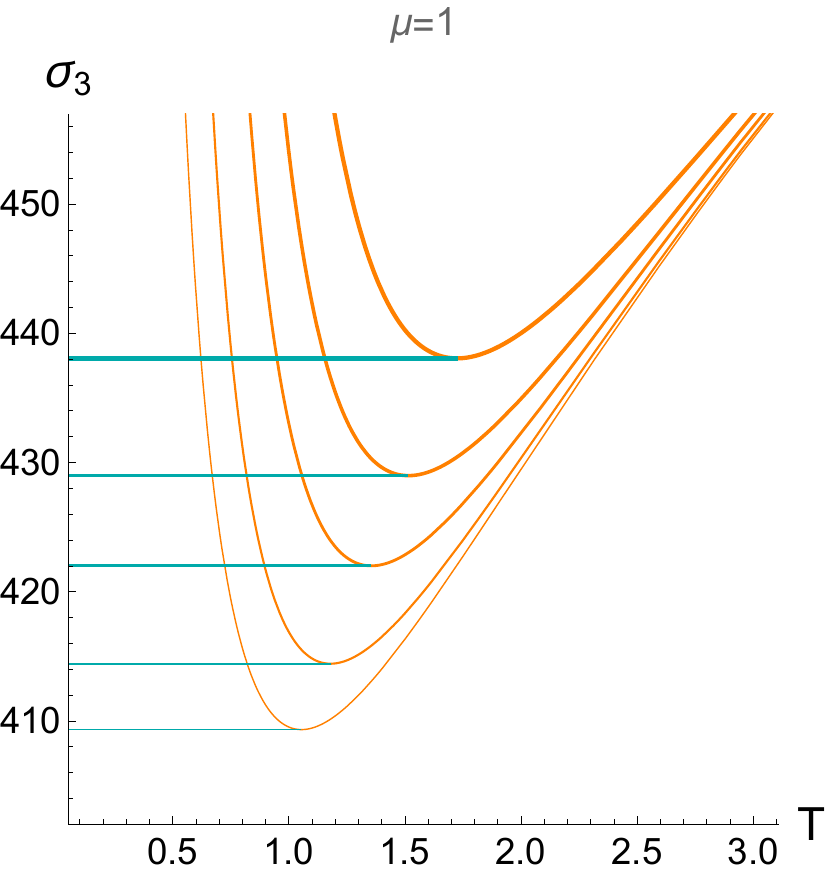}
  \qquad \qquad \quad
  \includegraphics[scale=0.49]{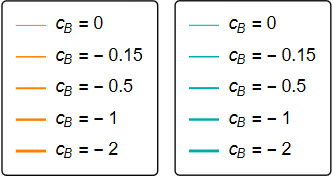}\\
  C \hspace{50 mm} 
  \caption{Spatial string tension $\sigma_3$ in the third orientation ${\cal{W}}_{y_{1}Y_{2}}$ as a function of  temperature $T$ for isotropic case $\nu=2$ at (A) $\mu = 0$, (B) $\mu = 0.5$ GeV, and (C) $\mu = 1$ GeV with different $c_B$ considering zero-boundary condition \eqref{zerobc}. The cyan line and orange curve show DW and horizon configuration, respectively; $[\sigma]^{\frac{1}{2}}=[T]=[\mu] = [c_B]^{\frac{1}{2}} =$ GeV.
  }
  \label{Fig:sigma-3-nu2}
\end{figure}

Fig.~\ref{Fig:sigma-3-nu2} shows the spatial string tension $\sigma_3$ as a function of temperature $T$ for the anisotropic case $\nu=2$ in the third SWL orientation ${\cal{W}}_{y_{1}Y_{2}}$ at $\mu = 0$, $\mu = 0.5$~GeV and $\mu = 1$ GeV and different magnetic fields $c_B$ considering the zero-boundary condition \eqref{zerobc}. The cyan lines present the dependence of $\sigma(z_{DW})$ on temperature in the DW configuration, while the orange lines present the dependence of $\sigma(z_h)$ on temperature in the horizon configuration. The spatial string tension exhibits a constant value in the DW configuration at $T<T_{cr}$ such as $\sigma_3 = 438$ GeV$^2$ at $\mu=0$ and $c_B=-\,2$~GeV$^2$. However, in the horizon configuration at $T>T_{cr}$ it increases monotonically.

\begin{figure}[h!]
  \centering
  \includegraphics[scale=0.41]{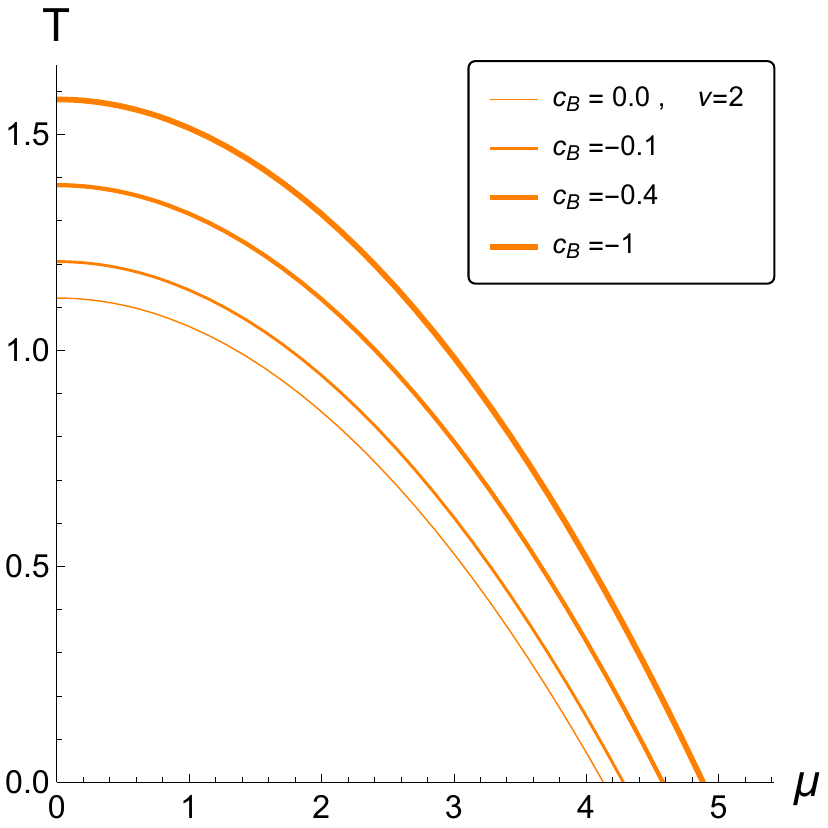}\qquad \qquad
  \includegraphics[scale=0.41]{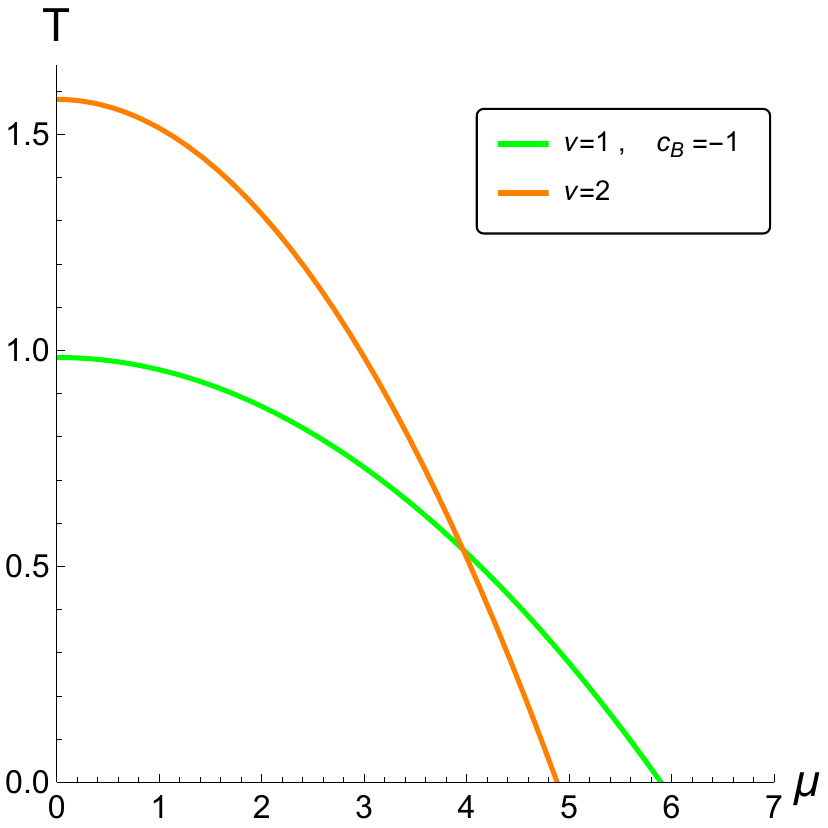}\\
   A \hspace{60 mm} B \\
  \caption{Phase diagrams in $(\mu,T)$-plane at (A) $\nu = 2$, and (B) comparison between $\nu = 1$ and $\nu = 2$ at fixed $c_B=-\,1$ GeV$^2$ in the third orientation ${\cal{W}}_{y_{1}Y_{2}}$, corresponding to the transition of $\sigma_3$ from the DW configuration to the horizon configuration, considering zero-boundary condition \eqref{zerobc}; $[T]=[\mu] = [c_B]^{\frac{1}{2}} =$ GeV.
  }
  \label{Fig:sigma-3-nu12}
\end{figure}

Fig.~\ref{Fig:sigma-3-nu12}A shows the phase diagrams in the $(\mu,T)$-plane at  $\nu = 2$, and Fig.~\ref{Fig:sigma-3-nu12}B presents the comparison between $\nu = 1$ and $\nu = 2$ with fixed magnetic field $c_B=-\,1$ GeV$^2$ in the third SWL orientation ${\cal{W}}_{y_{1}Y_{2}}$, corresponding to the transition from the DW configuration to the horizon configuration of $\sigma_3$, considering the zero-boundary condition \eqref{zerobc}. Fig.~\ref{Fig:sigma-3-nu12}A confirms the magnetic catalysis behavior at $\nu = 2$, and Fig.~\ref{Fig:sigma-3-nu12}B shows that inclusion of anisotropy increases or decreases the transition temperature depending on $\mu$ in this case.

\newpage
$$\,$$
\newpage

\subsection{Phase diagrams in particular SWL orientations} \label{com123}

\begin{figure}[h!]
  \centering
  \includegraphics[scale=0.41]{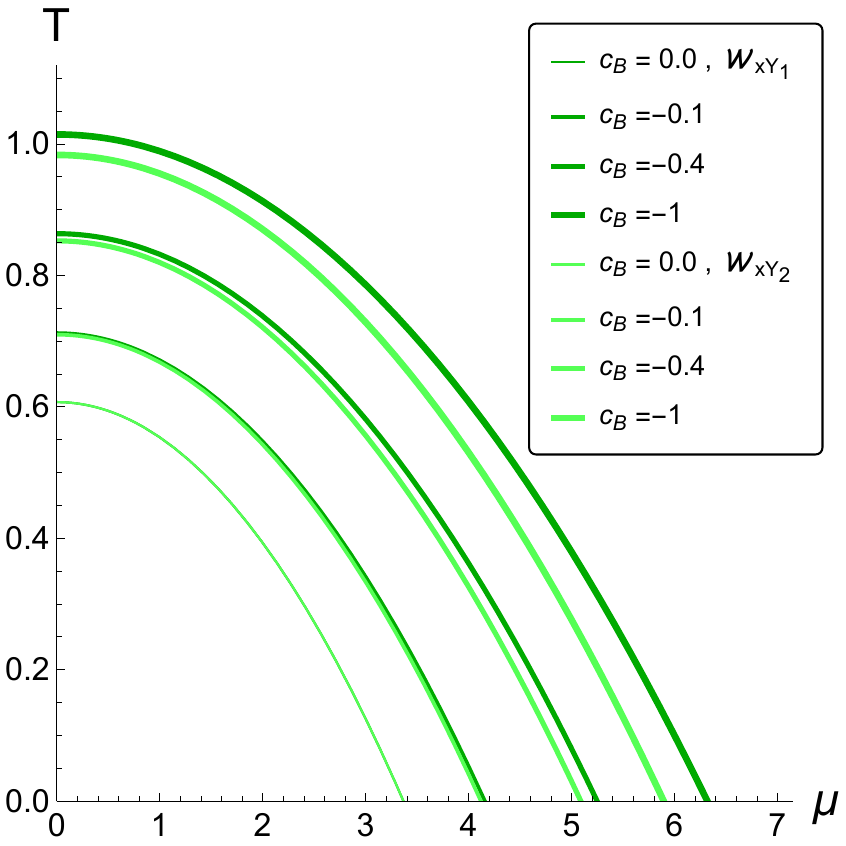}\qquad
  \includegraphics[scale=0.41]{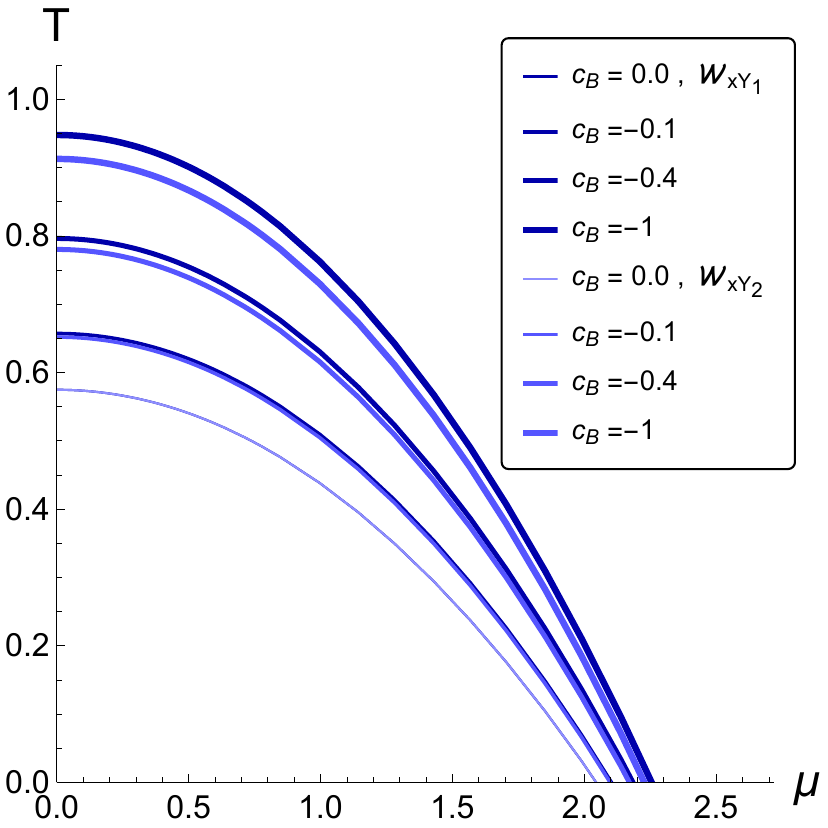}\\
   A \hspace{60 mm} B 
  \caption{Phase diagrams in $(\mu,T)$-plane at (A) $\nu = 1$, and (B) $\nu = 4.5$ in the first orientation ${\cal{W}}_{x Y_{1}}$ and the second orientation ${\cal{W}}_{x Y_{2}}$, corresponding to the transition of $\sigma_3$ from the DW configuration to the horizon configuration considering zero-boundary condition \eqref{zerobc}; $[T]=[\mu] = [c_B]^{\frac{1}{2}} =$ GeV.
  }
  \label{Fig:sigma-12-nu145}
\end{figure}

Different orientations of SWL mean different embeddings of the string in the bulk. In addition, different embeddings lead to different physics \cite{Casalderrey-Solana:2011dxg}, that is presented in Fig.~\ref{Fig:sigma-12-nu145}. 

To compare the phase diagrams in the $(\mu,T)$-plane in the first ${\cal{W}}_{x Y_{1}}$ and second orientation ${\cal{W}}_{x Y_{2}}$ at $\nu = 1$, and (B) $\nu = 4.5$ Figs.~\ref{Fig:sigma-12-nu145}A and \ref{Fig:sigma-12-nu145}B are presented, respectively. The transition temperature characterizes the phase transition between the DW configuration and the horizon configuration of $\sigma_3$ considering the zero-boundary condition \eqref{zerobc}. Our results reveal, that for the isotropic $\nu=1$ and anisotropic $\nu=4.5$ cases the transition temperature  has lower values for the second SWL orientation ${\cal{W}}_{x Y_{2}}$ compared to the first orientation. Note that for $c_B=0$ the phase transition curves coincide exactly as the values of $\sigma$ for both orientations are the same, see \eqref{sigmaxY1} and \eqref{sigmaxY2}.\\

\begin{figure}[h!]
  \centering
 \includegraphics[scale=0.50]{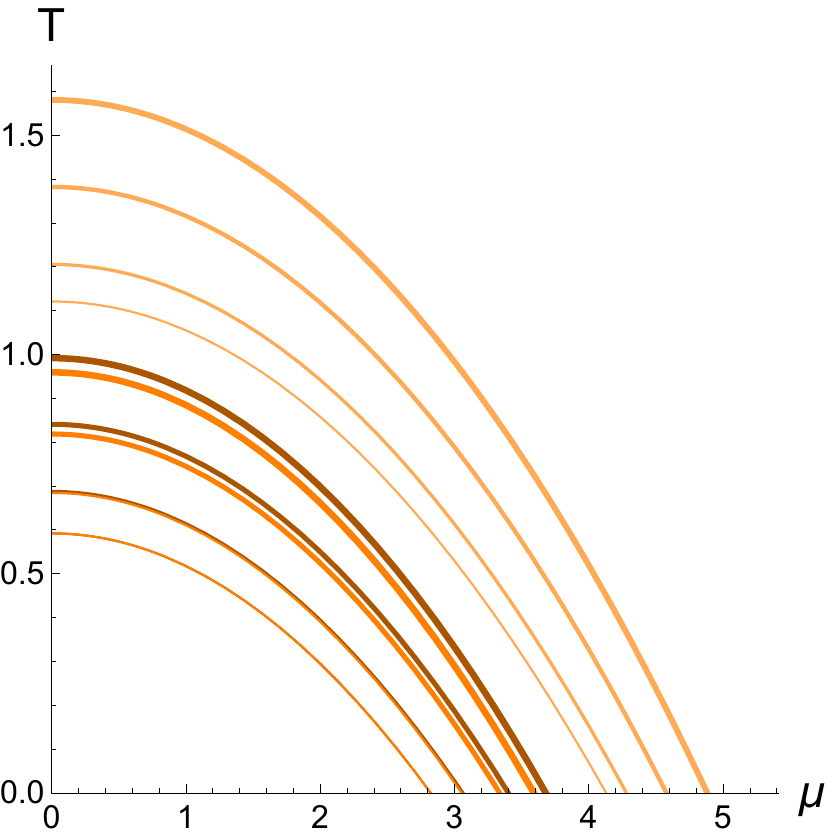}
 \qquad   \quad
 \includegraphics[scale=0.50]{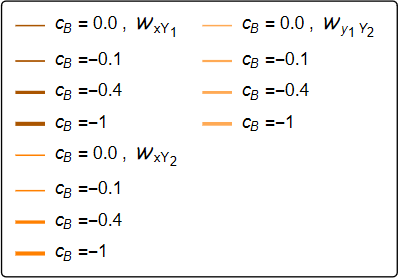}\\
  C \hspace{70 mm}
 \caption{Phase transition diagrams in $(\mu,T)$-plane at $\nu = 2$ for the first ${\cal{W}}_{x Y_{1}}$, and the second ${\cal{W}}_{x Y_{2}}$ with the third orientation ${\cal{W}}_{y_{1}Y_{2}}$, corresponding to the transition of $\sigma_3$ from the DW configuration to the horizon configuration at zero-boundary condition \eqref{zerobc}; $[T]=[\mu] = [c_B]^{\frac{1}{2}} =$ GeV.
      }
  \label{Fig:sigma-123-nu2}
\end{figure}

In Fig.~\ref{Fig:sigma-123-nu2} the comparison of phase transition diagrams in the $(\mu,T)$-plane at $\nu = 2$ among all three orientations, i.e. the first ${\cal{W}}_{x Y_{1}}$, the second ${\cal{W}}_{x Y_{2}}$ and the third orientation ${\cal{W}}_{y_{1}Y_{2}}$, corresponding to the transition of $\sigma_3$ from the DW configuration to the horizon configuration at the zero-boundary condition \eqref{zerobc} is shown. The result of Fig.~\ref{Fig:sigma-12-nu145} is confirmed for the first ${\cal{W}}_{x Y_{1}}$ and the second ${\cal{W}}_{x Y_{2}}$ orientations. In addition, Fig.~\ref{Fig:sigma-123-nu2} shows that the transition temperature  has higher values for the third SWL orientation ${\cal{W}}_{y_{1}Y_{2}}$ in comparison to the first and second ones for all values of the magnetic field $c_B$.

\newpage
$$\,$$
\newpage


\newpage
$$\,$$

\section{Comparing the spatial string tension with lattice calculations}\label{drag}

The connection between spatial string tensions and drag forces is remarkable.
The drag force specifies the resistance that the heavy quark undergoes while moving through the hot and dense QGP. The drag force on the stretched string can describe the energy loss of the heavy quark \cite{Herzog:2006gh,Gubser:2006bz}.
It was shown that in the strong coupling regime the drag force is proportional to the spatial string tension $\sigma \propto T^2$, where $T$ denotes temperature \cite{Sin:2006yz,Andreev:2017bvr}. In addition, the temperature dependence of the spatial string tension is studied in more detail in~\cite{Andreev:2006eh}. 

Note that for the general metric \eqref{Gbackgr} considering $\fg_1=1$ the drag forces on the heavy quarks in the $x$, $y_1$, and $y_2$ directions have already been calculated in the anisotropic model \cite{Arefeva:2020vhf} as 
\bea
  p_x&=& v_x\,\frac{\fb_s(z_h)}{z_h^2} 
  \label{px}, \label{px}\\
  p_{y_1}&=& v_{y_1}\,\frac{\fb_s(z_h)}{z_h^2}\,\fg_2 (z_h)
  \label{py1}, \label{py1} \\
  p_{y_2}&=& v_{y_2}\,\frac{\fb_s(z_h)}{z_h^2}\,\fg_3(z_h)
  \label{py2}.
\eea
If we take into account $L^2/\alpha'=\sqrt{\lambda}$ and identify 
\bea
  v_x&=&v \,\sqrt{\fg_2}, \\
  v_{y_1}&=&v \,\frac{\sqrt{\fg_3}}{\fg_2}, \\
  v_{y_2}&=&v \,\frac{\sqrt{\fg_2}}{\sqrt{\fg_3}},
\eea
the drag forces \eqref{px}--\eqref{py2} can reproduce spatial string tensions \eqref{sigmaxY1}, \eqref{sigmaxY2}, and \eqref{sigmayY2} at $z=z_h$:
\bea\label{sigmaxY1nd}
  \sigma_{xY_1}\Big|_{z = z_h}
  &=&\frac{\sqrt{\lambda}}{2\pi }\left(\frac{ \fb_s(z_h)}{z_h^2}\right) \sqrt{\fg_{1}\fg_{2}},\\
  \sigma_{xY_2}\Big|_{z = z_h}
  &=&\frac{\sqrt{\lambda}}{2\pi}\left(\frac{ \fb_s(z_h)}{z_h^2}\right) \sqrt{\fg_{1}\fg_{3}},
  \\
  \sigma_{y_1Y_2}\Big|_{z = z_h}
  &=&\frac{\sqrt{\lambda}}{2\pi}\left(\frac{\fb_s(z_h)}{z_h^2}\right) \sqrt{\fg_{2}\fg_{3}}
  \label{sigmayY2nd}
\eea
for some constant $v$ and $\fg_1=1$. Therefore, the spatial string tension is proportional to the drag force in the anisotropic medium.

It is interesting to investigate the temperature dependence of the spatial string tension. Taking $\fg_1=1$, $\fg_2(z_h) = (z_h/L)^{2-2/\nu}$, and $\fg_3(z_h) = (z_h/L)^{2-2/\nu}e^{c_Bz_h^2}$, we~have 
\bea\label{sigmaxY1nd}
  \sigma_{1}\equiv\,
  \sigma_{xY_1}\Big|_{z = z_h}
  &=&
  \frac{1}{2\pi \alpha'}\left(\frac{L^{1+1/\nu}    \fb_s(z_h)}{z_h^{1+1/\nu}}\right),\\
  \sigma_{2}\equiv\,
  \sigma_{xY_2}\Big|_{z = z_h}
  &=&
  \frac{1}{2\pi \alpha'}\left(\frac{L^{1+1/\nu} \fb_s(z_h)}{z_h^{1+1/\nu}}\right)e^{c_Bz_h^2/2}, \label{sigmaxY2nd} \\
  \sigma_{3}\equiv\,   
  \sigma_{y_1Y_2}\Big|_{z = z_h}
  &=&
  \frac{1}{2\pi \alpha'}\left(\frac{L^{2/\nu} \fb_s(z_h)}{z_h^{2/\nu}}\right)e^{c_Bz_h^2/2}. \label{sigmayY2nd}
\eea

\begin{figure}[h!]
  \centering
  \includegraphics[scale=0.44]{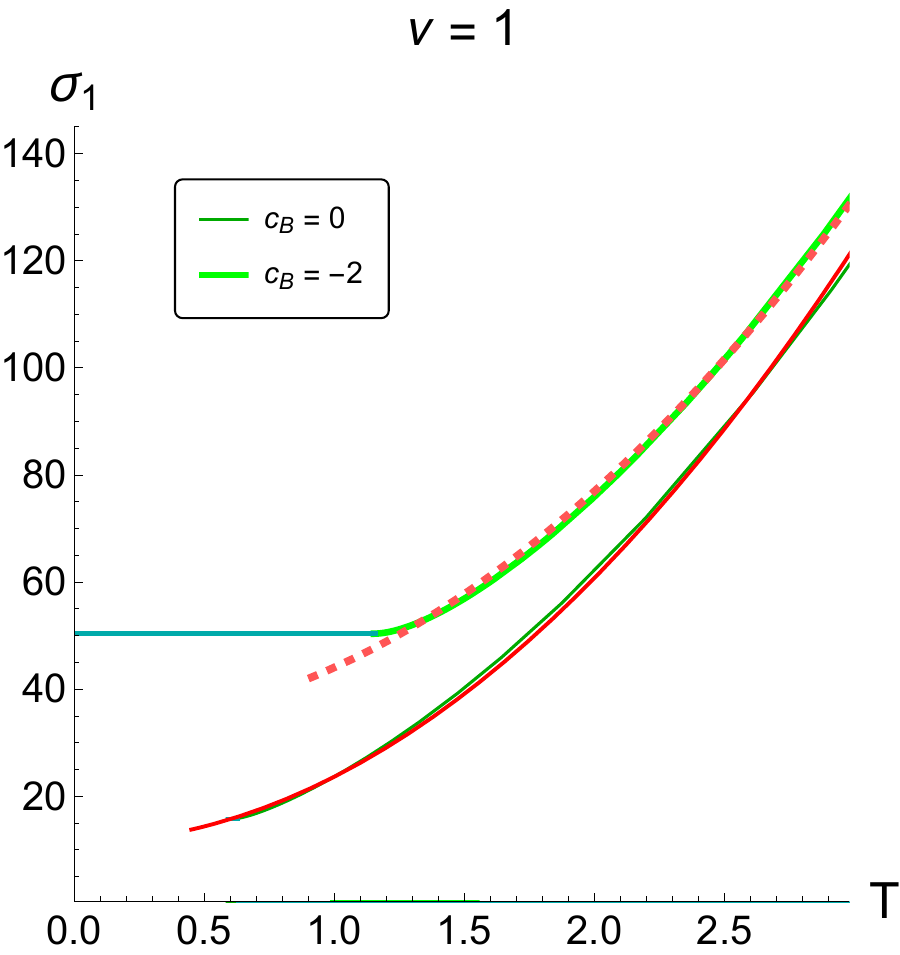}
  \qquad
  \includegraphics[scale=0.44]{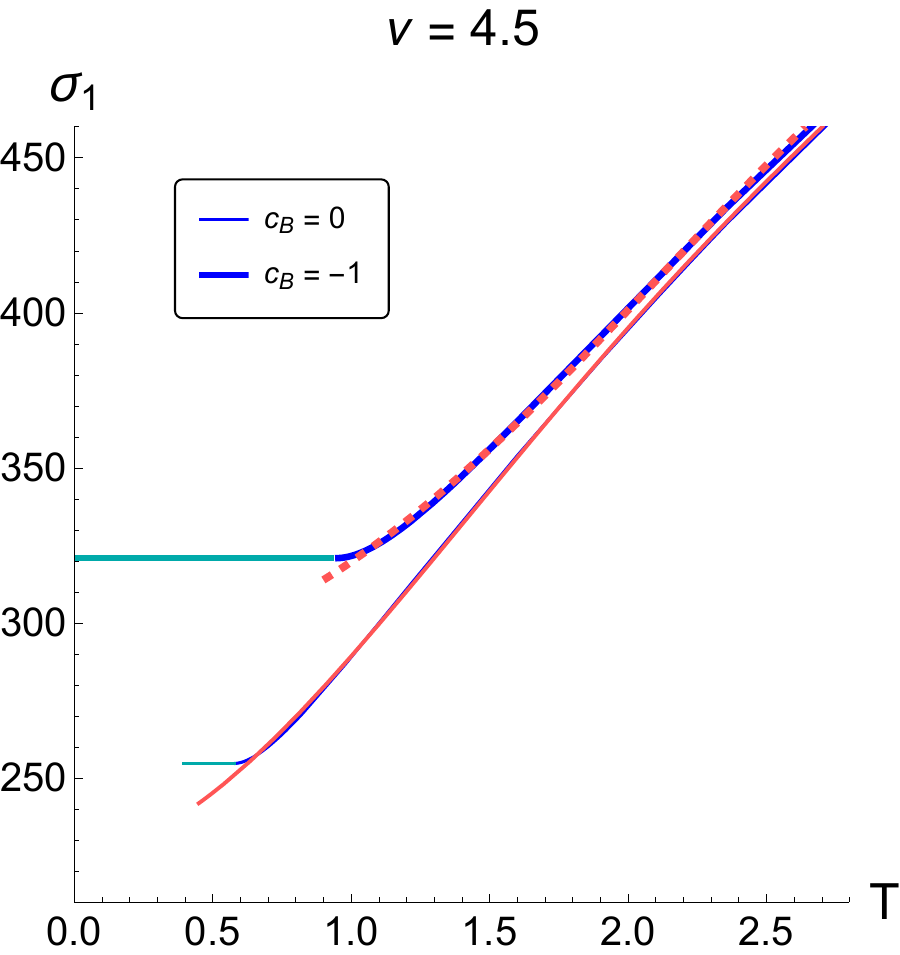}\\
  A \hspace{70 mm}B
  \caption{Spatial string tension $\sigma_1$ in the first orientation ${\cal{W}}_{x Y_{1}}$ as a function of temperature $T$ at $\mu = 0$ for (A) isotropic case $\nu=1$ that the solid red curve is $11.2 + 12.4\,T^2$, the dashed red curve is $33.1 + 10.9\,T^2$, and for (B) anisotropic case $\nu=4.5$, the solid red curve is $226.1 + 91.3\,T^2 - 31.9\,T^3 + 3.7\,T^4$ and the dashed red curve is $288.2 + 31.9\,T^2 + 0.8\,T^3 - 1.4\,T^4$;  $[\sigma]^{\frac{1}{2}}=[T]=[\mu] = [c_B]^{\frac{1}{2}} =$ GeV.
  }
  \label{Fig:sigma-T-v1}
\end{figure}

Fig.~\ref{Fig:sigma-T-v1}A presents the  string tension $\sigma_1$ in the first orientation ${\cal{W}}_{x Y_{1}}$ as a function of the temperature $T$ at $\mu = 0$ for the isotropic case $\nu=1$. The fitted solid red curve is $11.2 + 12.4\,T^2$, and the fitted dashed red curve is $33.1 + 10.9\,T^2$. Fig.~\ref{Fig:sigma-T-v1}B depicts the anisotropic case $\nu=4.5$ at $\mu = 0$. The fitted solid red curve is $226.1 + 91.3\,T^2 - 31.9\,T^3 + 3.7\,T^4$, and the fitted dashed red curve is $288.2 + 31.9\,T^2 + 0.8\,T^3 - 1.4\,T^4$. Our result shows, that for the fully isotropic case $\nu=1$ and $c_B=0$ we have $\sigma_1 \propto T^2$ which is compatible with the result reported in \cite{Gubser:2006bz,Sin:2006yz,Andreev:2006eh} and also is qualitatively consistent with the lattice results \cite{Bala:2025ilf,Maezawa:2007fc}. Inclusion of spatial anisotropy $\nu=4.5$ clearly shows that the behavior of $\sigma(T)$ deviates from the quadratic term and requires higher terms. 

\begin{figure}[t]
  \centering
  \includegraphics[scale=0.44]{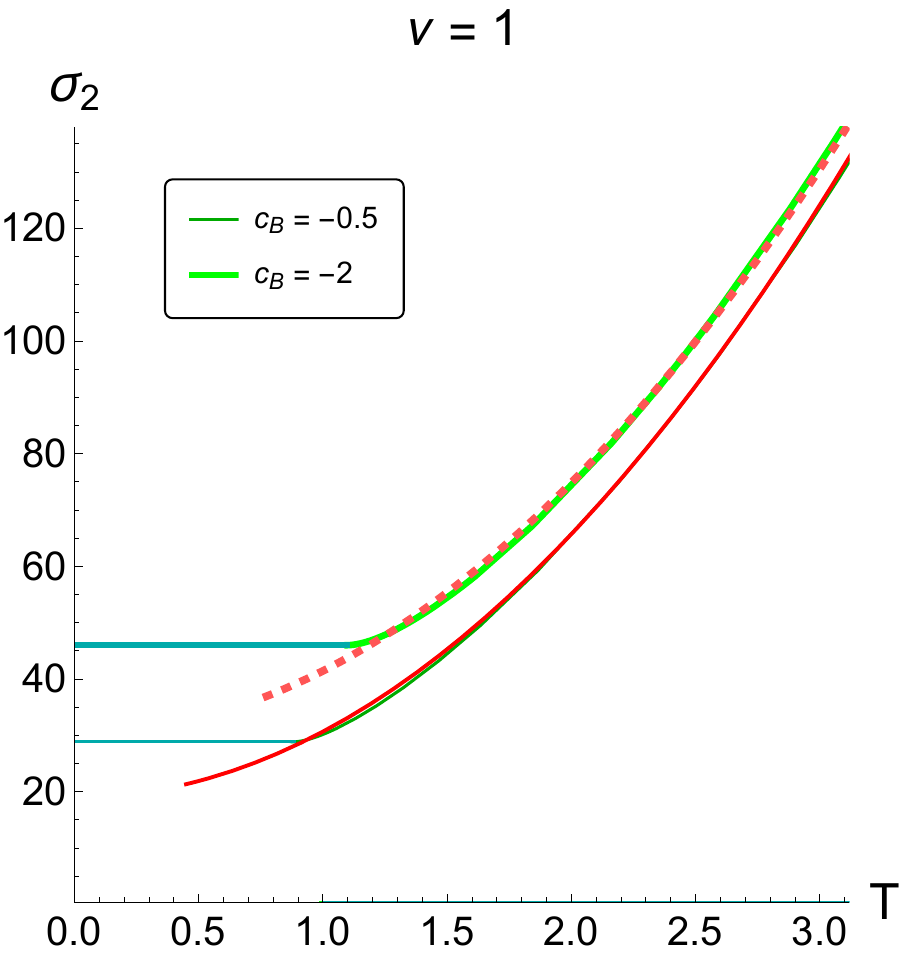}
  \qquad
  \includegraphics[scale=0.44]{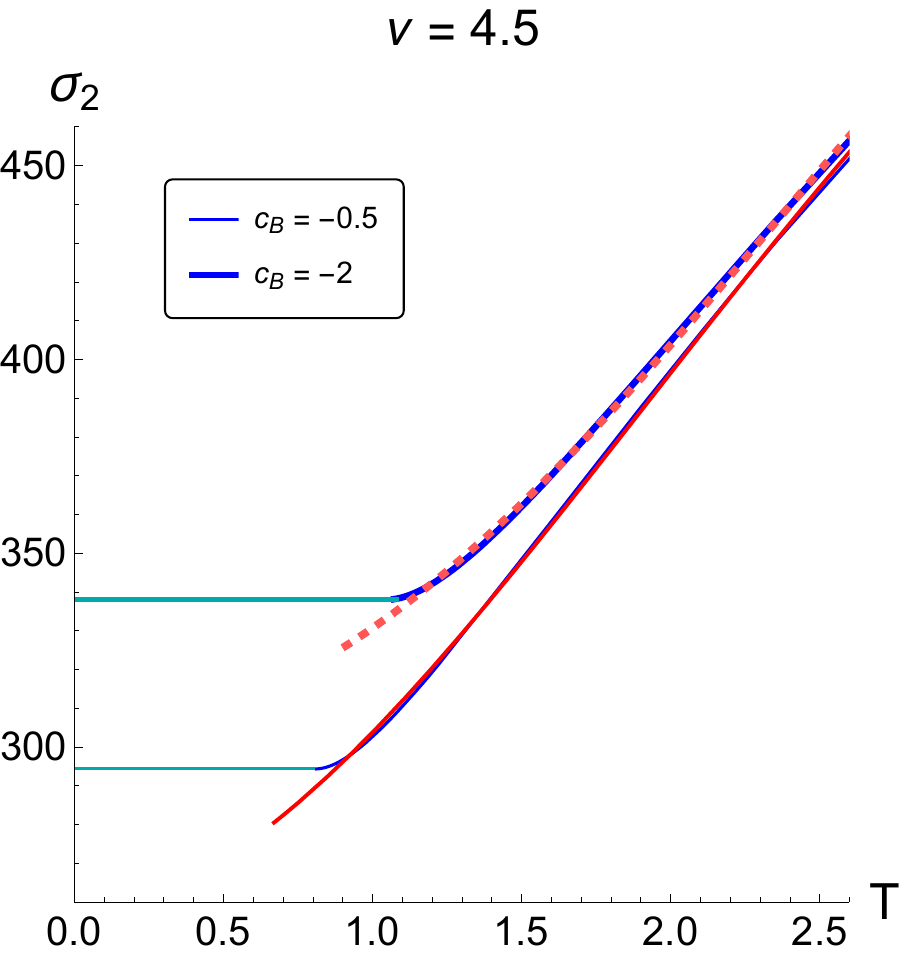}\\
  A \hspace{70 mm}B
  \caption{Spatial string tension $\sigma_2$ in the second orientation ${\cal{W}}_{x Y_{2}}$ as a function of  temperature $T$ at $\mu = 0$ for (A) isotropic case $\nu=1$ that the solid red curve is $18.9 + 11.7 \,T^2$, the dashed red curve is $30.2 + 11.1\,T^2$, and for (B) anisotropic case $\nu=4.5$ the solid red curve is $258.3 + 58.1\, T^2 - 13.6\, T^3 + 0.9\, T^4$, and the dashed red curve is $303.6 + 26.7\, T^2 + 1.3 \,T^3 - 1.1\, T^4$;  $[\sigma]^{\frac{1}{2}}=[T]=[\mu] = [c_B]^{\frac{1}{2}} =$ GeV. 
  }
  \label{Fig:sigma-T-v2}
\end{figure}

Fig.~\ref{Fig:sigma-T-v2}A shows the temperature dependence of the string tension $\sigma_2$ in the second orientation ${\cal{W}}_{x Y_{2}}$ at $\mu = 0$ for the isotropic case $\nu=1$. Fig.~\ref{Fig:sigma-T-v2}B presents the fitted solid red curve is $18.9 + 11.7 \,T^2$ and the fitted dashed red curve is $30.2 + 11.1\,T^2$ for the anisotropic case $\nu=4.5$ at $\mu = 0$, the fitted solid red curve is $258.3 + 58.1\, T^2 - 13.6\, T^3 + 0.9\, T^4$ and the fitted dashed red curve is $303.6 + 26.7\, T^2 + 1.3 \,T^3 - 1.1\, T^4$. In the isotropic case $\nu=1$ at different values of the magnetic field $c_B$ we get $\sigma_2 \propto T^2$. Note that our results for non-zero magnetic field $c_B\neq0$ are our prediction and can probably be checked with lattice calculations. For the anisotropic case $\nu=4.5$ $\sigma(T)$ is proportional to the higher terms of $T$.\\

\begin{figure}[h!]
  \centering
  \includegraphics[scale=0.44]{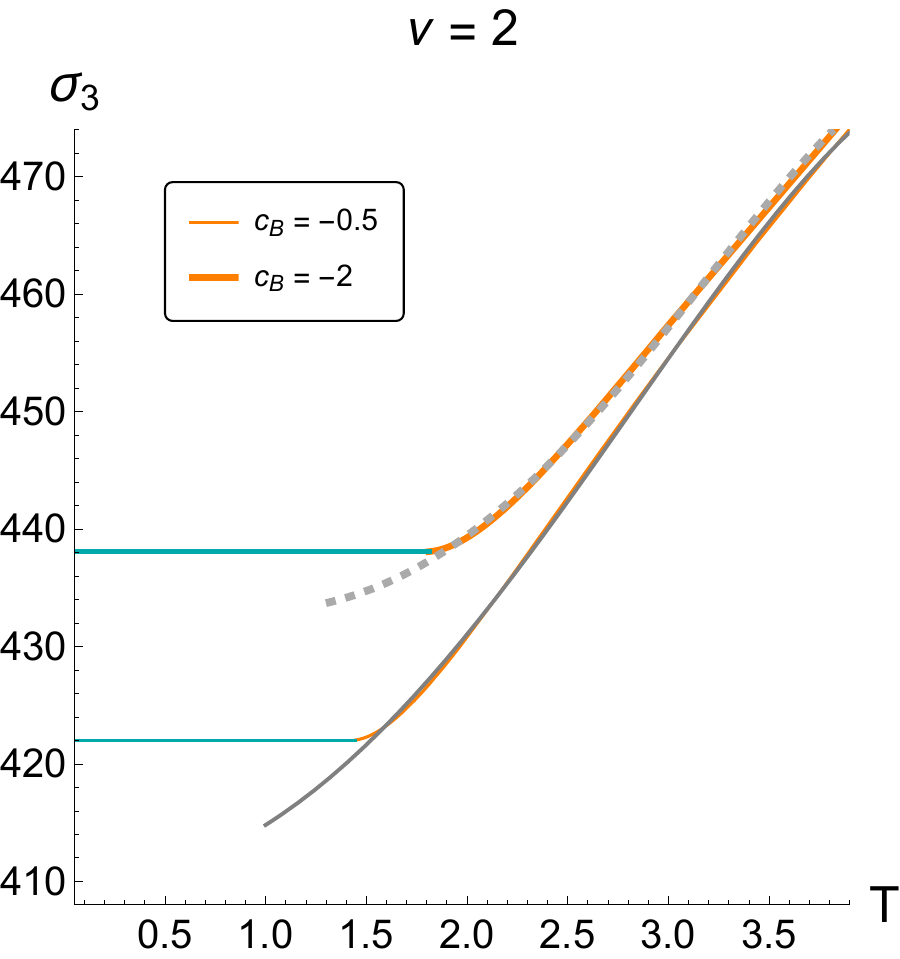}\\
  \caption{Spatial string tension $\sigma_3$ in the third orientation ${\cal{W}}_{y_{1}Y_{2}}$ as a function of  temperature $T$ at $\mu = 0$ for  anisotropic case $\nu=2$ the solid gray curve is $409.6 + 4.5\,T^2 + 0.9\,T^3 - 0.3\,T^4$, and the dashed gray curve is $433.8 - 4.3\,T^2 + 3.9\,T^3 - 0.6\, T^4$;  $[\sigma]^{\frac{1}{2}}=[T]=[\mu] = [c_B]^{\frac{1}{2}} =$ GeV. 
  }
  \label{Fig:sigma-T-v3}
\end{figure}

Applying equations \eqref{sigmaxY1nd}-\eqref{sigmayY2nd} to the third orientation ${\cal{W}}_{y_{1}Y_{2}}$ shows, that in the isotropic case $\nu=1$ with different $c_B$ the results for $\sigma(T)$ are the same as in the second orientation. Therefore, in Fig.~\ref{Fig:sigma-T-v3} the temperature dependence of $\sigma_3$ is investigated only for the anisotropic case $\nu=2$ at $\mu=0$ and different $c_B$. The solid gray curve is $409.6 + 4.5\,T^2 + 0.9\,T^3 - 0.3\,T^4$, and the dashed gray curve is $433.8 - 4.3\,T^2 + 3.9\,T^3 - 0.6\, T^4$. Therefore, in this case $\sigma(T)$ deviates from the quadratic behavior in the presence of spatial anisotropy.

It is important to note that for all orientations inclusion of the external magnetic field and the spatial anisotropy enhances the string tension in the horizon configuration, i.e. the drag force. Our results reveal, that for $T<T_{cr}$ the string tension value is temperature independent and gets a constant value. Similar behavior is reported in \cite{Andreev:2006eh} and also in lattice calculations \cite{Bala:2025ilf,Maezawa:2007fc}. In our model for all orientations at $T<T_{cr}$ the string tension profile is realized on the DW configuration and at $T>T_{cr}$ it is  realized on the horizon configuration  and it   increases monotonically.  \\

\newpage
$$\,$$

\section{Conclusion}
\label{Sect:conclusion}

In this paper, we investigate the effective potential and the spatial string tension $\sigma$ and its phase transitions between the DW configuration and the horizon configuration using spatial Wilson loops. To do this, we used the Einstein-dilaton-three-Maxwell action and incorporated a 5-dimensional metric with a special warp factor to construct an anisotropic holographic model under a strong magnetic field for the heavy-quark case \cite{Arefeva:2023jjh}. In addition, the effect of spatial anisotropy $\nu$, that reflects the noncentrality of HIC, on the string tension for three particular orientations is studied.

 We obtained the SWL and the effective potential for general orientations, namely the dependence of $\sigma$ on temperature $T$ is studied for three particular SWL orientations, i.e. $\sigma_i$, $i = 1, 2, 3$ corresponding to ${\cal{W}}_{x Y_{1}}$, ${\cal{W}}_{x Y_{2}}$, and ${\cal{W}}_{y_{1}Y_{2}}$, respectively.

Let us summarize our findings, depending on different orientations.

\begin{itemize}
\item \textit{The first SWL orientation ${\cal{W}}_{x Y_{1}}$:}
  \begin{itemize}
  \item some similarities between the zero-boundary condition \eqref{zerobc} and the physical-boundary condition \eqref{phi-fz-LQ}:
    \begin{itemize}
    \item the phase diagram and the DW coordinates do not depend on the boundary condition,
    \item inclusion of spatial anisotropy $\nu=4.5$ decreases the values of the critical transition $T$ between the DW and the horizon configuration in our model;
    \end{itemize}
  \item the main difference between the zero- and physical-boundary conditions:
    \begin{itemize}
    \item inclusion of spatial anisotropy $\nu=4.5$ and increasing the magnetic field $c_B$ causes the spatial string tension $\sigma_1$ to increase/decrease at zero/physical-boundary conditions,
    \item at very low $T$ in the case of $\mu=0$ with $c_B\neq0$, the $\sigma_1$ exists/does not exist at zero/physical-boundary conditions due to the absence of the DW coordinate.
    \end{itemize}  
  \end{itemize}
  \item \textit{The second SWL orientation ${\cal{W}}_{x Y_{2}}$:}
    \begin{itemize}
    \item the spatial string tension $\sigma$ increases with the spatial anisotropy $\nu$ and the external magnetic field $c_B$;
    \item the transition temperature occurs at lower values in the second SWL orientation, ${\cal{W}}_{x Y_{2}}$, compared to the first orientation, ${\cal{W}}_{x Y_{1}}$, for both the isotropic case $\nu=1$ and the anisotropic case $\nu=4.5$.
    \end{itemize}
  \item \textit{The third SWL orientation ${\cal{W}}_{y_{1}Y_{2}}$:}
    \begin{itemize}
    \item for $\nu=1$ the effective potential values and the DW coordinates are exactly the same as the results for the second SWL orientation ${\cal{W}}_{xY_{2}}$, see \eqref{sigmaxY2} and \eqref{sigmayY2};
    \item for $\nu=4.5$ there are no DW coordinates; in fact, there is a critical value of anisotropy $\nu_{cr} = 2.5$, i.e. at $\nu\geq 2.5$ the DW coordinates disappear;
    \item for $\nu=2$ the transition temperature has higher values in the third SWL orientation, ${\cal{W}}_{y_{1}Y_{2}}$, compared to the first ${\cal{W}}_{x Y_{1}}$ and second orientations ${\cal{W}}_{x Y_{2}}$ for all values of the magnetic field $c_B$.
  \end{itemize}
\end{itemize}

There also are some properties, that are common and independent of the different SWL orientations choice, which are summarized as follows.\\

\textit{Universal characteristics}

\begin{itemize}
    \item The phase transition has magnetic catalysis behavior, which means that the critical temperature $T_{cr}$ increases as the magnetic field $c_B$ increases. 
    \item As the magnetic field $c_B$ increases, the value of the DW coordinate decreases, meaning the phase transition occurs at a smaller holographic coordinate $z$ closer to the boundary.
    \item DW coordinates are independent of the chemical potential value.
    \item At lower temperature $T<T_{cr}$ in the DW configuration the string tension is independent of the temperature.
    \item At higher temperature $T>T_{cr}$ in the horizon configuration the string tension increases monotonically. 
    \item Inclusion of the external magnetic field and spatial anisotropy enhance the string tension in the horizon configuration, i.e. the drag force.
    \item For the fully isotropic case $\nu=1$ and $c_B=0$ the temperature dependence of the string tension $\sigma \propto T^2$ is qualitatively consistent with the lattice results \cite{Bala:2025ilf,Maezawa:2007fc}.
     
\end{itemize}

\begin{figure}[h]
  \centering
  \qquad{$c_B= 0$}\hspace{7pt} \qquad  \quad {$c_B=-0.005$ GeV${}^2$} \hspace{13pt}  {$c_B=-0.05$ GeV${}^2$}\hspace{11pt}  \quad{$c_B=-0.5$ GeV${}^2$}\\
   $\nu=1$\\
  \includegraphics[scale=0.28]{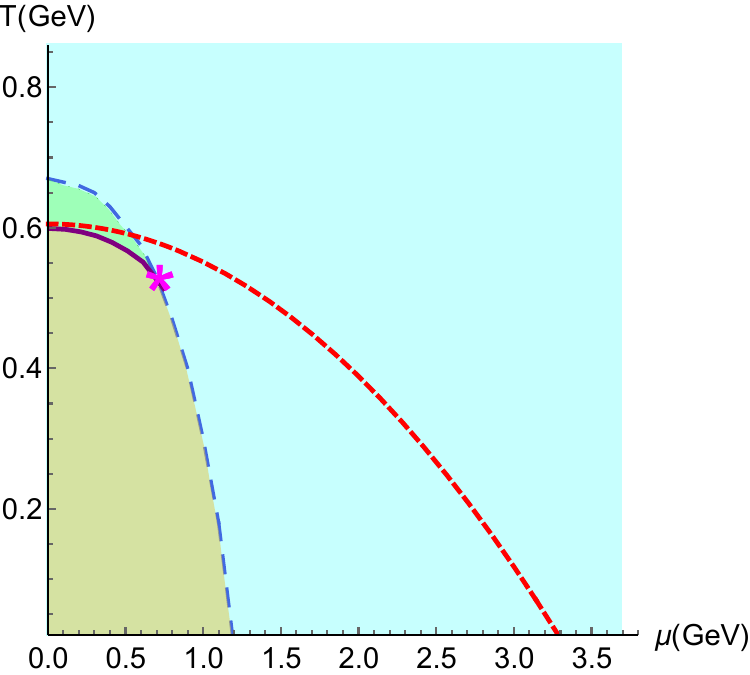}
  \includegraphics[scale=0.28]{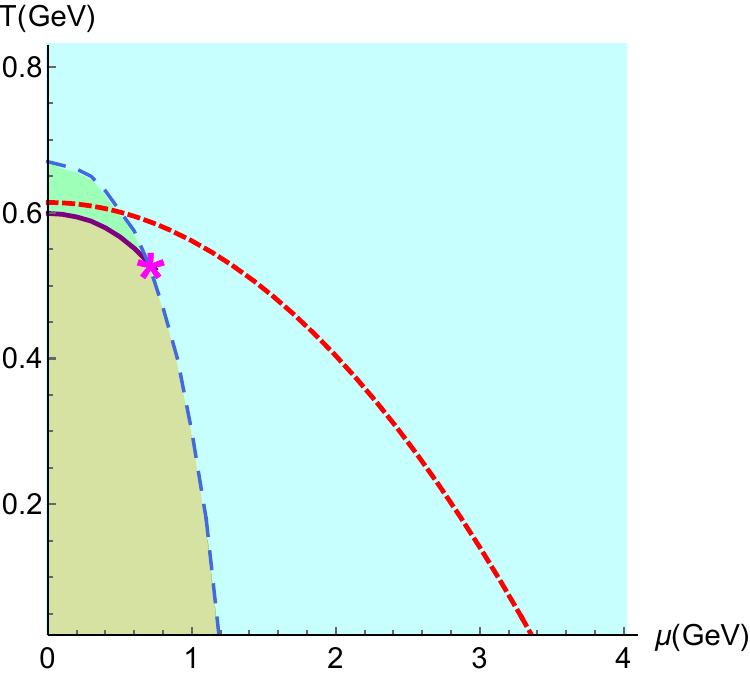}
  \includegraphics[scale=0.28]{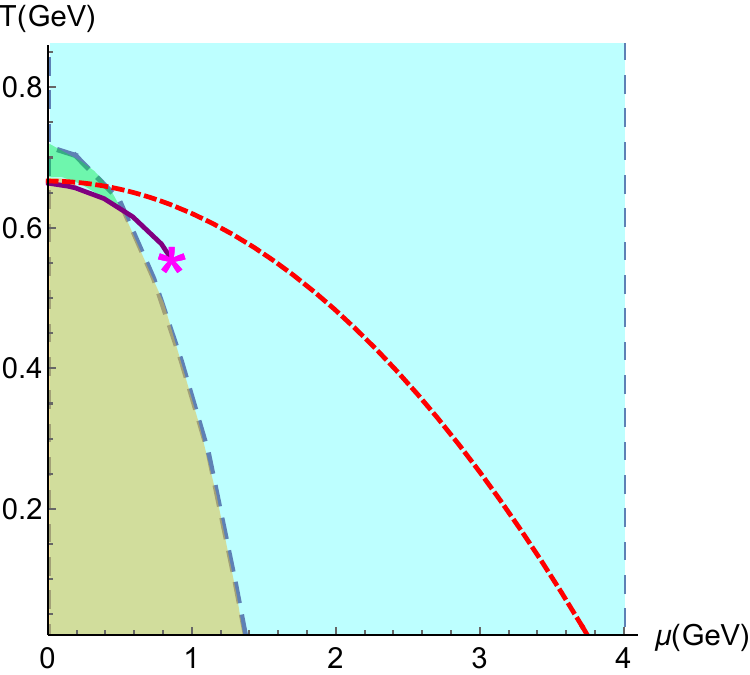}
  \includegraphics[scale=0.28]{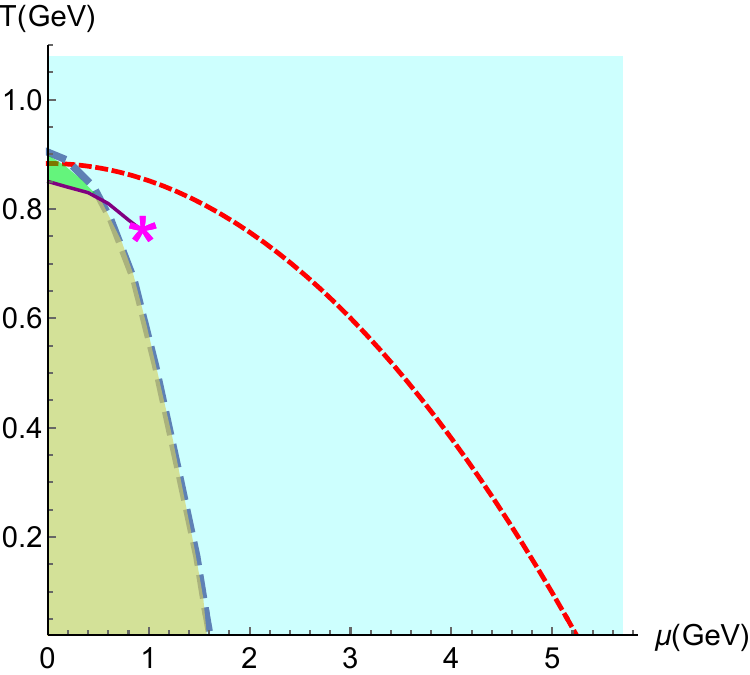} \\
  $\nu=1.5$\\
  \includegraphics[scale=0.28]{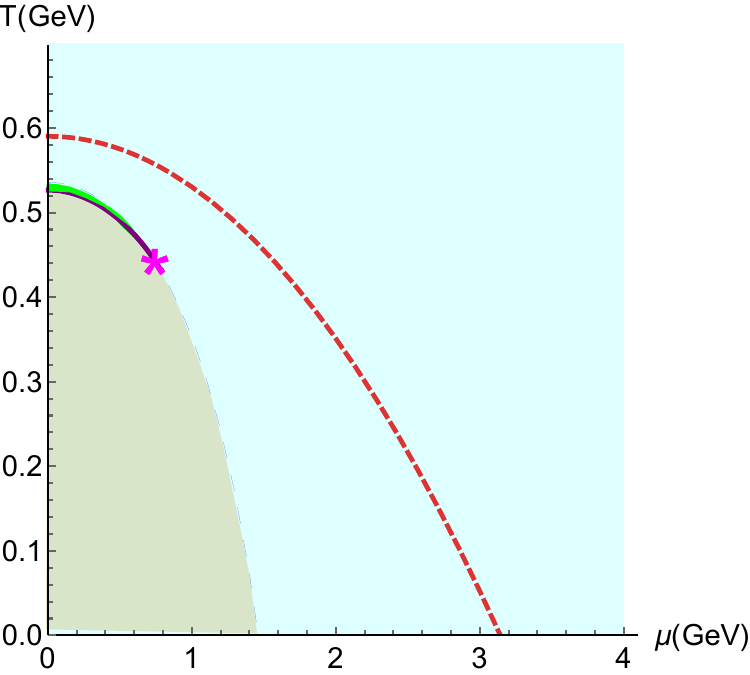}
  \includegraphics[scale=0.28]{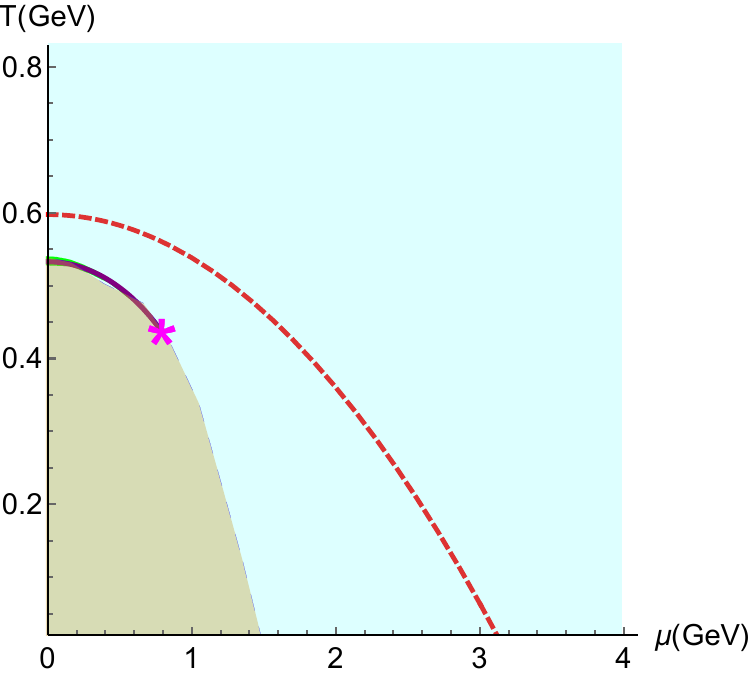}
  \includegraphics[scale=0.28]{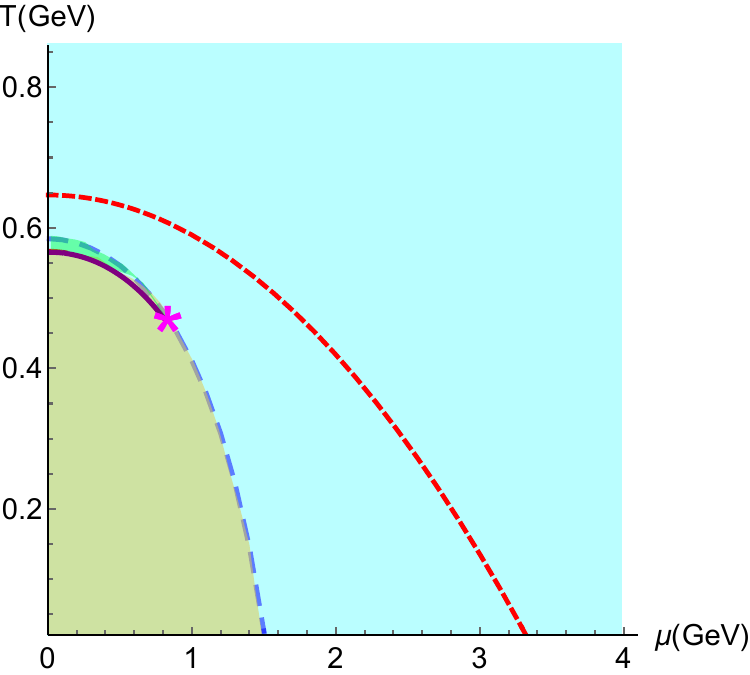}
  \includegraphics[scale=0.28]{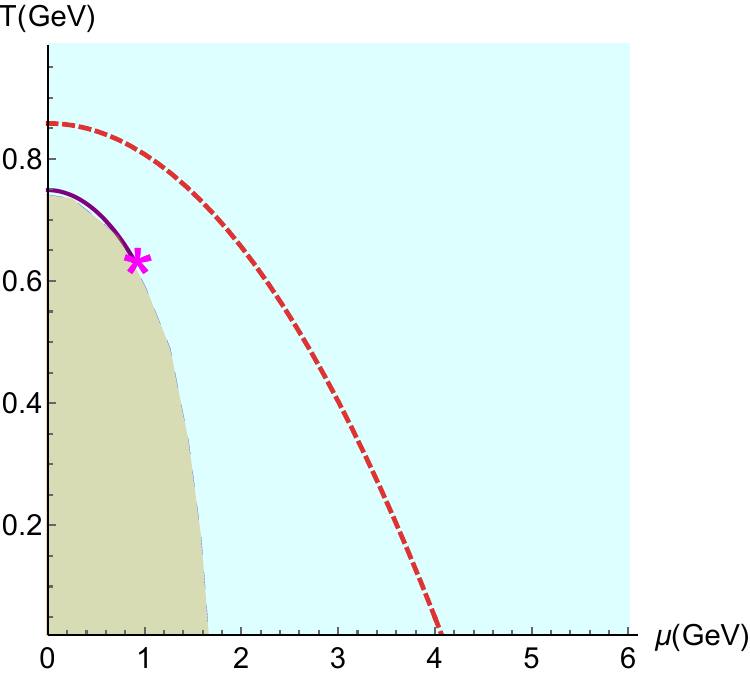}\\
  $\nu=3$\\
  \includegraphics[scale=0.28]{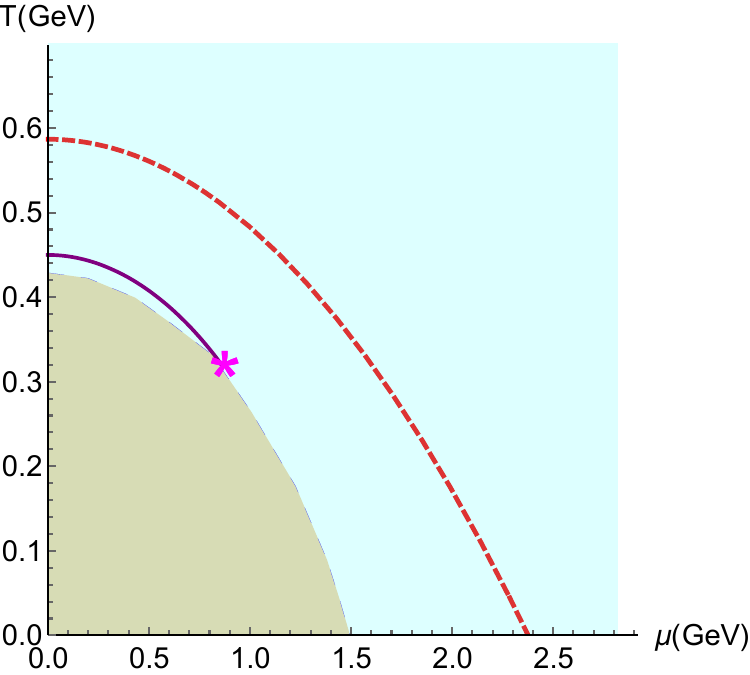}
  \includegraphics[scale=0.28]{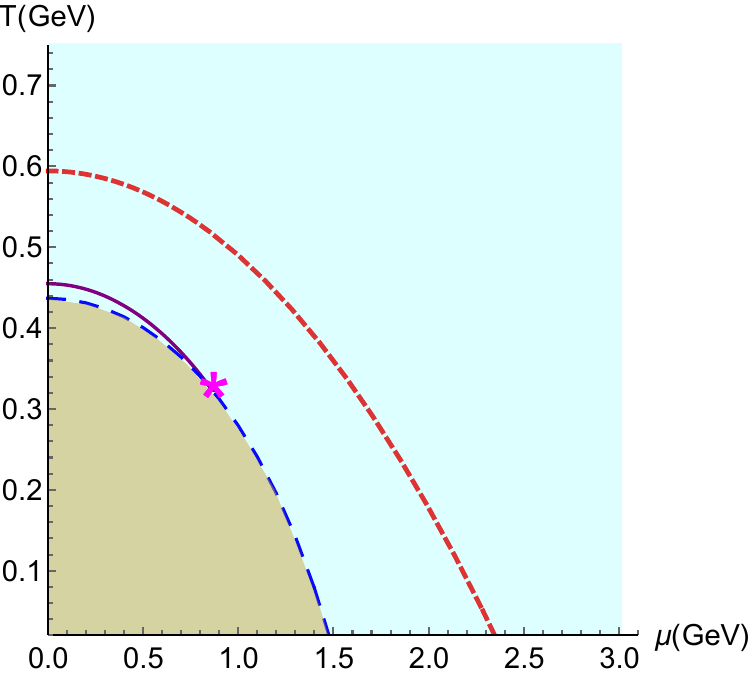}
  \includegraphics[scale=0.28]{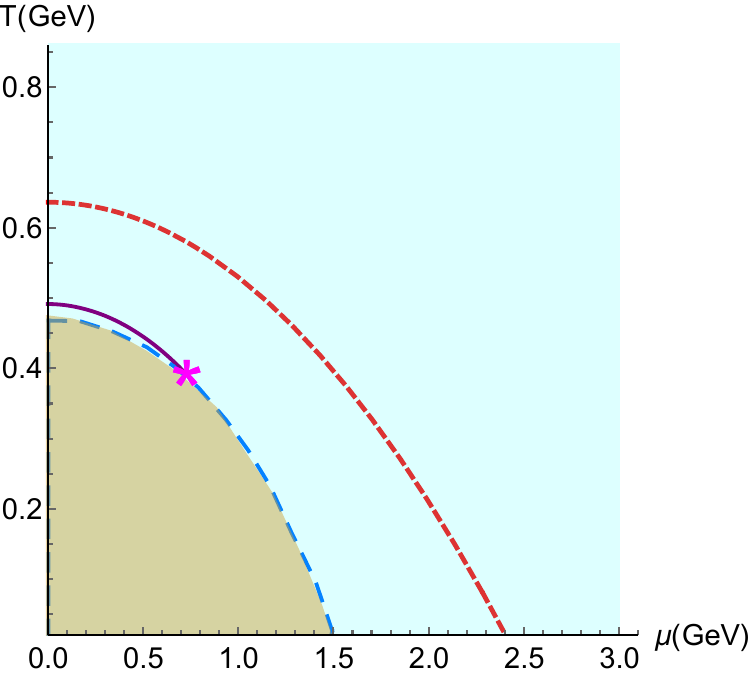}
  \includegraphics[scale=0.28]{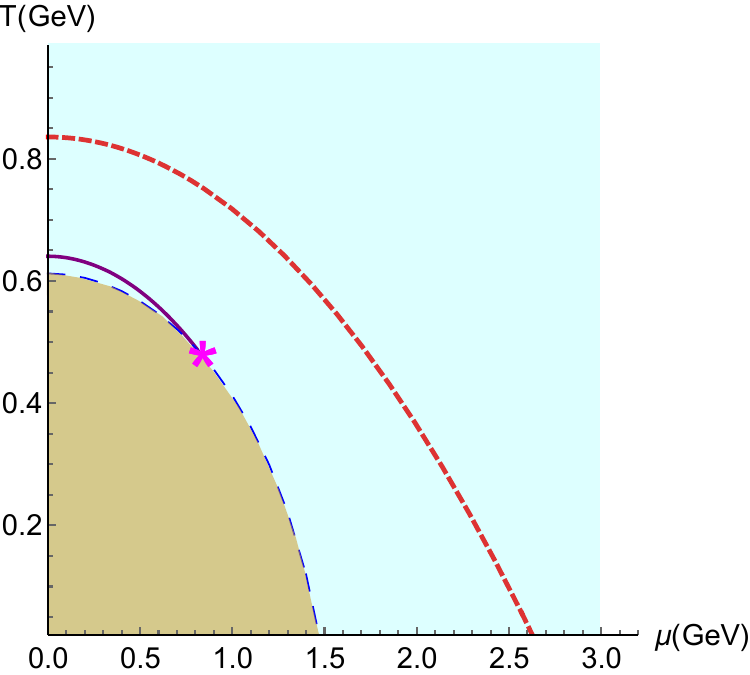} \\
  $\nu=4.5$\\
  \includegraphics[scale=0.28]{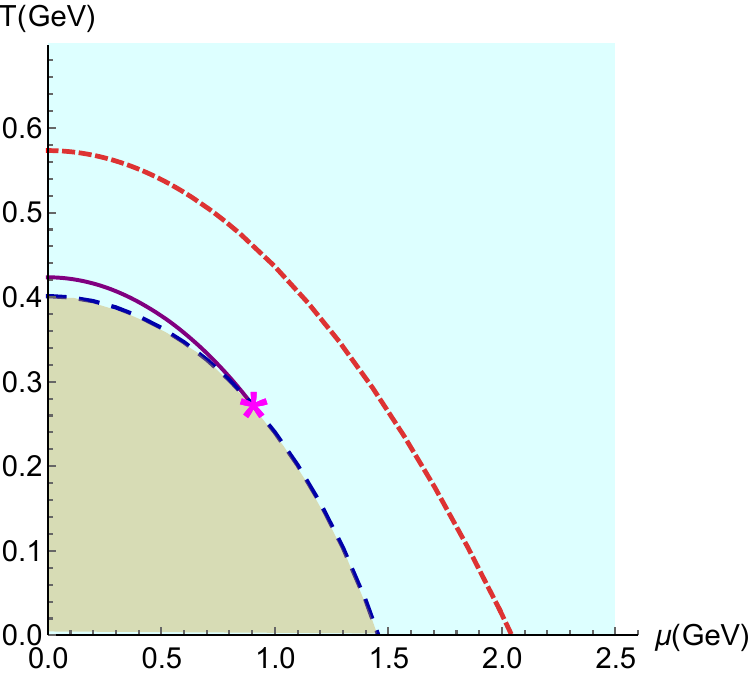}
  \includegraphics[scale=0.28]{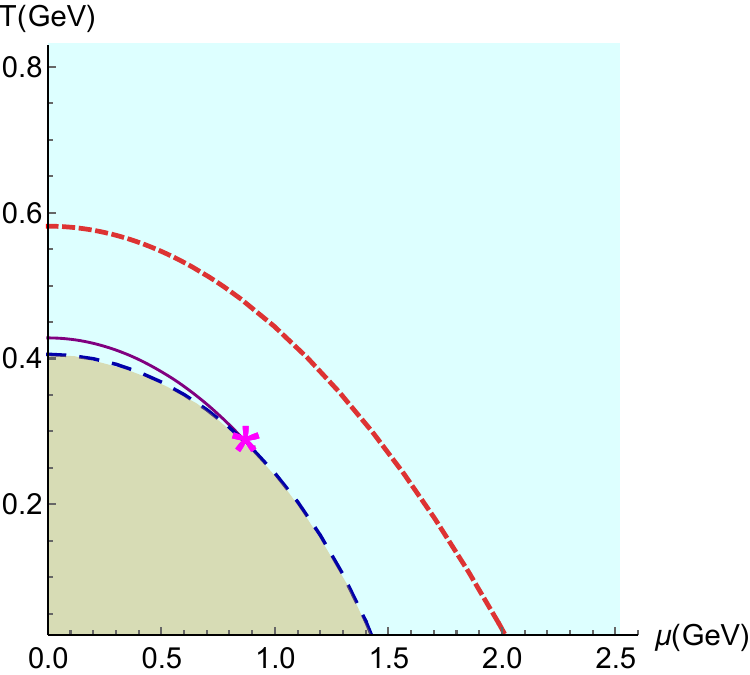}
  \includegraphics[scale=0.28]{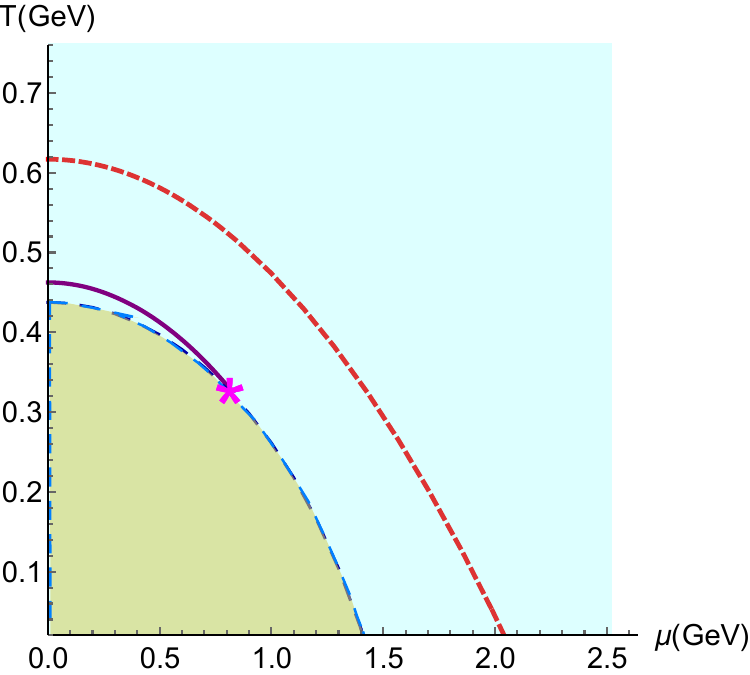}
  \includegraphics[scale=0.28]{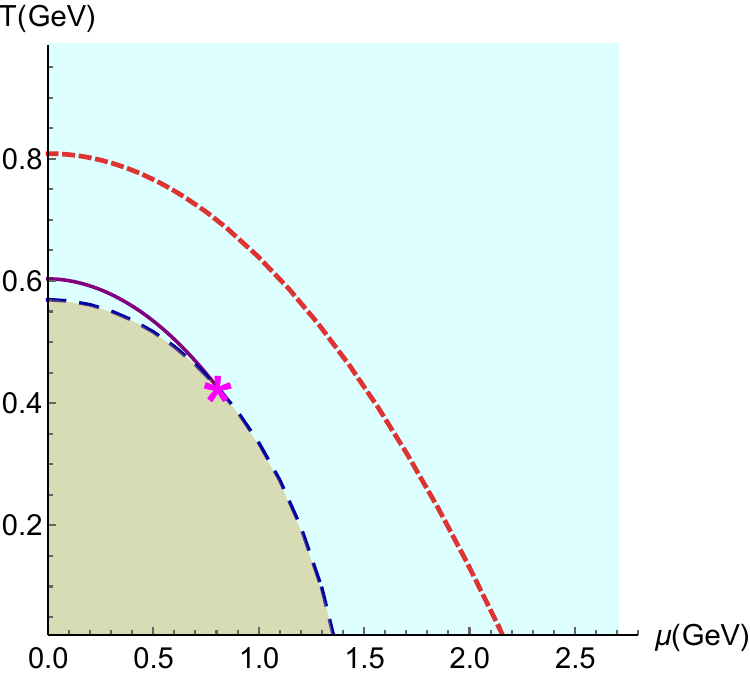}\\
  A\hspace{100pt}B\hspace{100pt}
  C\hspace{100pt}D
  \caption{Phase diagrams in ($\mu$,$T$)-plane for HQ models in columns (A) $c_B=0$, (B) $c_B=-\,0.005$ GeV${}^2$, (C) $c_B=-\,0.05$ GeV${}^2$, and (D) $c_B=-\,0.5$ GeV${}^2$. The first, second, third, and forth raw shows $\nu=1$, $\nu=1.5$, $\nu=3$, and $\nu=4.5$, respectively.
  }\label{fig:PDswl}
\end{figure}

It is very interesting to compare the phase transition between the DW and the horizon configuration, obtained by SWL, with the confinement/deconfinement phase transition line, obtained from temporal WL and the 1st-order phase transition line following from the thermodynamics of the model. Fig.~\ref{fig:PDswl} shows this comparison for different values of magnetic field $c_B=0, -\,0.005, -\,0.05, -\,0.5$ GeV$^2$, and different spatial anisotropies $\nu=1, 1.5, 3, 4.5$. In this comparison the 2nd orientation of SWL, $xY_{2}$, is used, while for temporal WL, $tY_{2}$ orientation is considered. The phase transition line obtained by SWL is depicted by a red dashed line, the confinement/deconfinement phase transition is depicted by a blue dashed line, and the first‑order phase transition is depicted by a solid magenta line. The blue regions represent the QGP phase, the green regions correspond to the quarkyonic phase, and the brown regions correspond to the confined phase.

Fig.~\ref{fig:PDswl} depicts, that at small $\mu$ and for spatially isotropic case $\nu=1$ the transition temperature via SWL is lower than the confinement/deconfinement transition temperature and higher than the 1st order phase transition temperature. However, including anisotropies $\nu=1.5$, $\nu=3$, and $\nu=4.5$, our result shows that for any $\mu$ the transition temperature via SWL is always higher than the confinement/deconfinement and the 1st order transition temperature.

Note that diffusion coefficient  is studied near the critical point in terms of temperature \cite{Andreev:2017bvr}, and also temperature dependence of diffusion coefficient is investigated in \cite{Zhu:2021nbl} near the phase transition in isotropic backgrounds. Studying and extending the diffusion coefficient in the anisotropic background that we introduced in this research will be considered for future research.

Furthermore, it would be interesting to take into account other anisotropic models with different warp factors  $\fb_s(z)$, corresponding to the light quark model \cite{Arefeva:2020byn} or the heavy quark model \cite{Bohra:2020qom}, to study the DW coordinates and string tension. Recently, this background has been used in holographic QCD to investigate the jet quenching parameter \cite{Arefeva:2025uym}.

\newpage
$$\,$$
\newpage

\section*{Acknowledgments}
The work of I.A. and P.S. was performed at the Steklov International Mathematical Center and supported by the Ministry of Science and Higher Education of the Russian Federation (agreement no. 075-15-2025-303). The work of A. H. was started at the Steklov International Mathematical Center and supported by the Ministry of Science and Higher Education of the Russian Federation (Agreement No. 075-15-2022-265). The work of I. A. and P. S. is also  supported by Theoretical Physics and Mathematics Advancement Foundation ``BASIS (grant No. 24-1-1-82-1, grant No. 23-1-4-43-1, respectively).

\appendix

\section*{Appendix}

\section{Solution of EOMs}\label{appendixA}

Varying Lagrangian (\ref{eq:2.01}) over the metric we get the Einstein equations of motion (EOMs):
\be \label{EEOM}
  G_{\mu \nu} =T_{\mu \nu}\,,
\ee
where
\be 
  G_{\mu \nu}=R_{\mu \nu} - \cfrac{1}{2} \, g_{\mu \nu} R\,~,\qquad 
  \cfrac{\delta S_m}{\delta g^{\mu \nu}} = T_{\mu \nu} \sqrt{-g},
\ee
and varying Lagrangian over the fields we get field equations
\bea\label{phiEOM}
  \quad 
  D_\mu D^\mu \phi+V'(\phi)+ \sum _{i=0,1,3}\cfrac{f_i'(\phi)}{4} \ F_{(i)}^2=0,\\
  \label{EMEOM}
  \partial_{\mu}(\sqrt{-g}\,f_i\, F_{(i)}^{\mu \nu})=0.
\eea  
The explicit forms of EOMs are given by
\begin{gather}
  \begin{split}  
    A_t'' + A_t' \left(
    \cfrac{\fb'}{2 \fb} + \cfrac{f_0'}{f_0} 
    + \cfrac{\nu - 2}{\nu z} + c_B z
    \right) = 0, \label{eq:2.17} 
  \end{split} \\
  \begin{split}  
    g'' + g' \left(
    \cfrac{3 \fb'}{2 \fb} - \cfrac{\nu + 2}{\nu z} + c_B z
    \right)
    - \left( \cfrac{z}{L} \right)^2 \cfrac{f_0 \,  (A_t')^2}{\fb}
    - \left(\cfrac{z}{L} \right)^{\frac{2}{\nu}} \cfrac{q_3^2 f_3}{\fb} = 0,
  \end{split} \label{eq:2.18} \\
  \begin{split}  
    \fb'' - \cfrac{3 (\fb')^2}{2 \fb} + \cfrac{2 \fb'}{z}
    - \cfrac{4 \fb}{3 \nu z^2} \left(
      1 - \cfrac{1}{\nu}
      + \left( 1 - \cfrac{3 \nu}{2} \right) c_B z^2
      - \cfrac{\nu c_B^2 z^4}{2}
    \right)
    + \cfrac{\fb \, (\phi')^2}{3} = 0,  
  \end{split} \label{eq:2.19} \\
  \begin{split}
    2 g' \left( 1 - \cfrac{1}{\nu} \right) 
    + 3 g \left( 1 - \cfrac{1}{\nu} \right) \left(
      \cfrac{\fb'}{\fb} - \cfrac{4 \left( \nu + 1 \right)}{3 \nu z}
      + \cfrac{2 c_B z}{3}
    \right)
    + \left( \cfrac{L}{z} \right)^{1-\frac{4}{\nu}} 
    \cfrac{L \, e^{-c_Bz^2} q_1^2 \, f_1}{\fb} = 0,
  \end{split}\label{eq:2.20}
\end{gather}
\begin{gather}
  \begin{split}
    2 g' &\left( 1 - \cfrac{1}{\nu} + c_B z^2 \right) 
    + 3 g \left[ \Big( 1 - \cfrac{1}{\nu} + c_B z^2 \Big) 
      \left(
        \cfrac{\fb'}{\fb} - \cfrac{4}{3 \nu z} + \cfrac{2 c_B z}{3}
      \right)
      - \cfrac{4 \left( \nu - 1 \right)}{3 \nu z} \right] + \\
    + &\left( \cfrac{L}{z} \right)^{1-\frac{4}{\nu}}
    \cfrac{L \, e^{-c_Bz^2} q_1^2 \, f_1}{\fb}
    - \left( \cfrac{z}{L} \right)^{1+\frac{2}{\nu}}
    \cfrac{L \, q_3^2 \, f_3}{\fb} = 0,
  \end{split}\label{eq:2.21} \\ 
  \begin{split}
    \cfrac{\fb''}{\fb} &+ \cfrac{(\fb')^2}{2 \fb^2}
    + \cfrac{3 \fb'}{\fb} \left(
      \cfrac{g'}{2 g} - \cfrac{\nu + 1}{\nu z} + \cfrac{2 c_B z}{3}
    \right)
    - \cfrac{g'}{3 z g} \left( 5 + \cfrac{4}{\nu} - 3 c_B z^2 \right)
    + \\
    &+ \cfrac{8}{3 z^2}
    \left( 1 + \cfrac{3}{2 \nu} + \cfrac{1}{2\nu^2} \right)
    - \cfrac{4 c_B}{3}
    \left( 1 + \cfrac{3}{2 \nu} - \cfrac{c_B z^2}{2} \right) 
    + \cfrac{g''}{3 g} 
    + \cfrac{2}{3} \left( \cfrac{L}{z} \right)^2 \cfrac{\fb V}{g} 
    = 0.  
  \end{split}\label{eq:2.22}
  \end{gather}
For more details, we refer the interested reader to \cite{Arefeva:2023jjh}.

\section{ Arbitrary orientations of Wilson loop}\label{appendixB}

In the holographic approach, the Nambu-Goto action in the background
\eqref{Gbackgr} for the probe string is
\be
  S = \frac{1}{2 \pi \alpha'} \int d\xi^{1} d\xi^{2} \sqrt{- \det
    h_{\alpha\beta}},
\ee
where the induced metric $h_{\alpha\beta}$ is given by
\be
  h_{\alpha\beta} = G_{\mu\nu} \partial_{\alpha}
  X^{\mu} \partial_{\beta} X^{\nu}
\ee
and is the orientation dependent. Here we use the same orientation to parametrize the rectangular Wilson loop as was considered to calculate the entanglement entropy using the rectangular parallelepiped in \cite{Arefeva:2020uec}.
 
\begin{figure}[h!]
  \centering
  \includegraphics[scale=0.57]{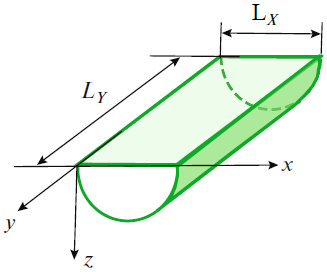}\qquad \qquad
  \includegraphics[scale=0.82]{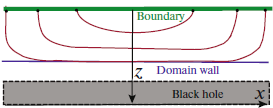} \\
  A \hspace{80 mm} B \\
  \includegraphics[scale=0.8]{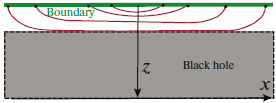} \\
  C
  \caption{(A) Spatial Wilson loop and its world-sheet. (B) Describing the DW configuration and the horizon configuration. (C) After the phase transition from DW configuration to horizon configuration \cite{Arefeva:2020vhf}; $[T]=[\mu] =$ GeV. 
  }
  \label{Fig:SWLs}
\end{figure}

\begin{figure}[h!]
  \centering
  \includegraphics[scale=0.57]{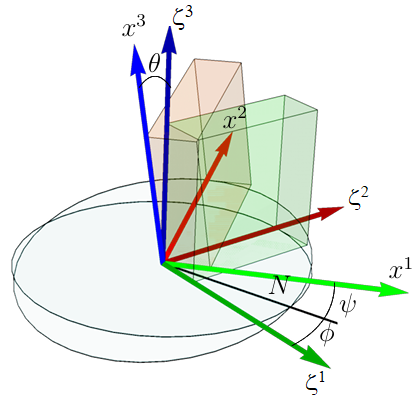}\\
  \caption{Rotating the green rectangular cube by the Euler angles $(\phi,\theta,\psi)$, we get the pink cube oriented along the axes $(x^1,x^2,x^3)$. The node line is denoted by $N$, and ($\zeta^1$,$\zeta^2$,$\zeta^3$) are axes before rotation \cite{Arefeva:2020uec}. 
  }
  \label{Fig:rotation}
\end{figure}

To parameterize the world-sheet of string, we use the rotation matrix $M(\phi, \theta,\psi )$ in a 3-dimensional space with the representation
\bea
x^i&=& \sum _{j=1,2,3} a_{ij}(\phi, \theta,\psi )\,\zeta^j,
\qquad i=1,2,3\,, \label{Emb}
\eea
where $a_{ij}$ in terms of the Euler angles $\phi, \theta,\psi $ are the components of the rotation matrix
\bea
  M(\phi, \theta,\psi )={\begin{pmatrix}a_{11}(\phi, \theta,\psi ) &\
    a_{12}(\phi, \theta,\psi ) &\ a_{13}(\phi, \theta,\psi ) \\
    a_{21}(\phi, \theta,\psi ) &\ a_{22}(\phi, \theta,\psi ) &\
    a_{23}(\phi, \theta,\psi ) \\
    a_{31}(\phi, \theta,\psi ) &\ a_{32}(\phi, \theta,\psi ) &\
    a_{33}(\phi, \theta,\psi ) 
  \end{pmatrix}},
  \label{Eulmat}
\eea
where
\bea
  \begin{array}{lll}
    &a_{11}(\phi,\theta,\psi) =\cos \phi \cos\psi -\cos \theta \sin \phi \sin\psi,\,\,\,  \\
    &a_{12}(\phi,\theta,\psi) =-\cos \psi \sin\phi -\cos\phi \cos \theta \sin\psi,\,\,\,\, \\
    &a_{13}(\phi,\theta,\psi) =\sin\theta \sin \psi, \\
    &a_{21}(\phi,\theta,\psi) =\cos \theta \cos \psi \sin\phi+\cos \phi \sin\psi, \\
    &a_{22}(\phi,\theta,\psi) =\cos \phi \cos \theta \cos\psi-\sin \phi \sin\psi,\\ 
    &a_{23}(\phi,\theta,\psi) =-\cos \psi \sin \theta, \\
    &a_{31}(\phi,\theta,\psi) =\sin\phi \sin \theta, \\
    &a_{32}(\phi,\theta,\psi) =\cos\phi\sin \theta, \\
    &a_{33}(\phi,\theta,\psi) =\cos \theta.
 \end{array} \label{EP}
\eea
Here $\phi$ is the angle between the $\zeta^1$-axis and the node line (N),
$\theta$ is the angle between the $\zeta^3$ and $x^3$-axes, and $\psi$ is the
angle between the node line N and $x^1$-axis; see Fig.~\ref{Fig:rotation}.

We use the following embedding to describe the 2-dimensional world-sheet of string in the 5-dimensional bulk:
\bea \label{GPar}
  X^0(\xi) &=& \const,\nn \\
  X^i(\xi) &=& \sum _{\alpha=1,2} a_{i\alpha }(\phi, \theta,\psi )\,\xi^\alpha,
  \quad i=1,2,3, \quad \alpha=1,2, \label{Emb} \\
  X^4(\xi) &=& z(\xi^1),\nn 
\eea
where $X^i$ are spatial coordinates.

The line element of the induced metric on the world-sheet is
\bea
  ds_{ws}^2 &=& g_{\alpha \beta }\,d\xi^\alpha d\xi^\beta, \qquad
  \alpha,\beta=1,2.
\eea
After applying the embedding relations \eqref{Emb} the induced metric can be written as
\bea
  ds_{ws}^2 &=& \frac{L^2 \fb_s(z)}{z^2} \left(
    \sum_{i=1,2,3}\fg_i(z)d(x^i)^2+\frac{d(x^4)^2}{g}
  \right)\nn\\
  &=&\frac{L^2 \fb_s(z)}{z^2}\left( 
    \sum_{i=1,2,3}\fg_i(z)\,\Big(
      \sum_{j=1,2}a_{ij}(\phi, \theta,\psi ) \, d\xi^{j}
    \Big)^2 
    +z'^{2} \ \frac{d(\xi^1)^2}{g(z)}
  \right).
\eea
We have
\bea
  g_{\alpha\beta} &=& \frac{L^2 \fb_s(z)}{z^2} \, 
  \bar{g}_{\alpha\beta},\\
  \bar g_{11}(z,\phi, \theta,\psi) &=& \fg_{1}a_{11}^2+\fg_{2}a_{21}^2+\fg_{3}a_{31}^2+\frac{z'^{2}}{g},\nn
  \\
  \bar g_{22}(z,\phi, \theta,\psi) &=& \fg_{1}a_{12}^2+\fg_{2}a_{22}^2+\fg_{3}a_{32}^2,\nn\\
  \bar g_{12}(z,\phi,\theta,\psi) &=& \fg_{1}a_{11} a_{12} + \fg_{2}a_{21}a_{22} + \fg_{3}a_{31}a_{32},\nn \\
  \bar g_{21} &=& \bar g_{12}.
  \label{barg}
\eea
The determinant of the induced metric is given by
\bea
  \det g_{\alpha\beta} &=& \left(\frac{L^2 \fb_s}{z^2}\right)^2 \Biggl(
    \left(
      \fg_{1}a_{11}^2 + \fg_{2}a_{21}^2 + \fg_{3}a_{31}^2 + \frac{z'^{2}}{g}
    \right) 
    (\fg_{1}a_{12}^2 + \fg_{2}a_{22}^2 + \fg_{3}a_{32}^2) - \nn \\  
    &-&(\fg_{1}a_{11} a_{12} + \fg_{2}a_{21}a_{22} +
    \fg_{3}a_{31}a_{32})^2 
  \Biggr).
\eea

The Nambu-Goto action for SWL is given by
\be
  {\cal S}_{SWL} = \frac{1}{2\pi \alpha'}\int _{{\cal W}} \left(\frac{L^2 \fb_s}{z^2}\right)
  \sqrt{\left(\fg_{1}\fg_{2}a_{33}^2 + \fg_{1}\fg_{3}a_{23}^2 +
    \fg_{2}\fg_{3}a_{13}^2 +
    \frac{z'^{2}}{g} \, \bar g_{22}\right)} \ d\xi^{1}d\xi^{2},
  \label{S_swl}
\ee
where the integration is over the world-sheet~${\cal W}$ components and $g$, $\fg_1$, $\fg_2$, $\fg_3$ are functions of $z$ and $\bar
g_{22}$, $\bar g_{33}$, $\bar g_{23}$ are functions of $z$ and Euler angles. The effective potential can be obtained as
\be
  {\cal V}(z(\xi)) = \frac{1}{2\pi \alpha'}\left(\frac{L^2 \fb_s}{z^2}\right) \,
  \sqrt{\fg_{1}\fg_{2}a_{33}^2 + \fg_{1}\fg_{3}a_{23}^2 +
  \fg_{2}\fg_{3}a_{13}^2}. \label{V_swl}
\ee

Note that the action \eqref{S_swl} and the effective potential \eqref{V_swl} depend on the angles of rotation and anisotropy.

\section{String tension at DW and horizon configurations}\label{appendixC}

Using equation \eqref{EfPot}, the first integral related to the action \eqref{BI} is given by
\bea
\label{FI}
  \frac{M(z(\xi)){\cal F}(z(\xi))}{\sqrt{{\cal F}(z(\xi))+(z'(\xi))^2}}={\cal I}. 
\eea
At the turning point $z=z_*$, where  the minimal surface has the closest distance to the horizon, we have ${z'(\xi) =0}$. Then, from \eqref{FI} one can obtain
\be \label{star}
  M(z_{*})\sqrt{{\cal F}(z_{*})}={\cal I}.
\ee
Utilizing \eqref{FI} and \eqref{star} to find $z'$, the length $\ell$ and the action ${\cal S}$ \eqref{BI} are given by general formulas
\bea\label{ell1}
  \frac\ell2 &=&
  \int_{\epsilon}^{z_*}\frac{1}{\sqrt{{\cal F}(z)}} \
  \frac{dz}{\sqrt{\frac{{\cal V}^2(z)}{{\cal V}^2(z_*)} -1 }},\quad \\
  \frac{{\cal S}}{2} &=&
  \int_\epsilon^{z_*}\frac{M(z)dz}{\sqrt{1-\frac{{\cal V}^2(z_*)}{{\cal V}^2(z)}}}\,,
  \label{calS1}
\eea
%
%
%
where $\epsilon$ is the regulator in  gravity theory and corresponds to the UV cut-off in gauge theory. Taking into account the limit $\ell\to \infty$, we have two categories.
\begin{itemize}
\item There is a stationary point of ${\cal V}(z)$ that
  \be
    {\cal V}^\prime\Big|_{z_{DW}} = 0,
  \ee
  where $z=z_{DW}$ is a DW coordinate that can be considered equal to the turning point $z_*=z_{DW}$. Near the turning point we have
  \be
    \sqrt{\frac{{\cal V}^2(z)}{{\cal V}^2(z_{DW})}-1} = \sqrt{\frac{
      {\cal V}''(z_{DW})}{{\cal V}(z_{DW})}}\,(z-z_*) + {\cal O}\bigl((z-z_*)^2\bigr).
  \ee
  The using \eqref{ell1} and \eqref{calS1} one gets
  \bea
    \ell &\underset{z\to z_*}{\sim }& \frac{1}{\sqrt{{\cal F}(z_{DW})}} \,
    \sqrt{\frac{{\cal V}(z_{DW})}{{\cal V}''(z_{DW})}} \, \log (z-z_*), \\
    \cS&\underset{z\to z_*}{\sim}& M(z_{DW}) \, \sqrt{\frac{{\cal V}(z_{DW})}{{\cal V}''(z_{DW})}} \log (z-z_*).
  \eea
  Therefore
  \bea
    \cS&\sim&  M(z_{DW})~\sqrt{{\cal F}(z_{DW})} ~ \ell.
  \eea
  Then we have
  \bea
    \sigma _{DW} &=& M(z_{DW}) \,\sqrt{{\cal F}(z_{DW})}, \label{ten}
  \eea
  where $\sigma _{DW}$ is the spatial string tension in the DW configuration.
  \item There is no stationary point in the region $0 < z < z_h$ for ${\cal V}(z)$. We suppose the near horizon expansion
  \be
    {\cal F}(z) = \fF(z_h) (z-z_h)+{\cal O}((z-z_h)^2).
  \ee
  In this configuration, the string stretches on the horizon and we take $z_*=z_h$. We propose two items: 
  \begin{itemize}
  \item if $M(z_h)\neq \infty$, we have
  \bea
    \ell&\to& \infty,\\
    S&\to&0;
  \eea
  \item if $M(z) \underset{z\to z_h}{\to}\infty$ as
  \be
    M(z)\underset{z\sim z_h}{\sim} \frac{\fm (z_h)}{\sqrt{z-z_h}};
  \ee
  then, using \eqref{ell1} and \eqref{calS1} we have
  \bea
    \ell &\underset{z\to z_h}{\sim }& \frac{1}{\sqrt{\fF (z_h)}} \
    \frac{1}{\sqrt{\frac{2 {\cal V}'(z_h)}{{\cal V}(z_h)}}} \, \log
    (z-z_h), \\
    \cS&\underset{z\to z_h}{\sim}& \fm(z_h) \,
    \frac{1}{\sqrt{\frac{2{\cal V}'(z_h)}{{\cal V}(z_h)}}} \, \log
    (z-z_h),
  \eea
  and therefore
  \be
    \sigma_{z_h}=\fm(z_h)\,\sqrt{\fF(z_h)},
  \ee
  \end{itemize}
  where $\sigma _{z_h}$ is the spatial string tension in the horizon  configuration.  
\end{itemize}

\end{document}